\documentclass[12pt,a4paper]{article}
\usepackage[parfill]{parskip}    % Activate to begin paragraphs with an empty line rather than an indent
\usepackage{graphicx}
\usepackage{amssymb}
\usepackage{amsmath}
\usepackage{fancyhdr}
\usepackage{flafter}
\usepackage{multicol}
\usepackage{setspace}
\usepackage{subfig}
\usepackage{lscape}
\title{Adiabatic Quantum Computing}
\author{Sebastian Pinski}
\begin{document}

%Title Page
\thispagestyle{empty}
\vspace*{1cm}
\begin{center}
{\Large\bf Adiabatic Quantum Computing\\}
\vspace{2cm}
{\large
MPhys Research Project\\
{\small by}\\
Sebastian Pinski\\
{\small supervised by}\\
Dr. John Samson\\
\vspace{0.5cm}
Department of Physics \\
Loughborough University}
\end{center}
\vspace*{2cm}

%Abstract
\begin{abstract}
This document contains a thorough introduction to the subject of conventional Quantum Computing, and the unconventional method known as Adiabatic Quantum Computation. Testing of a specifically written general simulation of the Adiabatic Quantum Computer with 1, 2 and 3 qubits gives realistic results. Finally the application of an NP hard problem (Max Independent Set) to 4 and 8 qubit cases yields not only the cardinality of Max Independent Sets but also a visual representation. Leading to the development of a general algorithm for the Max Independent Set Problem that is executable on an Adiabatic Quantum Computer.
\end{abstract}

\newpage
\pagenumbering{roman}
\setcounter{page}{1}
\thispagestyle{empty}
\tableofcontents

\newpage
\thispagestyle{empty}
\listoffigures
\listoftables

\newpage
\pagenumbering{arabic}
\setcounter{page}{1}

\section{Introduction} \label{sect:Intro}
Adiabatic Quantum Computing (AQC) is a relatively new subject in the world of quantum computing, let alone Physics. Inspiration for this project has come from recent controversy around D-Wave Systems in British Columbia, Canada, who claim to have built a working AQC which is now commercially available and hope to be distributing a 1024 qubit chip by the end of 2008. Their 16 qubit chip was demonstrated online for the Supercomputing 2007 conference within which a few small problems were solved; although the explanations that journalists and critics received were minimal and very little was divulged in the question and answer session. This `unconvincing' demonstration\footnote{http://www.scottaaronson.com/blog/?p=225} has caused physicists and computer scientists to hit back at D-Wave. The aim of this project is to give an introduction to the historic advances in classical and quantum computing and to explore the methods of AQC.

Initially Section \ref{subsect:CC} contains a short introduction to classical computation with very little concentration on the methods of implementing classical algorithms. The main topic in this section is the complexity theory used to classify problems which in essence leads to problems that cannot be efficiently solved on classical computers, essentially justifying the need to develop quantum computers.

Section \ref{subsect:QC} is a full introduction to the method of conventional quantum computing with explanations of basic principles including the DiVincenzo checklist and leading onto the implementation and failures of this method. Section \ref{sect:AQC} introduces the new unconventional method of quantum computing (AQC) with a thorough introduction of the problem that is extensively investigated in Section \ref{sect:Results}.

A brief summary of results with a final conclusion is available in Section \ref{sect:Con}. Suggestions of further research topics for future students are available in Section \ref{sect:Further} and finally, the appendix containing only the program code of the simulation used throughout this project is viewable in Section \ref{sect:App}.

\section{Classical Computation} \label{subsect:CC}

\begin{figure} 
\centering 
\includegraphics[width=\textwidth]{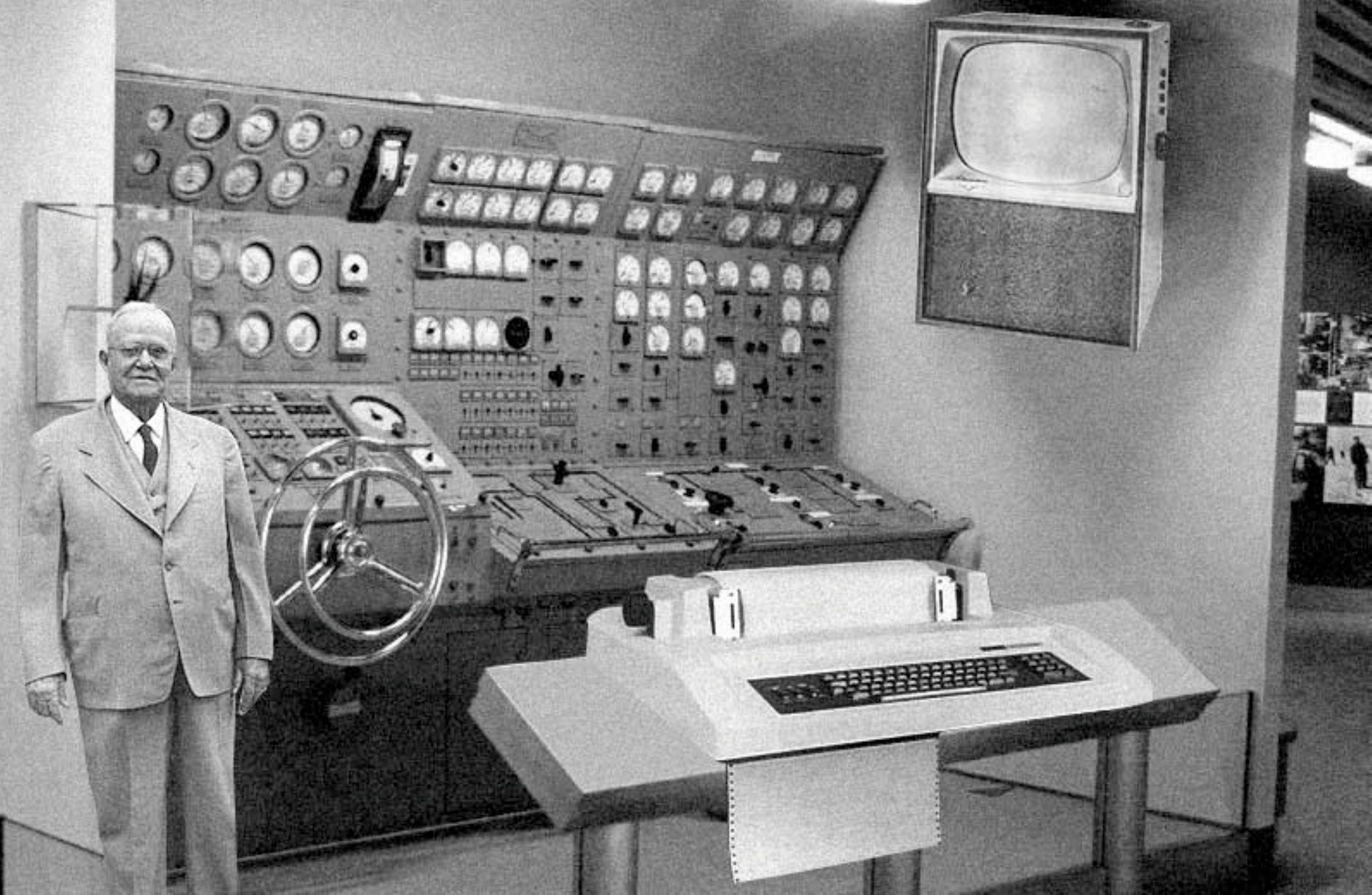} 
\begin{quote}
\it 
`Scientists from the RAND corporation have created this model to illustrate how a ``home computer" could look in the year 2004. However the needed technology will not be economically feasible for the average home. Also the scientists readily admit that the computer will require not yet invented technology to actually work, but 50 years from now scientific progress is expected to solve these problems. With teletype interface and the Fortran language, it will be easy to use.'
\end{quote}
\textbf{\caption{Mock-Up of How a Computer Would Look in 2004 from 1954\label{fig:MockUp} }}
\end{figure}

Computers have come a long way since the very beginning. There have been countless physical realisations of the computer itself, moving from gears to relays to valves to transistors and finally to integrated circuits. All of these advances in technology have been extremely rapid and computers of this day and age are far more advanced than the early veterans of computing could have ever imagined. See Figure \ref{fig:MockUp} for an eye-opening example.

The perception of the slow and steady progression of computing evolution held until the publication of one paper in 1965, written by Intel co-founder Gordon E. Moore\cite{Moore:1965}. The general outcome of this paper was what is known now as Moore's Law which describes the trend in the history of computer hardware, that:
\begin{quote} `the number of transistors that can be inexpensively placed on an integrated circuit is doubling approximately every two years.'\end{quote} Although now the figure of 18 months is more widely accepted, the law looks set to continue for at least another decade.

Even with the advances in technology and the continuation of Moore's Law, the only real change to computing is the greatly increased speed of computation. The capability of the machines has yet to be altered.

\subsection{Complexity Classes} \label{subsubsect:Complex}

Even with the greatly increased speed of computation, many problems are still unsolvable on the computers used today. For a computer to solve a problem, it must follow a specific set of instructions that can be computationally applied to yield the solution to any given instance of that problem. This is known as an algorithm. The more efficient the algorithm the faster the computer can solve that problem. Problems are categorised into classes of complexity, the class of complexity is determined by taking the most efficient known algorithm for a specific problem and calculating the number of computational steps needed to solve the problem as a function of the input size. Depending on how the number of computational steps relates to the size of the problem, the complexity class is found. This method is used, as to a computer scientist the crucial information about a problem is how quickly the time needed to solve it grows as the problem size increases.

Below is a short introduction to various classical complexity classes, with official classifications taken from the National Institute of Standards and Technology\footnote{http://www.nist.gov}:

\paragraph{NP Problems} \label{subsect:NP}
\begin{quote}
``The complexity class of decision problems for which answers can be checked by an algorithm whose run time is polynomial in the size of the input. Note that this does not require or imply that an answer can be found quickly, only that any claimed solution can be verified quickly. ``NP'' is the class that a Non-deterministic Turing machine accepts in Polynomial time.''
\end{quote}

The class NP encapsulates all P and NP-Complete problems. The simple explanation for the problems contained in the NP class is that they can be checked in polynomial time. Meaning that all problems in the sub-classes of NP can also be checked in polynomial time.

\paragraph{P Problems} \label{subsect:P}
\begin{quote}
``The complexity class of languages that can be accepted by a deterministic Turing machine in polynomial time.''
\end{quote}

Problems in the class `P' are the type that a classical computer can solve in a `reasonable' amount of time. The `P' stands for polynomial time, which means that the time taken to solve the problem with the most efficient algorithm is the size of the problem raised to some fixed power. One of the simplest polynomial time algorithms is multiplication, where if we take two $n$-digit numbers the amount of time required to multiply those numbers is the total number of digits squared ($n^2$).

\paragraph{NP-Complete Problems} \label{subsect:NPcomplete}
\begin{quote}
``The complexity class of decision problems for which answers can be checked for correctness, given a certificate, by an algorithm whose run time is polynomial in the size of the input (that is, it is NP) and no other NP problem is more than a polynomial factor harder. Informally, a problem is NP-Complete if answers can be verified quickly, and a quick algorithm to solve this problem can be used to solve all other NP problems quickly.''
\end{quote}

Problems in class NP-Complete are very different to NP problems, whereby every known algorithm for their solution will take an amount of time that increases exponentially with the problem size. Therefore solutions to these problems can be checked in polynomial time, but not necessarily solved in polynomial time. That is not to say that a polynomial time algorithm does not exist for these problems, it may yet be found, although for now this is the most widely accepted conjecture. The theory of NP completeness was developed by Cook, Karp and Levin in the 1970s\cite{SA:2008}, who have theorised that if an efficient algorithm for any one of these problems were to be found, it could be adapted to solve all other NP problems\cite{Farhi:2001}. The implication of this is phenomenal, as developing a polynomial time algorithm for just one of these problems is enough to relieve many mathematicians of many years of anguish.

A subset of NP-Complete is NP-Hard, these are NP-Complete problems in optimisation form. For example the travelling salesman problem is to find a route that visits all possible towns on a travelling salesman's list that induces a mileage less than some nominal value, which is NP-Complete. The optimisation form of this problem is to find the shortest route possible whilst visiting all towns.
\newline
\newline
One must stress that the explanations of complexity classes above are just conjecture. Results to date indicate that $NP\neq P$ although this has never been proven true or false, therefore the question still remains open. In the future a classical polynomial time algorithm may be found for an NP-Complete problem, this would indicate that all NP problems are solvable in polynomial time. 

\section{Quantum Computation} \label{subsect:QC}

The idea of quantum computing itself was loosely created by renowned physicist Richard Feynman in 1982\cite{Brooks:1999}. His idea of simulating quantum mechanical problems with other more easily accessible quantum devices is the basis of this new branch of computing. Although initially these ideas did not spur much excitement between physicists in the form of actual quantum computation, although some research groups started investigating simple physical quantum systems that could possibly be used for this type of application.

It wasn't until 1985 when David Deutsch published his theoretical paper on the universal quantum computer and what it might achieve that physicists started to take the subject more seriously. Deutsch indicated the possibility of simultaneous operations due to quantum superposition and the computational efficiency of such a facility. He went on to develop his own algorithm, known as the Deutsch Algorithm that in simple terms can determine whether a (quantum) coin is bias with a single toss. Though it had little practical use, the race was now on to find a practical algorithm for use on a quantum computer.

Again, interest in the subject soon dwindled until 1994 when Peter Shor devised an algorithm for efficient factorisation on a quantum computer that was published in 1996\cite{Shor:1996}. This regenerated interest in the subject and ultimately led to the the next major discovery in 1997 by Lov Grover who devised an algorithm that can search a database containing $N$ entries with approximately $\sqrt N$ steps\cite{Grover:2008}.

Now thousands of computer scientists and physicists are attempting to develop new algorithms for use on physical quantum computers that are being developed by many companies all over the world. IBM are working on their NMR quantum computer, the Innsbruck group with their Ion-trap quantum computer, D-Wave systems with their superconducting qubits realisation, and many more.

\subsection{How Quantum Computing Differs}\label{subsubsect:Differs}

New technologies will always allow classical computers to advance, although whether it is done by cramming more transistors onto a silicon chip or by increasing the clock-speed of processors the same fundamental fact that only the speed of computation is increasing still holds. The major fundamental difference with quantum computers is that the classical bit is replaced by the qubit ($\mathbf{qu}antum \mathbf{bit}$). This has created a variety of new possibilities. 

As a crude example, 4 classical bits can be re-arranged into 16 ($2^n$, $n$ being the number of bits) different combinations (e.g. 0001, 0010, 0011, etc.). For a classical computer, these 4 bits combined can only be in one of the 16 possible states at any given time. The difference for a quantum computer containing 4 qubits is that the qubits can be in a linear super-position off all sixteen possible states at any one time. Now say that both the classical and quantum computer perform a computation whereby all of these combinations must be explored. The classical computer will initialise the bits and perform the required operation to each individual state. Whereas the quantum computer can prepare the qubits in a linear superposition of all 16 states and perform the operation only once.

At first glance this does not appear to be terribly impressive, until you realise the implications of how many more parallel operations can be achieved in one clock cycle by simply adding more qubits. Just by adding a single additional qubit to any system, the number of possible states for that system doubles as does the number of possible simultaneous operations. For a classical computer to match this computational power, for every bit that is added, you would have to double the number of physical processors within the computer.

\subsection{Quantum Complexity Class} \label{subsubsect:QComp}

\begin{figure} 
\centering 
\includegraphics[width=\textwidth]{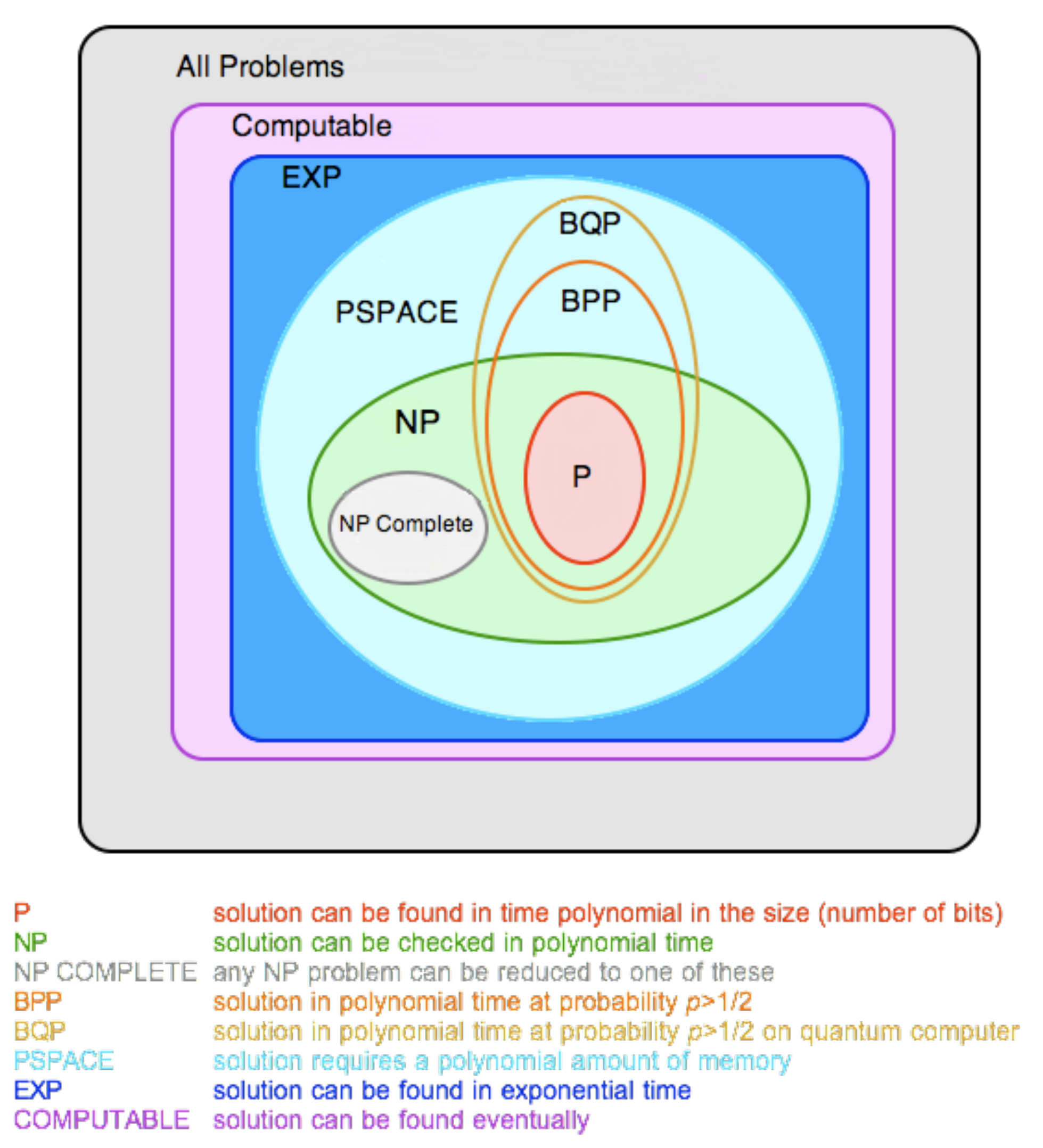} 
\textbf{\caption{Venn Diagram of Complexity Classes in Currently Believed Structure\label{fig:Comp} }}
\end{figure}

The quantum complexity class is known as BQP (bounded error, quantum, polynomial time), this includes all problems that can be solved with a probability of more than half in polynomial time on a quantum computer. Where BQP fits in with other complexity classes is best viewed pictorially in Figure \ref{fig:Comp}. It is widely believed that quantum computers will only prove useful if there are important problems in BQP that are not in P\cite{Samson:2008}. An example of a problem that has theoretically already entered this class is prime factorisation by Shor's Algorithm. Interestingly, this is probably the problem that is encouraging the development of quantum computers, as having the ability to perform fast prime factorisation will enable the deciphering of the RSA encryption used for the majority of monetary transactions across the internet.

\subsection{The Qubit}\label{subsubsect:Qubit}

The qubit can be of many physical forms. Generally the qubit is a two state (or two states in a larger spectrum) quantum system, some examples include the vertical and horizontal polarisations of single photons, or the two spin states of a spin half particle. The state of which is described by $\vert\psi\rangle$:

\begin{equation}
\vert\psi\rangle = e^{i\gamma}\left(\cos\left(\textstyle{\theta\over 2}\right)\vert0\rangle + e^{i\phi} \sin\left(\textstyle{\theta\over 2}\right) \vert1\rangle\right) \label{equ:State}
\end{equation}

Where $\gamma$, $\theta$ ${}_{(0\leq\theta\leq\pi)}$ and $\phi$ ${}_{(0\leq\phi\leq\ 2\pi)}$ are real. Although the overall phase factor ($e^{i \gamma}$) is excluded as it has no observable effect\cite{Nielsen:2000}. Therefore Equation \ref{equ:State} can be re-written in its normalised form as follows:
\begin{equation}
\vert\psi\rangle = \lambda \vert 0\rangle + \mu\vert 1\rangle
\end{equation}
Where:
\begin{equation}
\vert\lambda\vert^2 + \vert\mu\vert^2 = 1\label{equ:Normal}
\end{equation}

\begin{figure} 
\centering 
\includegraphics[width=\textwidth]{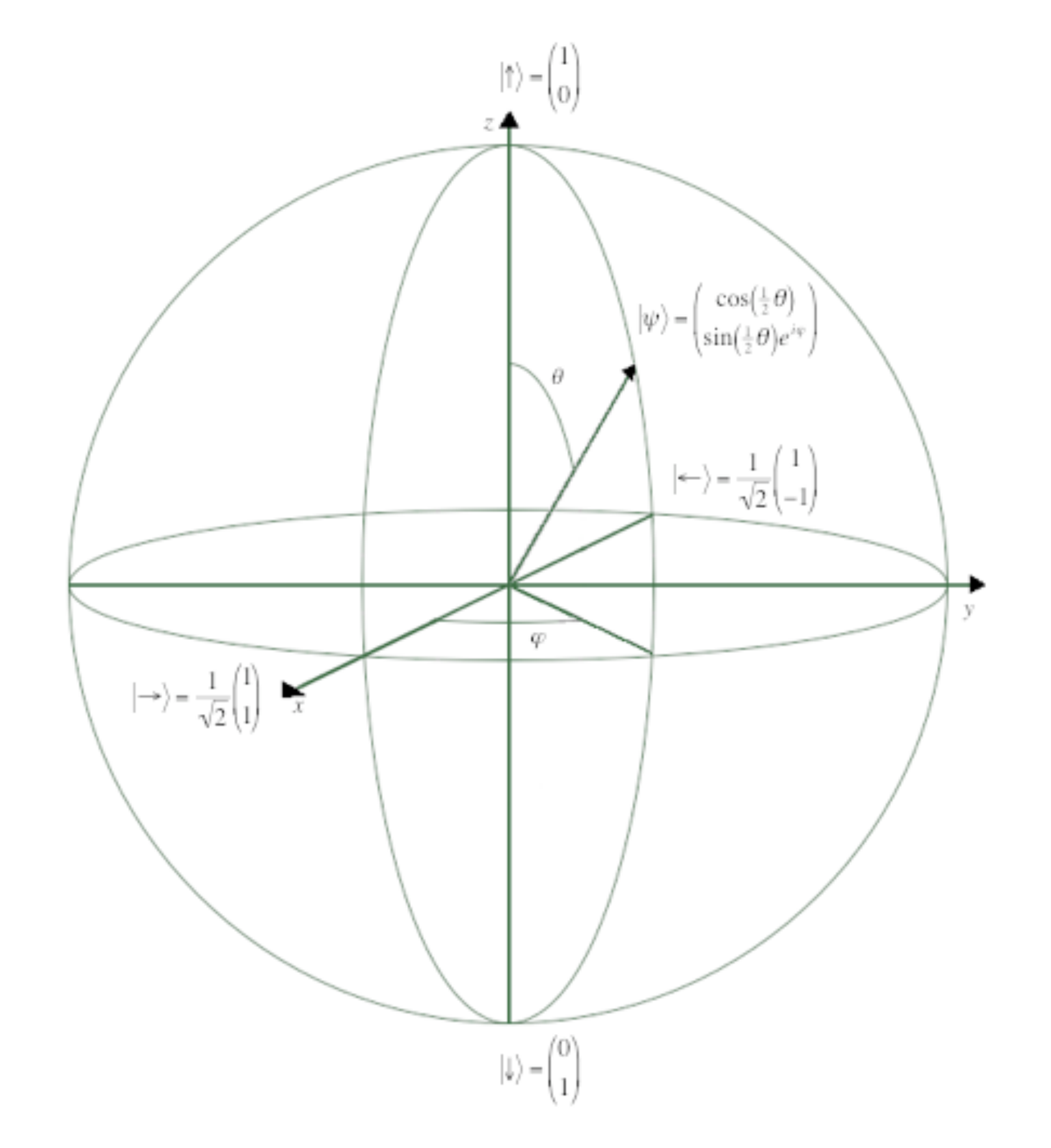} 
\textbf{\caption{The Bloch Sphere\label{fig:Bloch} }}
\end{figure}

This allows for a superposition of the two states $\vert0\rangle$ and $\vert1\rangle$, with $\vert\lambda\vert^2$ and $\vert\mu\vert^2$ as probabilities of measuring the classical states of 0 and 1 respectively. The normalisation condition in Equation \ref{equ:Normal} permits the use of a geometric representation known as the Bloch sphere (see Figure \ref{fig:Bloch}). This is a unit sphere where any point on the surface of the sphere is a possible pure state of the qubit it represents.

As a qubit can only be measured in one of the two possible states, they are taken as givens. This has lead many to adopt the simpler vector notation for the state. The general state vector is given as:

\begin{equation}
\vert\psi\rangle = \left( \begin{array}{c} \cos(\textstyle{\theta\over 2}) \\ \sin(\textstyle{\theta\over 2}) e^{i\phi} \end{array} \right)
\end{equation}

Where the top value is the linear coefficient for state $\vert 0\rangle$ and the bottom $\vert 1\rangle$. This representation will be used from now on.

Another representation to consider is the density matrix. This is a far more general representation of the state of the system. The density matrix ($\rho$) can be used to represent mixed states (statistical mixtures of various states or a quantum system that has decohered) as appose to the ability of state vectors for sole representation of pure states only. The density matrix is formulated as follows:

\begin{equation}
\rho= \sum_i p_i \vert\psi_i\rangle \langle\psi_i\vert
\end{equation}

Where $\vert\psi_i\rangle$ is the state of qubit $i$. This representation will not be used for the majority of this document as the quantum computer system will be modelled without noise and decoherence for which the state vector representation will suffice.

\subsection{The DiVincenzo Criterion} \label{subsubsect:DiV}
Many various two state systems could be classed as qubits, although for them to be useful in quantum computation other factors must be explored. The DiVincenzo criterion\cite{DiVincenzo:2000} is a basic checklist outlining the properties that a quantum device must incorporate for it to be classed as a working quantum computer. The most basic checklist contains five criteria that are needed for quantum computing to be implemented although another two necessities are listed for the creation of an all-in-one quantum computing/communication device.

\begin{enumerate}
\item\textit{A Scalable Physical System with Well Characterised Qubits}
\paragraph{}The basic principles of the qubit have already been mentioned in Section \ref{subsubsect:Qubit}. Generally if the qubit used satisfies the outlined description it can be classed as well characterised and may be useful for some implementation of quantum computing. Although the scalability is one of the greatest issues that is yet to be addressed for many physical realisations of a quantum computer, such as the Ion-Trap quantum computer, as currently this realisation has yet to reach double figures for the number of coherent qubits. This requirement is needed to perform useful operations, as the number of accessible qubits limits the input size of the problem.\newline

\item\textit{The Ability to Initialise the State of the Qubit to a Simple State $($e.g. $\vert 000\ldots 0 \rangle)$}
\paragraph{}In most cases the means of initialisation means cooling the system so that it is initialised in its known ground state, this must be achievable repeatedly with high fidelity, as from this state any other state can be reached by using the universal set of quantum gates (requirement 4). It is also possible to measure the qubits and once their state is known, operations can be performed to prepare them in their required initialisation. This requirement is very important as without knowing the initial state of the system before computation, no sense can be made of an output after application of some algorithm.\newline

\item\textit{Adequately Long Decoherence Times}
\paragraph{}The greatest worry for conventional quantum computers is decoherence. The largest source of decoherence is the entanglement of the system with the surrounding environment. These effects are detrimental to the operation of a quantum computer. In many cases decoherence is described as a loss of information, although the information is only locally lost. In fact the quantum information stored in the system and environment is now shared, causing the qubits in a quantum computer to behave almost classically. In some cases decoherence has even been described as measurement of the system by the environment, although more worryingly it has even been shown that decoherence leads to an exponential increase of error rate with the input size\cite{Brooks:1999}.\newline

It is widely believed that a sufficiently long time to perform a useful algorithm is $\gtrsim 10^{4-5}$ clock cycles. Meaning that the quantum gates must be able to perform 10,000 operations within a single decoherence time. Many systems simply cannot achieve these expectations, although measures have been made to increase the decoherence times of the quantum computer to an acceptable standard. It has been shown that quantum error correction codes can effectively be implemented\cite{DiVincenzo:2000}. Although systems with error correction capabilities need a minimum of three qubits to represent only a single  one (to filter phase errors a minimum of 5 qubits is necessary), this not only triples the number of qubits but significantly  increases the number of resources needed to successfully complete an operation, including the physical `gate' mechanisms.

\item\textit{A ``Universal" Set of Quantum Gates}
\paragraph{}Quantum systems evolve unitarily according to their Hamiltonians, although for quantum computation to be performed on a specific system, the user must have full control of the qubits by manipulation of the applied Hamiltonians. This is due to the fact that the unitary operations build up the quantum algorithm, which are the specific set of instructions given to the qubits for any given problem before the result is read (just like classical computing). It is widely believed that any unitary transform can be formed from a series of single-spin operations and CNOT (controlled-NOT) gates\cite{DiVincenzo:1994}. Therefore even if only these operations can be performed, this criterion will be satisfied.\newline

\item\textit{A Qubit-Specific Measurement Capability}
\paragraph{}The final requirement is somewhat obvious. The need to be able to read ones answer (final state of the system) post computation is an ability that all quantum computers must possess. This requires a coupling between the quantum system and a classical measurement system, although this coupling cannot be continually apparent as this would introduce decoherence or even a complete collapse of the superposition state during the computation, deeming the computation a failure. 

Ideally the system will be in a pure superposition state after computation, where all possible collapses of the state into a correct answer to the problem have a cumulative probability of (or very close to) 1. Yet this will not always be the case. It has been suggested that a trade-off between fidelity and other resources which result in reliable computation be made\cite{DiVincenzo:2000}. This is due to the simple fact that the target 100\% efficiency will probably never be realised, therefore with systems that have a reliability/efficiency of 90\% but need a higher figure (for example 97\%) should simply be re-run to increase the accuracy of the results.\\
\end{enumerate}

To a lesser extent, the following two criteria would be useful for a universal quantum computing/communication device. If quantum computers ever become a household product, the dawn of a new era of hackers will be very real. With quantum computers the RSA encryption we all rely on in this day and age for monetary transactions across the internet will become easily decipherable. For this reason, quantum key-distribution/cryptography will become necessary as will the need for a universal quantum computing/communication device:
\begin{enumerate}
\item\textit{The Ability To Interconvert Stationary and Flying Qubits}
\item\textit{The Ability to Transmit Flying Qubits Between Specified Locations}
\end{enumerate}

\subsection{Conventional Quantum Computing} \label{subsubsect:Conv}

As with all computers, classical and quantum, the bits or qubits are initialised to some state. A series of logic gates are applied depending on the algorithm that the user wishes to implement and finally the output is read. Due to the quantum properties of the qubit, there are many more types of one or two qubit gates that can be implemented in a quantum system than possible in a classical system. These additional gates are designed to exploit the quantum properties of the system. For example, the Hadamard gate that is specifically designed to initialise a qubit into a superposition of its two classical states. 

Take a circuit diagram containing classical logic gates, to implement some algorithm on a desired input, where the input bit is represented by some voltage passing through a wire and the current state of this bit is dependent on the voltage. The bit of information travels through the wire and is acted upon by the obstacles (logic gates) in its path. The circuit diagrams for a  quantum algorithm acting on a selection of qubits look similar, although the qubits do not physically move throughout the system, the diagram is more of a timeline of operations that are applied to the qubits. Therefore a line on one of these diagrams will never make a U-turn for obvious reasons.

There is currently no systematic way of programming a quantum computer and developing an algorithm for a specific task requires extreme `cleverness', therefore there are very few useful algorithms that have been developed for implementation on a conventional quantum computer. The first and simplest quantum algorithm is known as the Deutsch Algorithm, it is used to check if a binary function is balanced with one evaluation of the function. If this were possible classically, we would be able to check if a coin was bias with a single toss.

\begin{figure} 
\centering 
\includegraphics[width=\textwidth]{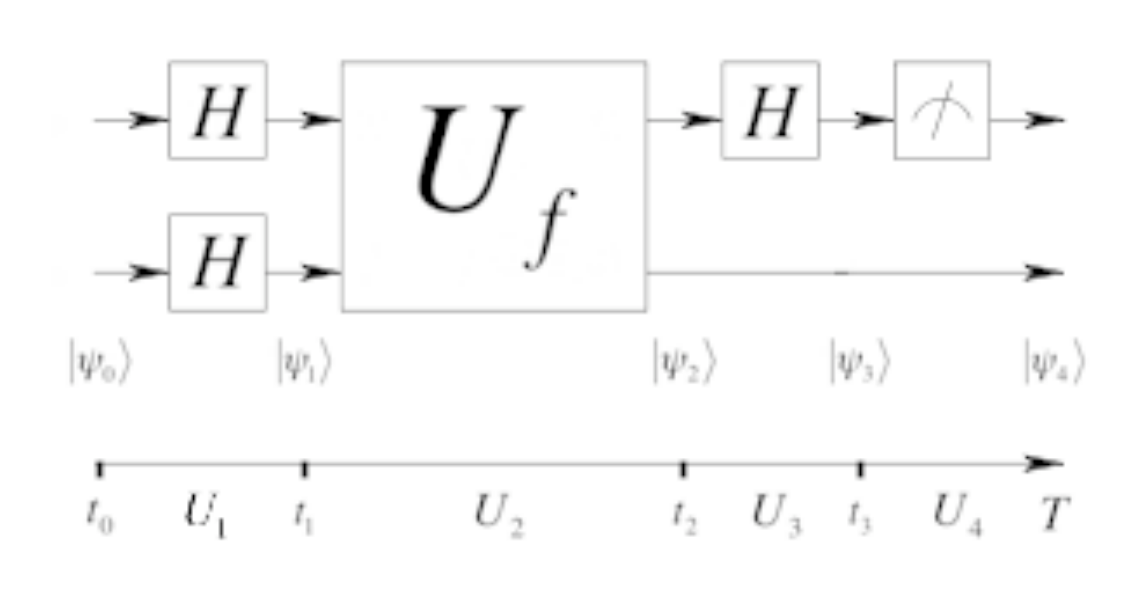} 
\textbf{\caption{The Deutsch Algorithm in Circuit Diagram Form\label{fig:Deu} }}
\end{figure}

The algorithm is a series of gates applied to the qubits as pictured in Figure \ref{fig:Deu}. The timeline is labelled with a series of unitary transformations (gates) that the qubits undergo, interestingly, all operations must be unitary so a quantum system is completely reversible unlike classical computers. The inputs (initialised states) needed to implement this algorithm are two qubits each initialised as one of the two measurable states of a single qubit. Therefore $\vert \Psi_0\rangle = \vert 01\rangle$, this state is then acted upon by two Hardamard gates which rotates both qubits into a superposition:

\begin{equation*}
\vert \Psi_1\rangle = H \otimes H\vert 01\rangle = \left({\vert0\rangle + \vert1\rangle}\over \sqrt{2}\right)\otimes\left({\vert0\rangle - \vert1\rangle}\over \sqrt{2}\right)
\end{equation*}

$\vert\Psi_1\rangle$ is then fed into the two-bit controlled-f gate ($U_f$):
\begin{equation*}
\vert \Psi_{2}\rangle = U_{f} \vert \Psi_1\rangle= {1\over 2}\sum_{x=0}^1{\vert x\rangle} \otimes \left( \vert f(x)\rangle - \overline{\vert f(x)\rangle}\right)
\end{equation*}
\begin{equation*}
\quad\quad= {1\over 2}\sum_{x=0}^1(-1)^{f(x)}\vert x\rangle \otimes \left( \vert 0\rangle - \vert 1\rangle\right)
\end{equation*}
\begin{equation*}
= \left \{ 
\begin{matrix}
\pm \left({\vert 0\rangle+\vert 1\rangle \over \sqrt 2}\right) \otimes \left({\vert 0\rangle -\vert 1\rangle \over \sqrt 2}\right)\quad
$if$\quad f(0) = f(1)\\ 
\pm \left({\vert 0\rangle -\vert 1\rangle \over \sqrt 2}\right) \otimes \left({\vert 0\rangle -\vert 1\rangle \over \sqrt 2}\right)\quad $if$\quad f(0) \neq f(1)\end{matrix} \right.
\end{equation*}

Now it is clear that $\vert f(x)\rangle -\overline{\vert f(x)\rangle} =\vert 0\rangle - \vert 1\rangle$ if $f(x) = 0$ and $\vert f(x)\rangle - \overline{\vert f(x)\rangle} = \vert 1\rangle - \vert 0\rangle$ if $f(x)= 1$. Applying the Hadamard gate to the first bit again yields:

\begin{equation*}
\vert\Psi_3\rangle = H \otimes I\vert\Psi_2\rangle = \left\{ 
\begin{matrix}
\pm \vert 0\rangle \otimes \left({\vert 0\rangle -\vert 1\rangle \over \sqrt 2}\right)\quad
$if$\quad f(0) = f(1)\\ 
\pm \vert 1\rangle \otimes \left({\vert 0\rangle -\vert 1\rangle \over \sqrt 2}\right)\quad $if$\quad f(0) \neq f(1)\end{matrix} \right.
\end{equation*}

By measuring the first bit we can deduce that $f$ is constant if the bit reads 0 or balanced if the bit reads 1.

As mentioned before, there are few useful algorithms that may be implemented on a physical conventional quantum computer. Most require far more qubits in the computation than the example above, and all involve many interesting sequences of gates to yield a final result. During a computation of this type, the applied gates transform the system into increasingly complex superposition states where a minimal disturbance can cause complete collapse of the system and as mentioned in Section \ref{subsubsect:DiV}, there is an exponential increase of error rate with input size. This has forced the development of new unconventional techniques.  

\section{Adiabatic Quantum Computing (AQC)} \label{sect:AQC}

AQC differs greatly from the standard method of quantum computation. Rather than performing sequences of discrete unitary operations on one or two qubits individually over a period of time, a single time-dependant Hamiltonian is applied to the whole system effectively reproducing the individual operations in a single evolution. Proofs of equivalence between the two methods exist and more information on these can be found in Section \ref{subsect:Equiv}.

To perform AQC, the system of qubits must evolve slowly in time from the ground state of an initial Hamiltonian with known eigenvalues, to a ground-state of the final Hamiltonian within which the answer to a specific problem is encoded, usually with unknown eigenvalues. There are many methods for this steady evolution although the most common is the slowly varying Hamiltonian\cite{Farhi:2000}:

\begin{equation} H(t)=\left(1-\textstyle{t\over{T}}\right)H_I+ \textstyle{t\over{T}} H_F \end{equation}

Where $H_I$ is the initial Hamiltonian whose ground state is known (normally a linear superposition of all admissible states\cite{Neven:WEB2}) and $H_F$ is the final Hamiltonian within which the problem is encoded. Time $t$ is the time during the evolution and $T$ is the total time of the evolution. Larger values of $T$ correspond to longer run times, and hence slower evolution. Thus the evolution should become increasingly adiabatic as $T$ becomes large. It is clear that when $t=0$ the Hamiltonian is solely $H_I$ and when $t=T$ the Hamiltonian is now $H_F$ with smooth linear interpolation between $H_I$ and $H_F$ while $0\leq t \leq T$. For simplicity the Hamiltonian is slightly altered, whereby $t/T$ is replaced with $\Omega t$ where $0\leq t\leq1$ and the constant $\Omega$ is a `rate' that can be changed for slower/faster evolution:

\begin{equation} H(t)=\left(1-\Omega t\right)H_I+ \Omega t H_F  \label{equ:Evolution}  \end{equation}

The system of qubits evolves with time according to the Schr\"{o}dinger Equation using the above time dependant Hamiltonian:

\begin{equation} -i\hbar {d\over{dt}}\vert\psi \left(t\right)\rangle = H\left(t\right)\vert\psi \left(t\right)\rangle \end{equation}

By representing the state of the system with state vector notation and the Hamiltonian in matrix form, the solution to this equation can be found by solving the above as a system of ordinary differential equations (ODEs). This is the basis of the simulation described in Section \ref{subsect:Simulation}.

AQC is thought to be a powerful method of solving NP-Hard minimisation problems\cite{Corato:2003}. The most common sector of mathematics to which these problems are related is Graph Theory, where nodes on a given graph are subjected to local biases as well as inter-node interactions. Therefore, in order for nodes to be represented by qubits there must be a method of implementing some local bias on the qubits as well as the inclusion of interactions between qubits. It is simply not feasible to experimentally implement problems of this type that involve fully connected graphs, therefore only nearest neighbour interactions can be modelled. The best known theory that describes these interactions is the Ising model\cite{Chandler:1987}. This model was initially developed to describe the inter-particle interactions in lattices to describe spontaneous magnetisation (amongst other effects). This model can be directly translated to be used on qubits as two-state magnetic spins of particles within lattices are directly analogous to the two states of any given qubit.

Given a graph G, with vertices V and edges E the total Hamiltonian needed to describe this system is:

\begin{equation} H_F=\sum_{i\in V} h_i\sigma^z_i-\sum_{(i,j\in E)} J_{ij}\sigma^z_i\sigma^z_j \label{equ:Ising}\end{equation}

Where $h_i$ is the local Hamiltonian for qubit $i$ and $J_{ij}$ is the Ising interaction (coupling constant) between nearest neighbours $i$ and $j$. Where if $J_{ij} > 0$ it is energetically more favourable for qubits $i$ and $j$ to be aligned. Finally, $\sigma^z$ is the Pauli spin matrix for the z-basis which is the classically measurable basis. More information on encoding specific problems into the form of the final Hamiltonian is available in Section \ref{subsect:MISP}. In most cases the local Hamiltonian is equal for all qubits, it may then be referred to as an external Hamiltonian ($H$):

\begin{equation} H_F=H \sum_{i\in V}\sigma^z_i-\sum_{(i,j\in E)} J_{ij}\sigma^z_i\sigma^z_j \label{equ:Isingn}\end{equation}

In almost all cases the initial Hamiltonian $H_I$ is taken to be of the form:

\begin{equation} H_I= \sum_i \sigma^x_i\end{equation}

As in all cases the ground state for this Hamiltonian is known or can be found easily. Where $\sigma^x$ is the Pauli spin matrix in the x-basis. This initial Hamiltonian is the reason why AQC is regarded as quantum, typically the ground state of qubits in this basis is a superposition state. This sets the method of AQC apart from classical annealing techniques which involve having an `always on' final Hamiltonian along with an external Hamiltonian of large magnitude in the same classical basis. Then slowly reducing the magnitude of the external Hamiltonian until only the problem Hamiltonian is applied. This method relies on the system switching between classical states to reduce the total energy of the system during the evolution until no further reduction of energy can occur. After annealing, measurement of the system will frequently give the ground state, although in many cases the system remains in a local energy minimum state. Without raising the energy of the system and neglecting possible quantum effects, there is no way of leaving a local energy minimum which is not the ground state of the system, and does not represent a solution to the encoded problem. Ruling this method of no use to any form of computation, simply due to its poor consistency.

By starting in a ground state of the total Hamiltonian which is specifically constructed so that the ground state is a linear superposition of all possible states. Then slowly evolving the system so that it remains in a superposition of the possible ground states, local minima cannot exist. In the quantum method, states become less or more probable and do not switch between one another. One way of visualising this is to simply take the state vector of the system and observe the coefficients during a simulated evolution. They will simply increase or decrease depending on the Hamiltonian that is applied (although this is not experimentally possible).

Assuming that a superposition state is easy to initialise and retain, the main source of error in an AQC is known as Landau-Zener tunnelling\cite{Amin:2008}. By evolving a system of qubits too quickly, it can tunnel into an energetic state that does not represent an answer to the encoded problem. The point in the evolution where this effect occurs most is at a so called `anti-crossing', although this is in fact the point in the evolution where the eigenvalue gap between the ground state and the first excited state is at a minimum. A relationship to find an adequate evolution time from the minimum gap of the system has been directly derived from the adiabatic theorem\cite{Farhi:2000}, and is as follows:

\begin{equation}
T \gg {\xi \over g_{min}^2} \label{equ:Min}
\end{equation}

Where $\xi$ is a constant for each system. Essentially, the size of T is governed by $g_{min}^{-2}$, although this is in no way helpful. With basic systems that can be effectively simulated by a classical computer, the eigenvalues can be pre-determined prior to simulation by checking all possible values for all possible times, although more complicated systems that are the sole reasons for the development of quantum computers will, in general, have unknown eigenvalues prior to experimentation. Without prior knowledge of the eigenvalues, an adequate run time cannot be determined.

Taking a single qubit evolution as an example, by evolving from a linear superposition of the two states (ground state of a Hamiltonian acting in the x-basis), to one of the two classical states (ground state of a Hamiltonian acting in either the positive or negative z-basis). Equation \ref{equ:Evolution} for the evolution becomes a straight line equation through the Bloch sphere, the closer the state is evolved past the centre of the sphere (fully mixed state), the higher the possibility of tunnelling to an excited state. This means that errors will occur most if qubits are evolved between orthogonal states, as the eigenvalues cross during such evolutions.

General decoherence that affects all quantum computers is due to entanglement with the environment. More information on the effect of this entanglement can be found in Section \ref{subsubsect:DiV}. Although it seems that this effect is less detrimental to an AQC than a conventional quantum computer. Many studies and simulations of entanglement to the environment have shown that the AQC has an inherent fault tolerance and robustness against decoherence. The stray coupling to the environment affects the intended path of the eigenvalues of the Hamiltonian, although as long as a gap is present between the ground and first excited state the system will still evolve adiabatically. In some cases this noise can even increase the fidelity of the system\cite{Amin:2008}\cite{Jordan:2008}\cite{Sarandy:2005}\cite{Aberg:2005}\cite{Roland:2005}\cite{Kam:2003}. For this reason, the effects of decoherence will be omitted from further study, and only perfect systems will be simulated.

With regards to the DiVincenzo checklist in Section \ref{subsubsect:DiV}, a system that can be used for AQC must as with conventional quantum computers, have scalable qubits, the ability to initialise and a measurement capability. As for the other criteria, the need for an adequately long decoherence time as explained above is not as necessary as it is for conventional quantum computers. Furthermore, the need for a universal set of quantum gates is specific to the operation of conventional quantum computers. Similarly, the ability to manufacture couplings between qubits and apply local Hamiltonians to individual qubits is necessary for AQC. This ability allows the AQC to simulate any algorithm that can be applied to a conventional quantum computer which is equivalent to possessing a universal set of gates (see Section \ref{subsect:Equiv}).

\subsection{Proof of Equivalence} \label{subsect:Equiv}

There is a somewhat simple proof of equivalence to conventional quantum computing by Mizel $et$ $al$\cite{Mizel:2007}, although the interested reader may prefer the thorough proof by Aharonov\cite{Aharonov:2008} who has showed that any quantum circuit can be simulated by an adiabatic quantum algorithm.

Interestingly these proofs show that an AQC is capable of performing all operations that are executable on a conventional quantum computer. Although no proof exists to show that a conventional quantum computer can reproduce the same operations as an AQC. Due to this phenomenal turn of events, much interest has been generated among physicists and computer scientists for research into the AQC. To the extent that it is now believed that the AQC is a form of the universal quantum computer theorised by Deutsch\cite{Jordan:2008}.

\subsection{Simulating the AQC} \label{subsect:Simulation}
The simulation of the AQC has been written with the use of Matlab and the full code can be found in the Appendix. The simulation in basic terms solves the Schr\"{o}dinger Equation as a system of ordinary differential equations. This is achieved by the user listing all Ising interactions between qubits and the local Hamiltonians on all the qubits as these are system dependent. The program will then compile the initial, problem and total Hamiltonians before finding and applying the initial ground state of the system. The program continues by compiling the ODEs and saving them to a separate `.m' file, these are then called and solved. The solver used is $ODE45$ which is based on an explicit Runge-Kutta formula known as the Dormand-Prince Pair, it is advised by Mathworks (creators of Matlab) that the $ODE45$ should always be used first and foremost.

The graphs that are plotted are dependent on the size of the inputted system. For all sized systems, the probability for individual states is plotted as a function of time, the eigenvalues of the system are plotted over time as well as an individual plot for the minimum gap energy between the two lowest eigenvalues. Furthermore, if the system is a single qubit, a plot of the evolution of the state trajectory through the Bloch sphere is provided, and for a two qubit system the concurrence (measure of entanglement) is represented graphically as a function of time. Once simulation is complete there is an output of all states and their final probabilities in the Matlab window.

\subsection{Max Independent Set Problem} \label{subsect:MISP}

A general algorithm for compiling a final Hamiltonian for any NP-Hard problem to use on an AQC has yet to be developed or may not even exist. For this reason a specific problem that is known to be NP-Hard has been selected for further study. The chosen problem is known as the Max Independent Set (MIS) problem. Further constraints and adaptations have been introduced so that it is ideal for the simulation in this situation. At time of printing the independent set problem remains an NP-Complete problem for cubic planar graphs\cite{Garey:1976}, and its optimisation counterpart the MIS problem remains NP-Hard. 

The independent set problem is best described with graph theory. Taking a graph G=(V,E) with vertices (nodes) V and edges E, does an independent set exist that has a cardinality of at least $k$ ($k$ = some integer)? An independent set is a subset of the system of which no vertices are directly connected by an edge E. The most basic algorithm to solve this problem is to take every subset in the system that is of size at least $k$ and examine each induvidually to determine if it is an independent set. The optimisation equivalent to this problem as mentioned above is the MIS problem, when the maximum $k$ for the given graph G must be determined. This is the problem that will be applied to the AQC simulation.

Using the Ising model it is NP-Hard to calculate the ground state of a planar\footnote{A graph is planar if the vertices do not intersect when drawn in a single two-dimensional plane.} two-dimensional system, whose inter qubit couplings favour anti-alignment of states whilst in the presence of an external Hamiltonian. If and only if each qubit is coupled to no more than three others\cite{Bar:1982}. At the same time it is NP-Hard to find the MIS of a cubic\footnote{The graph is known as cubic if all its vertices have three edges.} planar graph. Due to the equivalence of these problems the Ising model can be isomorphically mapped onto the MIS problem\cite{Kam:2002}. Whereby finding the ground state of the Ising model will also find an example of a MIS. For most problems there is a selection of MISs, but obtaining any one is NP-Hard\cite{Kam:2002}. On paper this ability yields nothing of interest, as both problems remain NP-Hard to solve. Although using a physical quantum system that behaves according to the Ising model, an instance of the MIS can be encoded into an applied Hamiltonian. This is useful as previously explained it is experimentally not feasible to implement a highly connected graph to any physical realisation of a quantum computer and secondly a planar graph can be implemented on a 2-dimensional chip containing qubits with only nearest neighbour interactions. Therefore, by measuring the ground state of the system, the MIS of thus system will also be determined. This is achieved by representing nodes on a graph with qubits, whereby the parity of the node is represented by qubit states $\vert 0\rangle$ and $\vert 1\rangle$. 

Without mathematical proof, the mapping of the Ising model with couplings that favour anti-alignment onto the MIS problem appears completely analogous (even without the presence of an external Hamiltonian). If it is energetically favourable for qubits to be anti-aligned, this in almost all cases would cause every qubit to be of opposite  parity to its nearest neighbours. If this were the case, finding the ground state of the system with the exclusion of an external Hamiltonian would still yield the MIS. Although this is not the case, many ground states of this form contain coupled pairs of qubits with the same parity that are opposite to other pairs (see Section \ref{subsect:4Q} and Figure \ref{fig:4Q_exp}). This in turn reveals no information about the MIS as no set of qubits of the same parity are completely isolated from one another deeming them non-independent. For this reason, the local Hamiltonian must be introduced on all qubits, to act as some form of penalty function, so that only qubits of one orientation can be aligned with their nearest neighbours.

\begin{figure} 
\centering 
\subfloat[{}]{
\label{fig:Situation1} 
\includegraphics[width=0.75\textwidth]{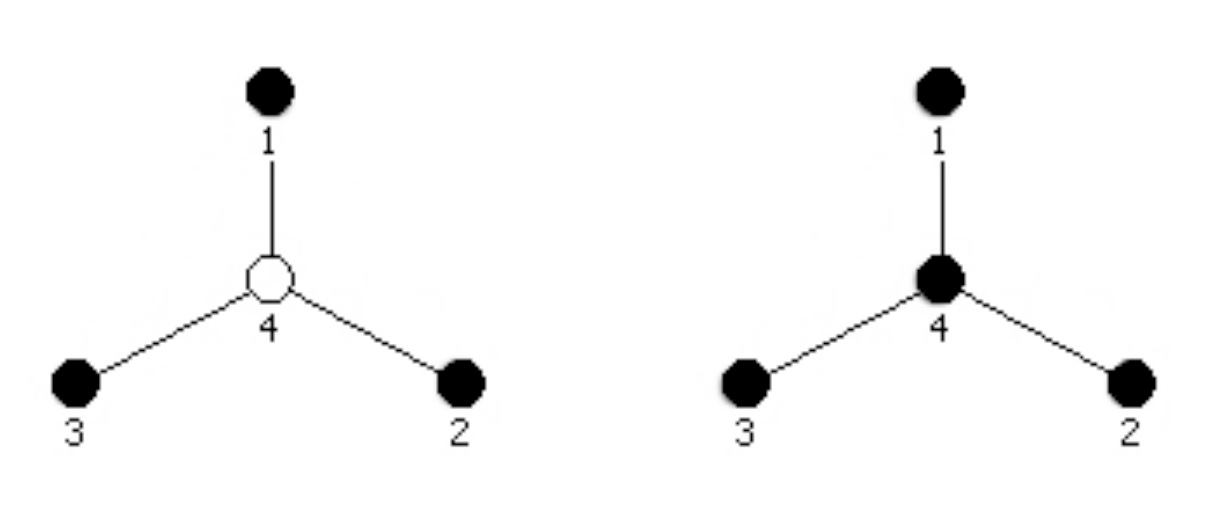} }\newline
\subfloat[{}]{
\label{fig:Situation2}
\includegraphics[width=0.75\textwidth]{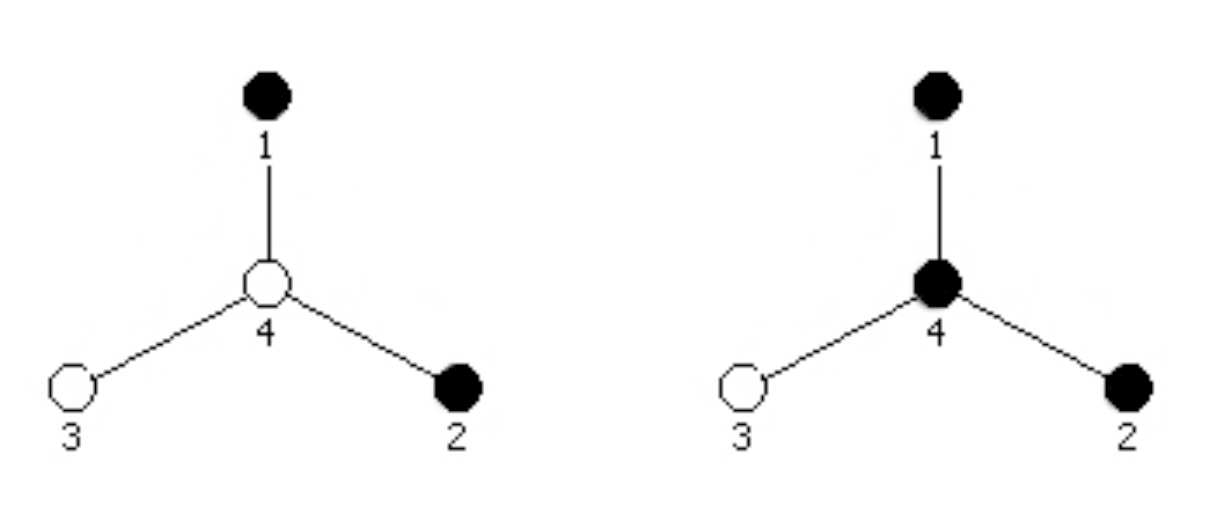} }\newline
\subfloat[{}]{
\label{fig:Situation3} 
\includegraphics[width=0.75\textwidth]{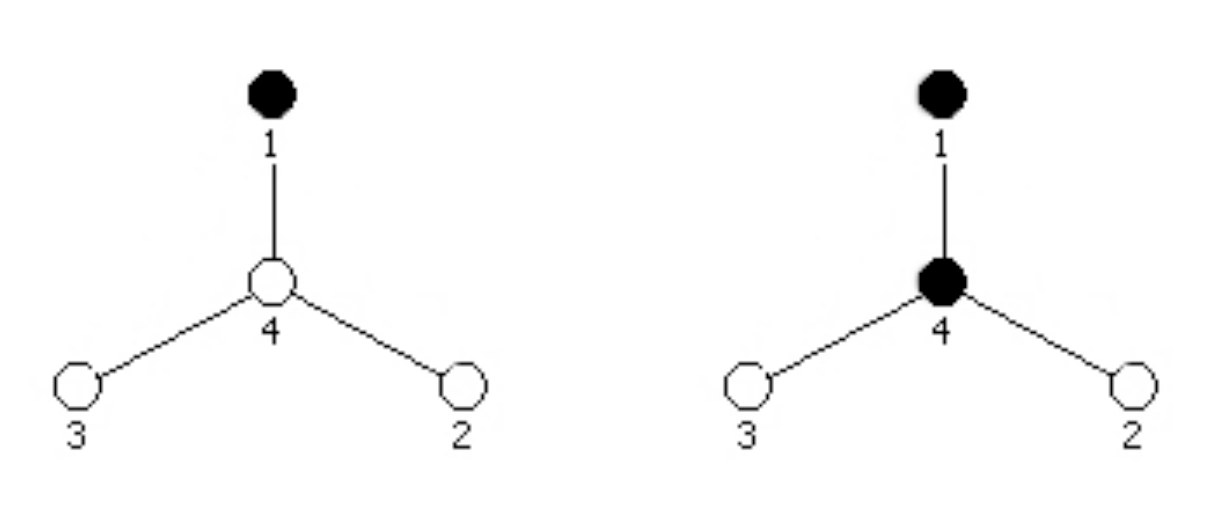} }\newline
\subfloat[{}]{
\label{fig:Situation4}
\includegraphics[width=0.75\textwidth]{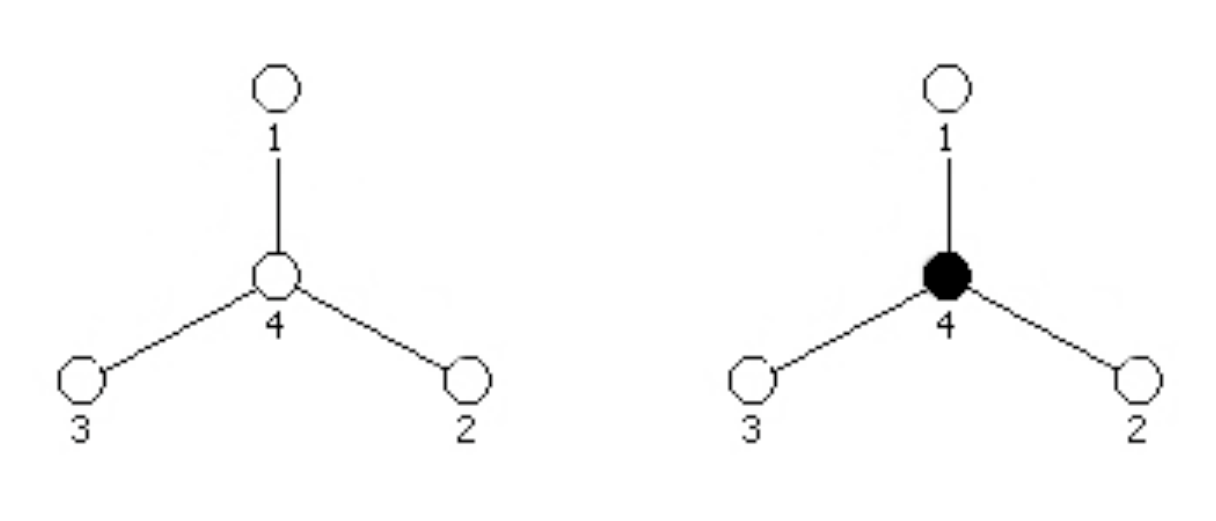} }\newline
\textbf{\caption{All Possible Situations that a Single Node in a Graph Can Undergo}}
\end{figure}

By exploring the energies of every possible combination of qubits that a single qubit can be connected to, a small range of Hamiltonian magnitudes can be determined for which the ground state of the system may represent a MIS of the system. For a cubic graph, whereby each qubit may only have three nearest neighbours there are four possible combinations of states that the nearest neighbours may occupy (see Figures \ref{fig:Situation1} to \ref{fig:Situation4} where filled circles represent 1s). By flipping the centre qubit between both states (0 and 1), and determining the total system energy of both orientations with Equation \ref{equ:Isingn} will allow inequalities to be developed.

Assuming the qubit that is part of the independent set will be of opposite direction to the applied Hamiltonian, the state 0 will represent independent qubits. The inequalities are determined by deciding which orientation of the qubit best describes an independent set, as the energy involving this orientation must be lowest of the two. Table \ref{table:ineqs} has a summary of the developed inequalities and the Hamiltonian value that satisfy the inequalities if $J_{ij}=-1$ for all interactions.

\begin{table}
\centering
\begin{tabular}{c|c|c}
Situation & Inequality & Hamiltonian\\
\hline
Fig. \ref{fig:Situation1} & $ -4H-3J > -2H+3J$ & $H < 3$ \\
Fig. \ref{fig:Situation2} & $ J > -2H-J$ & $H > 1$\\
Fig. \ref{fig:Situation3} & $ 2H-J > J$ & $H> -1$\\
Fig. \ref{fig:Situation4} & $ 4H-3J > 2H+3J$ & $H> -3$\\
\end{tabular}
\textbf{\caption{Table of Inequalities to Determine a Hamiltonian Range\label{table:ineqs}}}
\end{table}

In summary $1<h_i<3$ for all $i$ if $J_{ij}=-1$ for all interactions. This also shows that no matter the size of the system, a Hamiltonian within this range will cause the ground state to be a MIS as it satisfies all possible interactions in that system.

\section{Results \& Discussion} \label{sect:Results}
Below is an extensive set of results for 1, 2, 3, 4 and 8 qubit cases. For one and two qubits, simple simulations to test the theory of adiabatic evolution as well as quantum properties of the system are explored. The three qubit case is used to investigate the effects of frustration in a quantum system. For 4 and 8 qubits the MIS problem is applied, with exploration into the input parameters and the final results they yield.

\subsection{1 Qubit Simulation}
The system in this case has two possible classical states: 1 or 0. The initial Hamiltonian will be a linear superposition of the two states, this will be of the form: $\vert\psi\rangle = {1\over\sqrt{2}}(\vert 1\rangle +\vert 0\rangle)$. As there cannot be any Ising interactions, only a local Hamiltonian can be applied to the system. The Hamiltonian can work as a bias in either direction so these possibilities are explored as well as having no Hamiltonian whatsoever. The simulation is run for various values of $\Omega$ of which the reciprocal is the number of units of time.

\begin{table}
\centering
\begin{tabular}{l||lllllll}
\multicolumn{8}{l}{Bias=1}\\
$\Omega$ / t & 10 / 0.1 & 5 / 0.2 & 1 / 1 & 0.5 / 2 & 0.1 / 10 & 0.05 / 20 & 0.01 / 100\\
\hline
$P_{\vert 0\rangle}$ & 0.499167 & 0.496676 & 0.422382 & 0.247851 & 0.002732 & 0.000127 & 0.000006\\
$P_{\vert 1\rangle}$ & 0.500833 & 0.503324 & 0.577618 & 0.752148 & 0.996818 & 0.998933 & 0.995063\\
\hline
$P_T$ & 1 & 1 & 1 & 0.999999 & 0.99955 & 0.99906 & 0.995069\\
\multicolumn{8}{l}{}\\
\multicolumn{8}{l}{Bias=0}\\
$\Omega$ / t & 10 / 0.1 & 5 / 0.2 & 1 / 1 & 0.5 / 2 & 0.1 / 10 & 0.05 / 20 & 0.01 / 100\\
\hline
$P_{\vert 0\rangle}$ & 0.5 & 0.5 & 0.5 & 0.5 & 0.499953 & 0.499842 & 0.498688\\
$P_{\vert 1\rangle}$ & 0.5 & 0.5 & 0.5 & 0.5 & 0.499953 & 0.499842 & 0.498688\\
\hline
$P_T$ & 1 & 1 & 1 & 1 & 0.999906 & 0.999684 & 0.997376\\
\multicolumn{8}{l}{}\\
\multicolumn{8}{l}{Bias=-1}\\
$\Omega$ / t & 10 / 0.1 & 5 / 0.2 & 1 / 1 & 0.5 / 2 & 0.1 / 10 & 0.05 / 20 & 0.01 / 100\\
\hline
$P_{\vert 0\rangle}$ & 0.500833 & 0.503324 & 0.577618 & 0.752148 & 0.996818 & 0.998933 & 0.995063\\
$P_{\vert 1\rangle}$ & 0.499167 & 0.496676 & 0.422382 & 0.247851 & 0.002732 & 0.000127 & 0.000006\\
\hline
$P_T$ & 1 & 1 & 1 & 0.999999 & 0.99955 & 0.99906 & 0.995069\\
\end{tabular}

\textbf{\caption{Table of Obtained Final Probabilities for Single Qubit States $\vert0\rangle$ and $\vert1\rangle$ in Specified Circumstances\label{table:1q}}}
\end{table}

It is clear from Table \ref{table:1q} that the simulation is providing the desired effects. When the local (problem) Hamiltonian is increased over time, the qubit aligns with the direction of the Hamiltonian as to remain in the ground state. This is also a good test of the adiabatic theorem; it is clear that with sufficiently slow evolutions the qubit is more likely to be measured in the ground state of the Hamiltonian at the end of the evolution. With a fast evolution that takes 0.1 units of time the qubit has an almost equal probability of being measured in either of the two states. Whereas after 10 units of time the likelihood of measuring the qubit in the ground state is greater with almost a probability of 1. This probability is further increased by running the simulation for 20 units of time, although there is an interesting decrease after 100 units. The combined probability of both states is therefore calculated for all simulations (with the values given in Table \ref{table:1q}). There seems to be a very small discrepancy of the total probability not accumulating to 1 after lengthier simulations. This effect appears to be dependent on time and the most legitimate reasoning is the rounding error introduced by the simulation's ODE solver. If a finite error is introduced into the probability after its calculation for every time interval, this error will become more prevalent with more time steps. Although in general the discrepancy is so small that no further action must be taken to obtain results of greater accuracy in future simulations.

\begin{figure} 
\centering 
\subfloat[Eigenvalues]{
\label{fig:1QEigs} 
\includegraphics[width=0.8\textwidth]{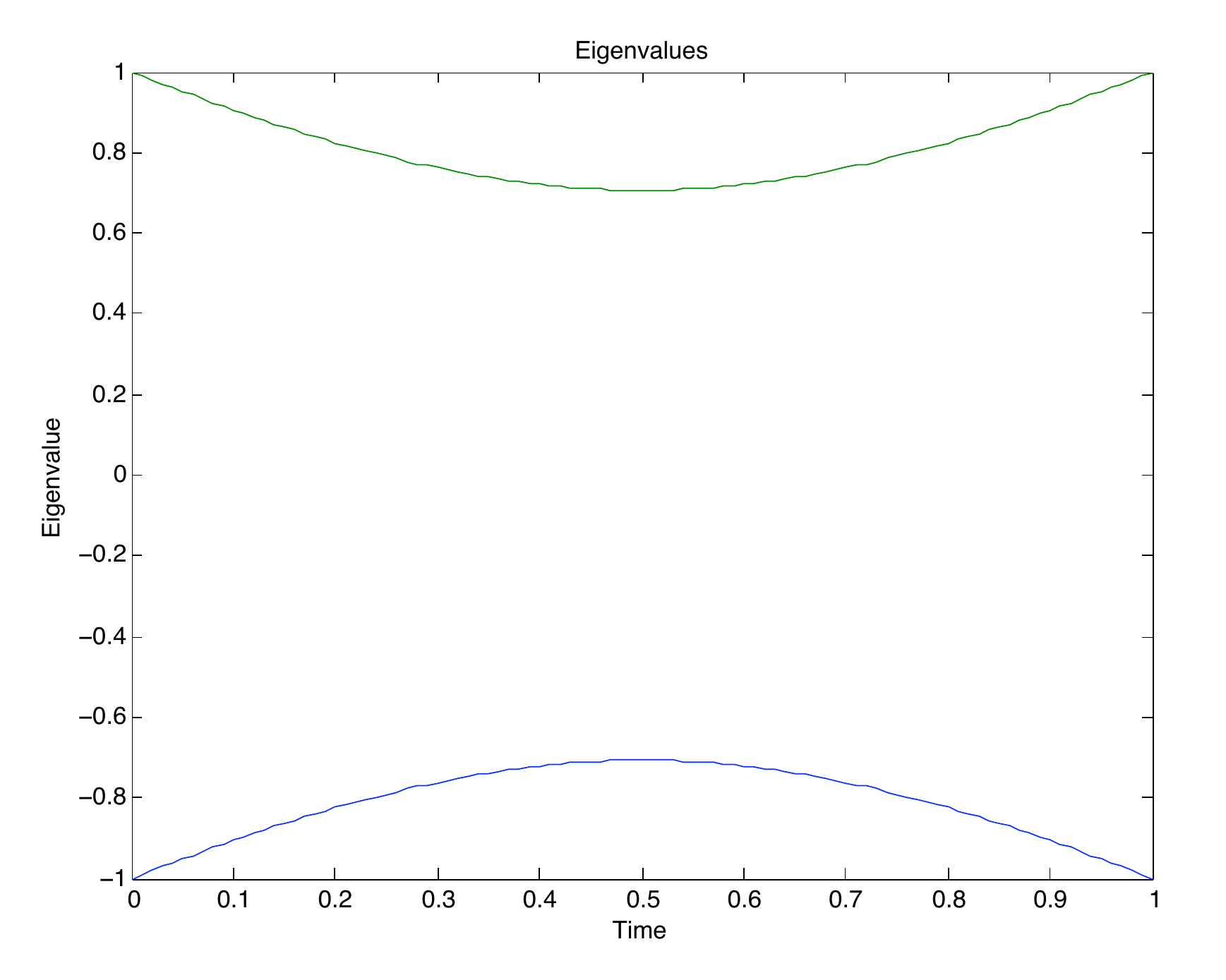} }\\
\subfloat[Gap Between Eigenvalues]{
\label{fig:1QGap}
\includegraphics[width=0.8\textwidth]{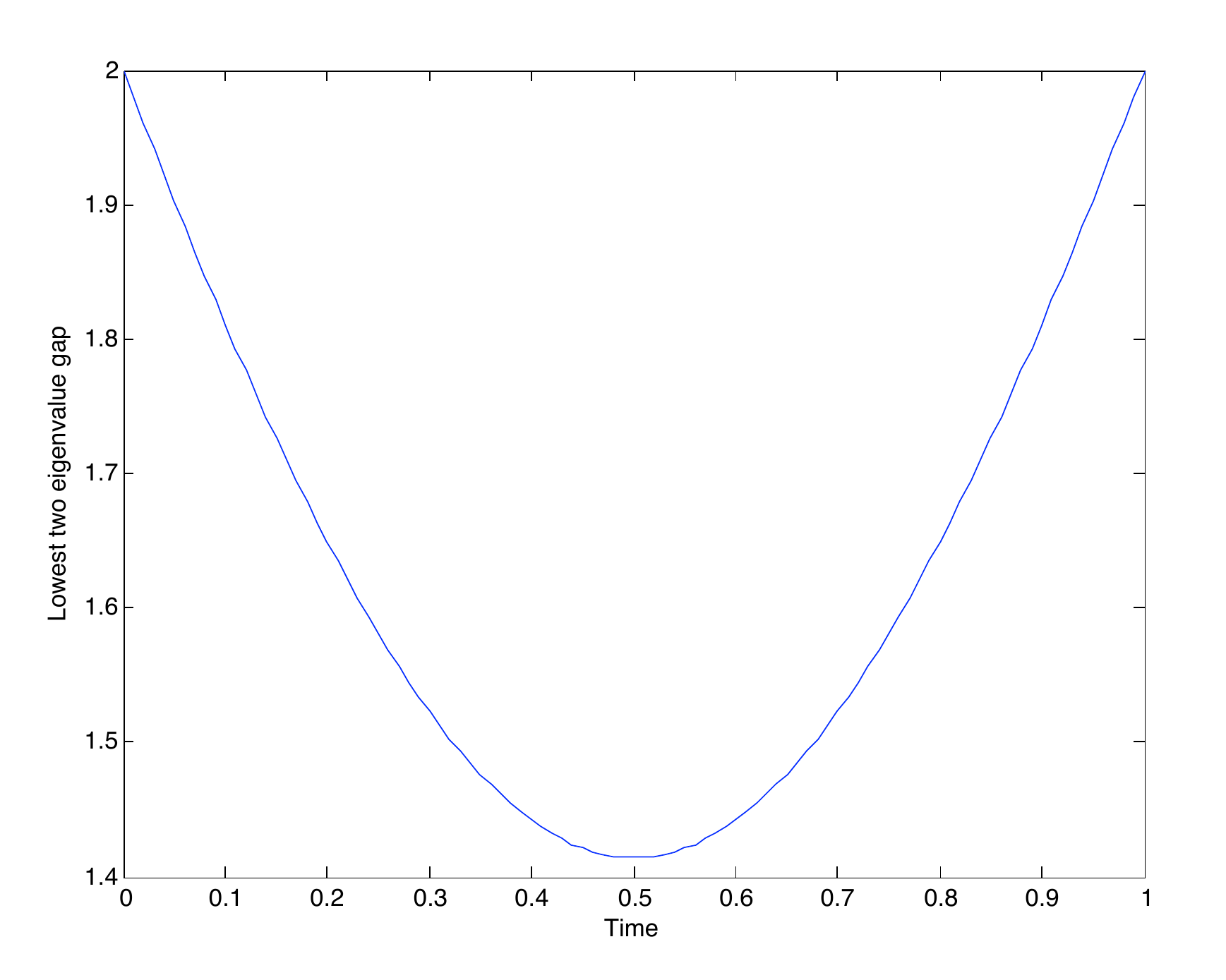} }
\textbf{\caption{Single Qubit Example of Eigenvalues and Their Gap as a Function of Time During the Evolution from $H_I$ to $H_F$}}
\end{figure}

\begin{figure} 
\centering 
\subfloat[$\Omega$ =0.1]{
\label{fig:1QProb0-1} 
\includegraphics[width=0.8\textwidth]{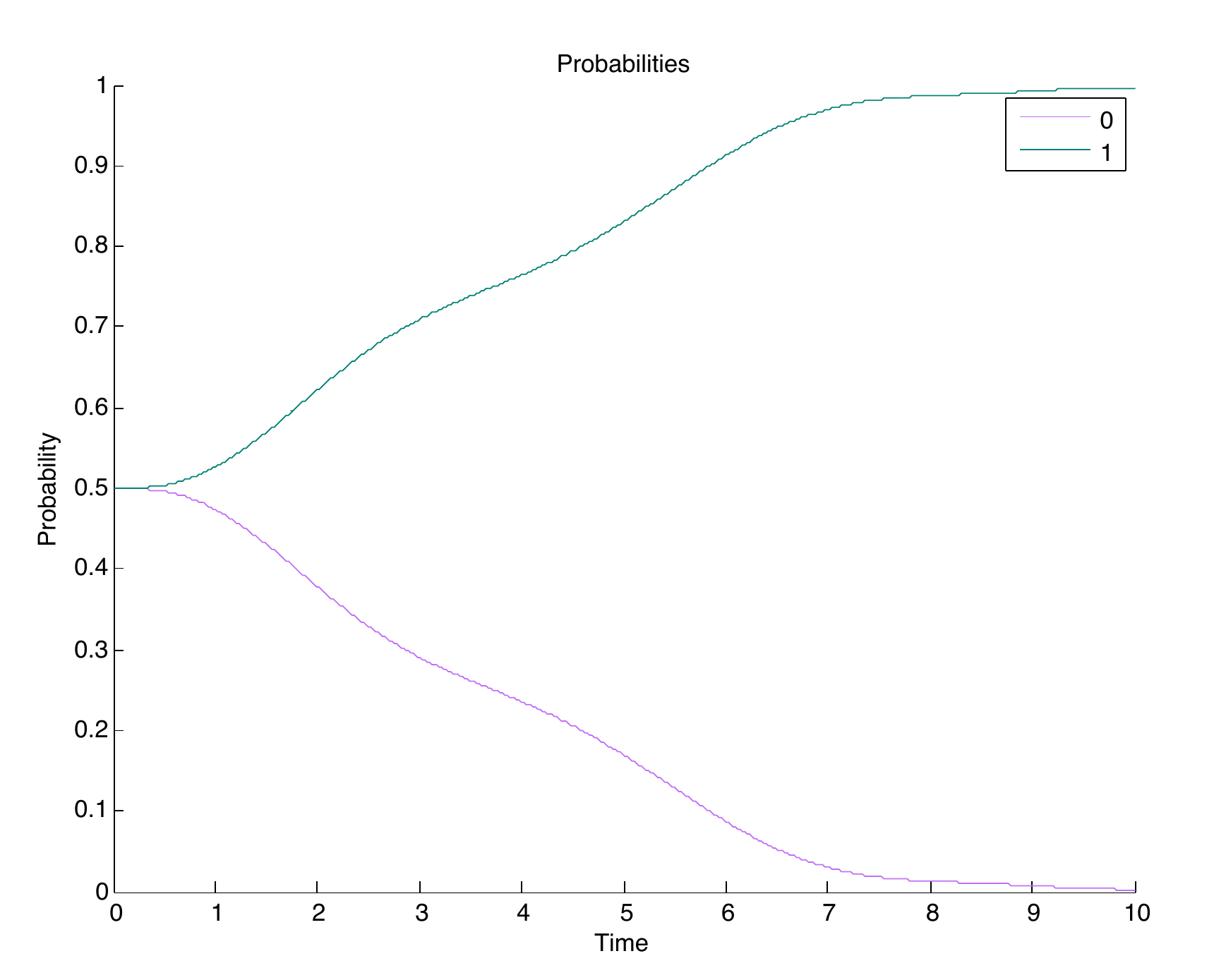} }\\
\subfloat[$\Omega$ =0.01]{
\label{fig:1QProb0-01}
\includegraphics[width=0.8\textwidth]{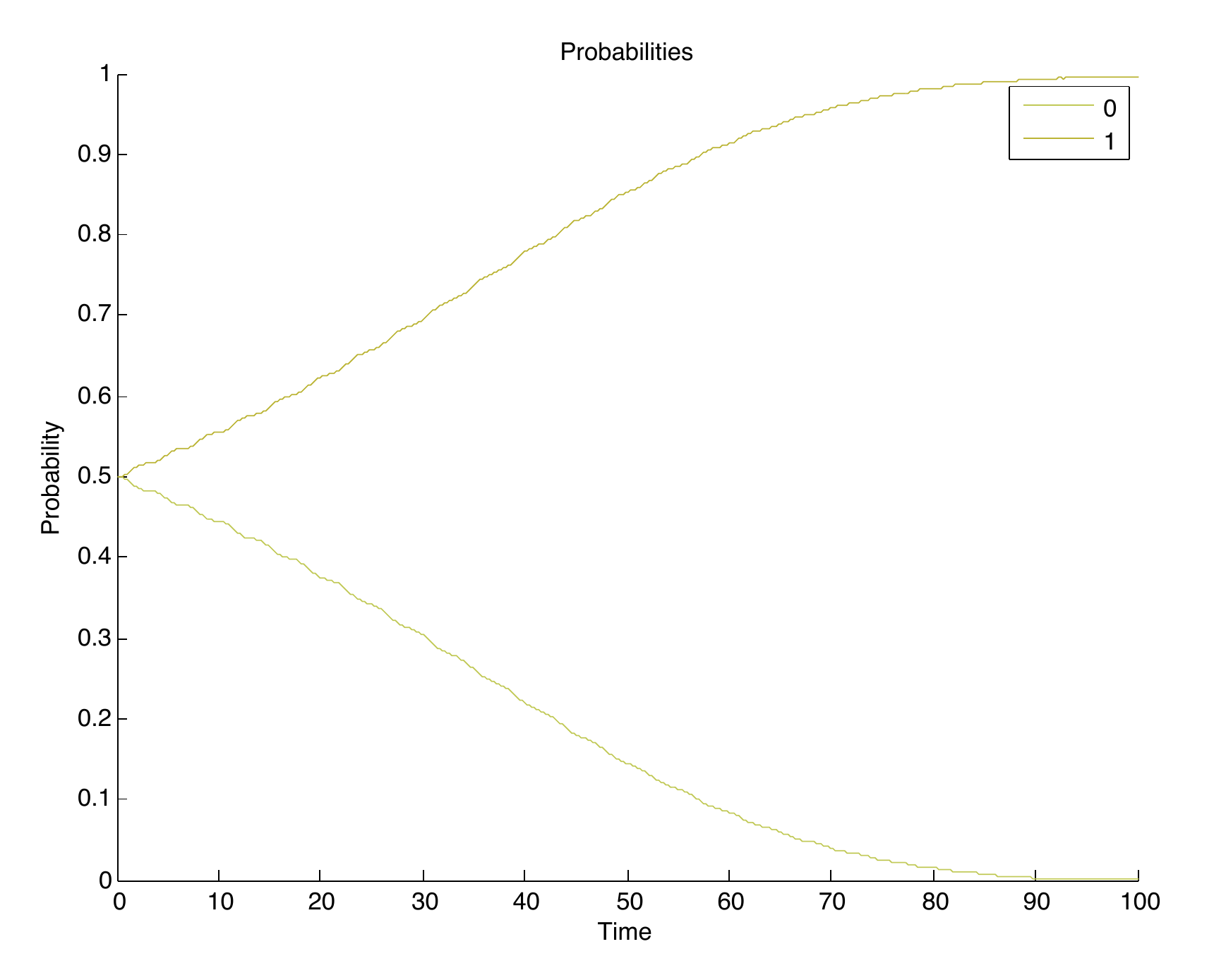} }
\textbf{\caption{Single Qubit Probabilities of States as Functions of Time After Slowly Applying a Bias in the Positive Z-Direction}}
\end{figure}

\begin{figure} 
\centering 
\subfloat[$\Omega$ =0.1]{
\label{fig:1QBloch0-1} 
\includegraphics[width=0.8\textwidth]{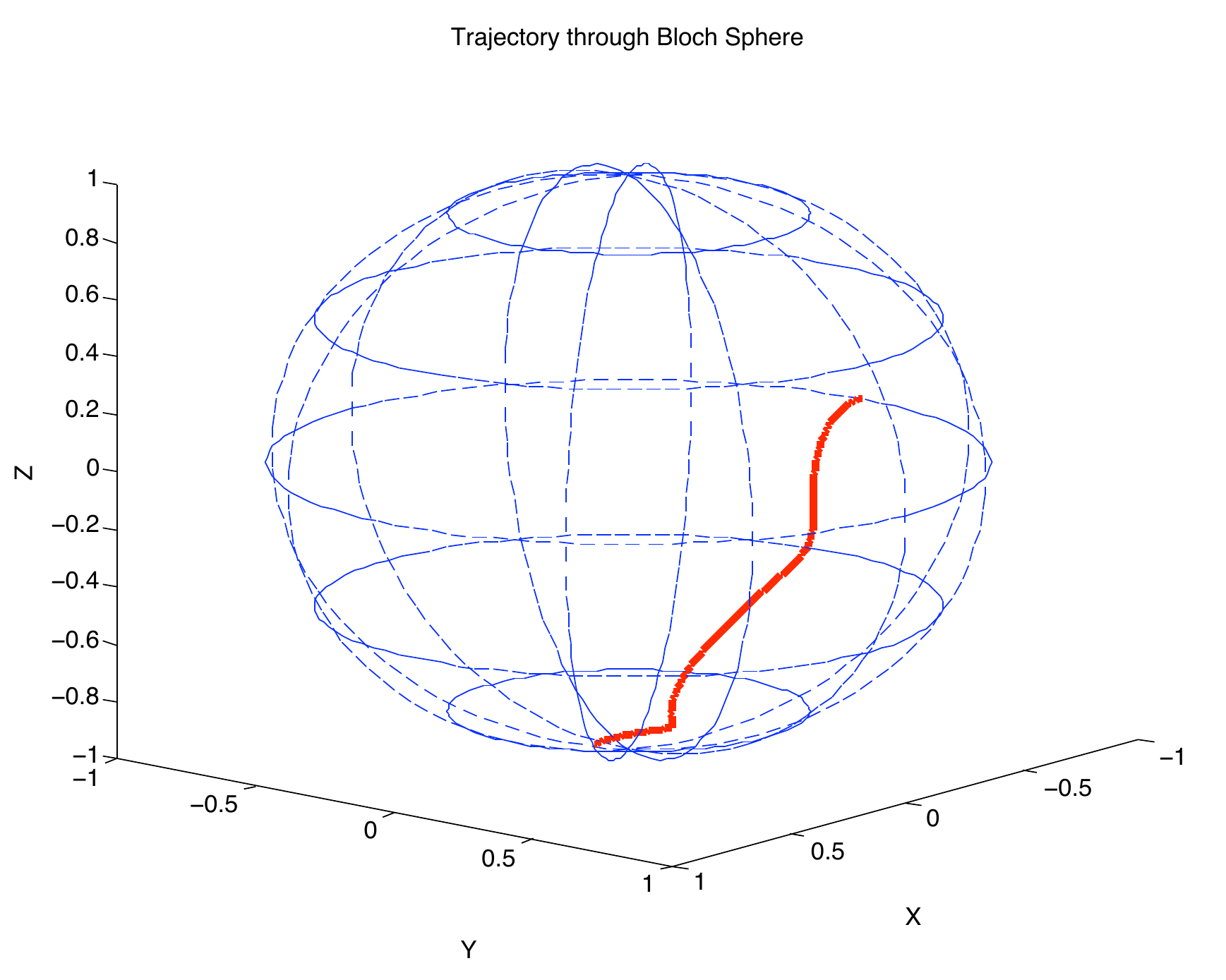} }\\
\subfloat[$\Omega$ =0.01]{
\label{fig:1QBloch0-01}
\includegraphics[width=0.8\textwidth]{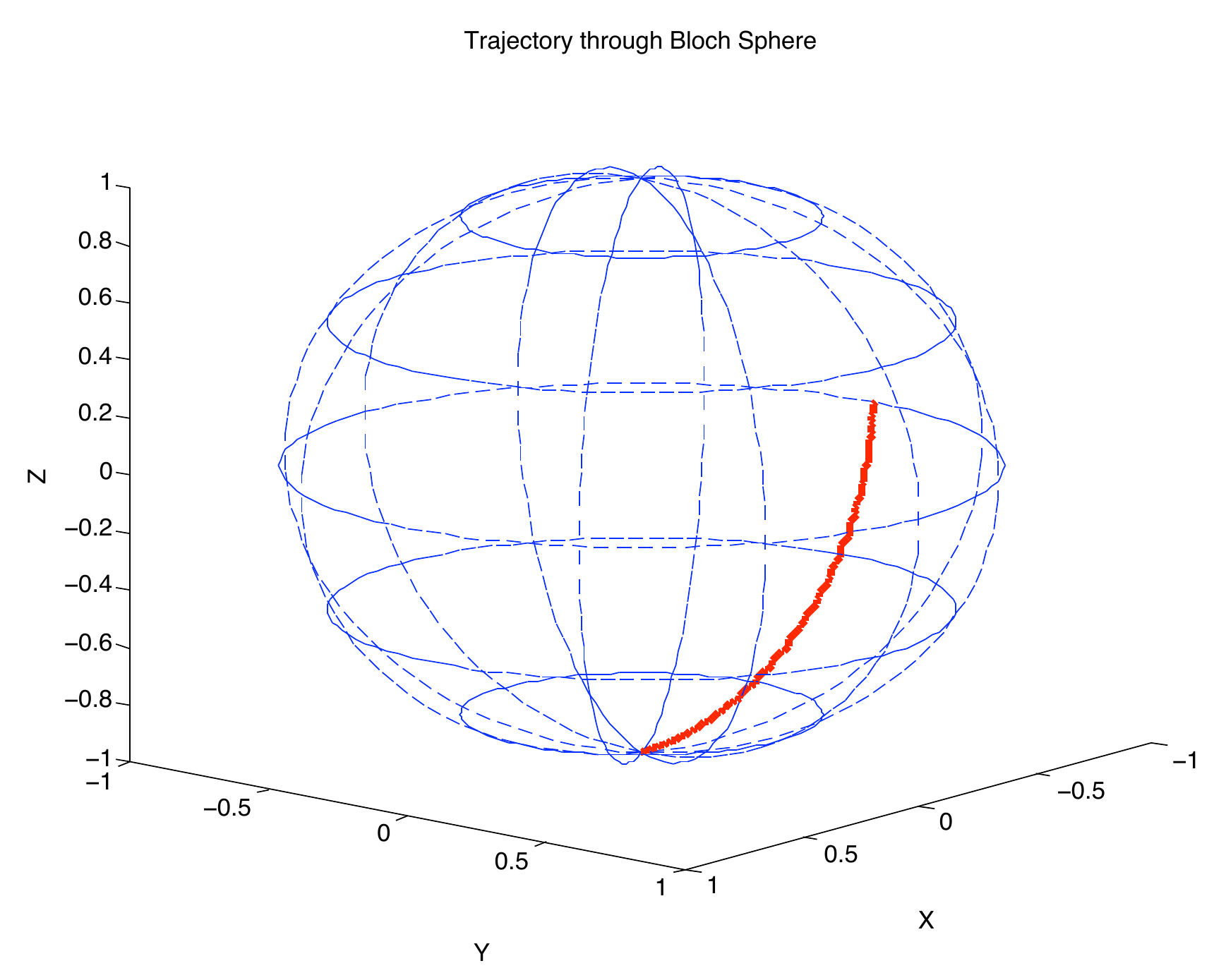} }
\textbf{\caption{Single Qubit Trajectories Through the Bloch Sphere as Functions of Time After Slowly Applying a Bias in the Positive Z-Direction}}
\end{figure}

To obtain further understanding of the qubits reaction to the applied Hamiltonians, plots of various aspects of the qubits evolution are produced for the application of the Hamiltonian that introduces a bias in the positive z-direction. This has not been done for the bias of -1 (as the plots are identical but with the states swapped). Nor have the plots been produced for application of no bias as this effect is only investigated to gather information on leakage and barely caused the qubit to evolve at all, graphically yielding nothing of interest. Plots are produced for the probabilities of states for various values of $\Omega$ (Figures \ref{fig:1QProb0-1} and \ref{fig:1QProb0-01}), eigenvalues of the two qubit states (Figure \ref{fig:1QEigs}) and the trajectory of the state of the qubit through the Bloch sphere as a function of time (Figures \ref{fig:1QBloch0-1} and \ref{fig:1QBloch0-01}).

\subsubsection{Larmor Precession}

By looking at the trajectories through the Bloch sphere for the single qubit case in Figures \ref{fig:1QBloch0-1}  and \ref{fig:1QBloch0-01} and reviewing the plots of probability over time in Figures \ref{fig:1QProb0-1}  and \ref{fig:1QProb0-01}. It is clear that some sort of oscillations are prominent in the probabilities, similarly there is a precession of the qubits orientation around the Hamiltonian direction in the Bloch spheres. This motion is similar to the trajectory of a point on the rim of a cone rolling on a flat surface. Further research reveals that the motion is due to a physical phenomena known as Larmor Precessions\cite{Fell:2008}.

Take the qubit to be a spin-$\textstyle {1\over2}$ magnetic moment where the Hamiltonian would be an applied magnetic field. If the magnetic moment is initialised at some direction and the field applied at an angle to this direction the field will exert some torque on the magnetic moment. Causing the moment to precess around the field direction, somewhat analogous to an ordinary compass with no oil (un-damped) in the earth's magnetic field or a spinning-top in the earth's gravitational field.

To test the simulation further, a simple scenario can be run to directly view the Larmor Precessions. By taking a constant Hamiltonian in the z-direction and initialising the qubit to a state other than its ground state, a precession around the field direction along the edge of the Bloch sphere should be observed, mapping out a plane across the sphere perpendicular to the z-basis. The qubit must be initialised to a state other than that of the ground state as otherwise the precession would not be visible. The precession would be occurring although it would be aligned with the field, effectively turning on a spot.

This raises the question as to why the precession is apparent during the initial evolutions in Figures \ref{fig:1QBloch0-1} and  \ref{fig:1QBloch0-01}, as the qubit is initialised in the ground state. This is due to the evolution from an initial to final Hamiltonian, for short periods of time the qubit will not be in the ground state as the Hamiltonian is constantly varying. This is why the precession is more visible on Figure \ref{fig:1QBloch0-1} compared to Figure \ref{fig:1QBloch0-01} as the Hamiltonian is varying at a greater rate.

This scenario can be solved easily by hand, here we have the Schr\"{o}dinger Equation with the constant Hamiltonian applied to the general state vector notation. Where $\hbar$ has been normalised out:

\begin{equation}
\left( \begin{array}{cc} 0 & 0 \\ 0 & 1\end{array}\right) \left( \begin{array}{c} \cos({\theta\over 2}) \\ \sin({\theta\over 2}) e^{i\phi} \end{array} \right) = i {d\over dt} \left( \begin{array}{c} \cos({\theta\over 2}) \\ \sin({\theta\over 2}) e^{i\phi} \end{array}\right)
\end{equation}

\begin{equation}
\left( \begin{array}{c} 0 \\ \sin({\theta\over 2}) e^{i\phi} \end{array} \right) = i {d\over dt} \left( \begin{array}{c} \cos({\theta\over 2}) \\ \sin({\theta\over 2}) e^{i\phi} \end{array}\right)
\end{equation}

By solving the system as two ODEs, a solution to the top half of the equation is easily found and is shown in Equation \ref{equ:Const}. This means that the coefficient for state $\vert 0\rangle$ is constant so the probability of measurement for both states remains constant as it solely relies of the value of $\theta$.

\begin{equation}
0 = i {d\over dt}\cos(\textstyle{\theta\over 2})
\end{equation}

\begin{equation}
\therefore \cos(\textstyle{\theta \over 2}) = constant \label{equ:Const}
\end{equation}

Going on to solve the second ODE, knowing that $cos({\theta \over 2})$ is constant the following can be determined:

\begin{equation}
\sin(\textstyle{\theta\over 2}) e^{i\phi} = i \displaystyle{d\over dt} \sin(\textstyle{\theta\over 2}) e^{i\phi} 
\end{equation}

\begin{equation}
\therefore e^{i\phi} = i{d\over dt}e^{i\phi} = -{d\phi \over dt}e^{i\phi}
\end{equation}

\begin{equation}
\therefore \phi= -t \label{equ:tDep}
\end{equation}

\begin{figure} 
\centering 
\subfloat[$\Omega$ =0.1, $\vert\psi\rangle= {1\over\sqrt{2}}(\sqrt{0.9}\vert 0\rangle +\sqrt{0.1}\vert 1\rangle)$]{
\label{fig:1QLarmoor} 
\includegraphics[width=0.8\textwidth]{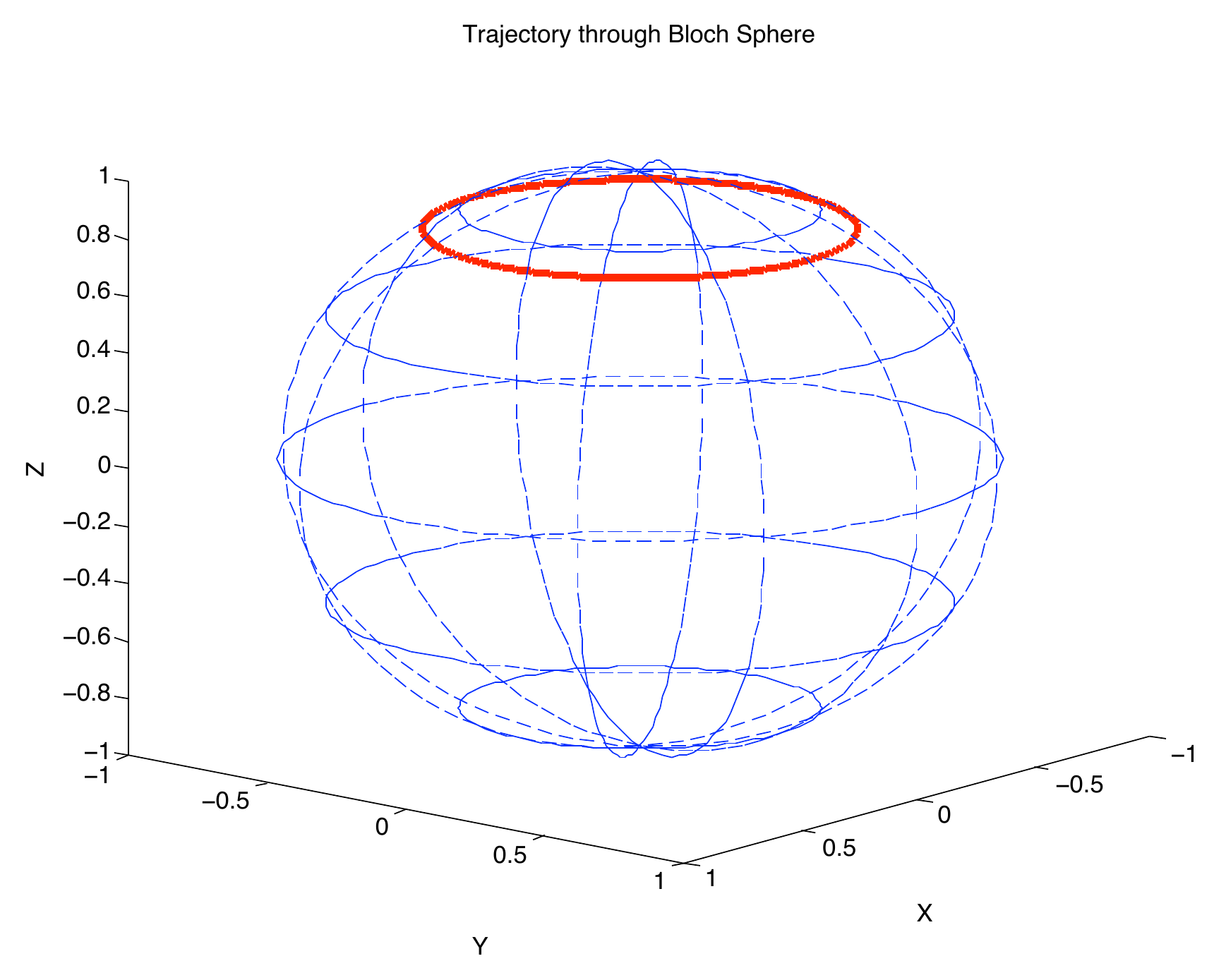} }\\
\subfloat[$\Omega$ =0.1, $\vert\psi\rangle= {1\over\sqrt{2}}(\sqrt{0.7}\vert 0\rangle +\sqrt{0.3}\vert 1\rangle)$]{
\label{fig:1QLarmoor1}
\includegraphics[width=0.8\textwidth]{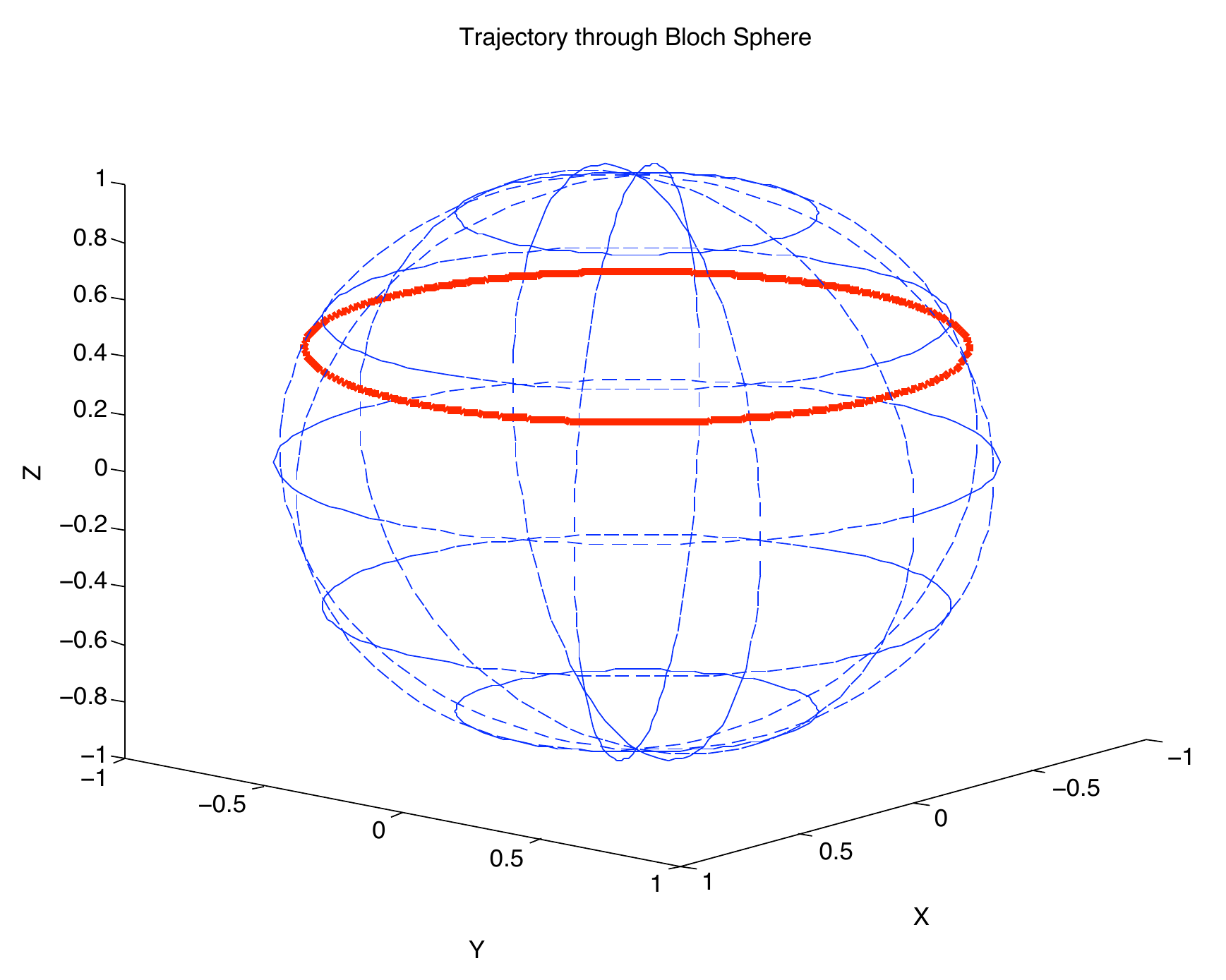} }
\textbf{\caption{State Precession Around Bloch Sphere with Application of a Constant Hamiltonian in the Z-Direction and Initialising the Qubit to the States Stated Above}}
\end{figure}

From Equation \ref{equ:tDep} it is clear that the phase $\phi$ is dependent on the time $t$. This time dependance is not visible in Figures \ref{fig:1QLarmoor} and \ref{fig:1QLarmoor1} as the precession overlaps many times to trace a circle on the edge of the Bloch sphere.

The Larmor frequency for full revolutions of the Bloch sphere is a known constant for many physically realisable qubits. This constant may be a good means of calibration for the unit of time in the simulation dependant on the type of qubit that is being modelled.

\subsection{2 Qubit Simulation}

The 2 qubit simulation is run to test various factors. Namely how the concurrence is affected by the applied final Hamiltonian (mainly the Ising interaction), a simple check of consistency of the calculated eigenvalues and finally if the final states have probabilities that resemble the expected most probable results.

From studying Table \ref{table:1q} for the single qubit simulation, it appears that the most adequate single time scale to run these simulations is with an $\Omega$ value of 0.1. This is the point that the probabilities give a clear indication of the correct results for the specific simulation and with this rate there is very little truncation error.

\begin{table}
\centering
\begin{tabular}{c||c|c||c|c||c|c}
{$h_i$=0} & \multicolumn{2}{c||}{$J_{ij}$=0} & \multicolumn{2}{c||}{$J_{ij}$=1} & \multicolumn{2}{c}{$J_{ij}$=-1}\\
{} & E & P  & E & P  & E & P \\
\hline
$\vert 00\rangle$ & 0 & 0.249923 & -1 & 0.498785 & 1 & 0.000910\\
$\vert 01\rangle$ & 0 & 0.249923 & 1 & 0.000910 & -1 & 0.498785\\
$\vert 10\rangle$ & 0 & 0.249923 & 1 & 0.000910 & -1 & 0.498785\\
$\vert 11\rangle$ & 0 & 0.249923 & -1 & 0.498785 & 1 & 0.000910\\
\multicolumn{7}{l}{}\\
{$h_i$=1} & \multicolumn{2}{c||}{$J_{ij}$=0} & \multicolumn{2}{c||}{$J_{ij}$=1} & \multicolumn{2}{c}{$J_{ij}$=-1}\\
{} & E & P  & E & P  & E & P \\
\hline
$\vert 00\rangle$ & 2 & 0.000007 & 1 & 0.000316 & 3 & 0.000159\\
$\vert 01\rangle$ & 0 & 0.002724 & 1 & 0.000126 & -1 & 0.248075\\
$\vert 10\rangle$ & 0 & 0.002724 & 1 & 0.000126 & -1 & 0.248075\\
$\vert 11\rangle$ & -2 & 0.993582 & -3 & 0.998174 & -1 & 0.503156\\
\end{tabular}
\textbf{\caption{Table of Obtained Final Eigenvalues and Probabilities for All 2 Qubit States in Specified Circumstances ($\Omega$=0.1)\label{table:2q}}}
\end{table}

Table \ref{table:2q} is a brief outline of the results obtained for Ising interactions -1, 0 and 1 with and without an (identical) local Hamiltonian acting on each of the qubits. The table contains both the final eigenvalues and probabilities of each of the states. By calculating the ground state energies of all 4 states in each individual set of conditions using Equation \ref{equ:Ising}, the eigenvalues have been confirmed. As for probabilities, the states with lowest eigenvalues (ground states) have the highest probabilities. In the cases where two or more states have the same lowest eigenvalue their probabilities are equal highest apart from the case where $H_i =1$ and $J_{ij}=-1$ where three states all have an identical final eigenvalue. In this case the probability is shared but not equally. As expected the two states related by symmetry ($\vert 01\rangle$ and $\vert 10\rangle$) have equal probability but the sum of their probabilities is equivalent to the single probability of state $\vert 11\rangle$. This is unexpected but the most plausible reason for this is the fact that state $\vert 11\rangle$ had the lowest eigenvalue during the majority of the evolution, at the end of the evolution when the two additional states joined the lowest energy group the possibility of tunnelling to one of these became less probable therefore the possibly of remaining in this state is higher.

\begin{figure} 
\centering 
\subfloat[$h_i$=0]{
\label{fig:2QconcJ-1H0} 
\includegraphics[width=0.8\textwidth]{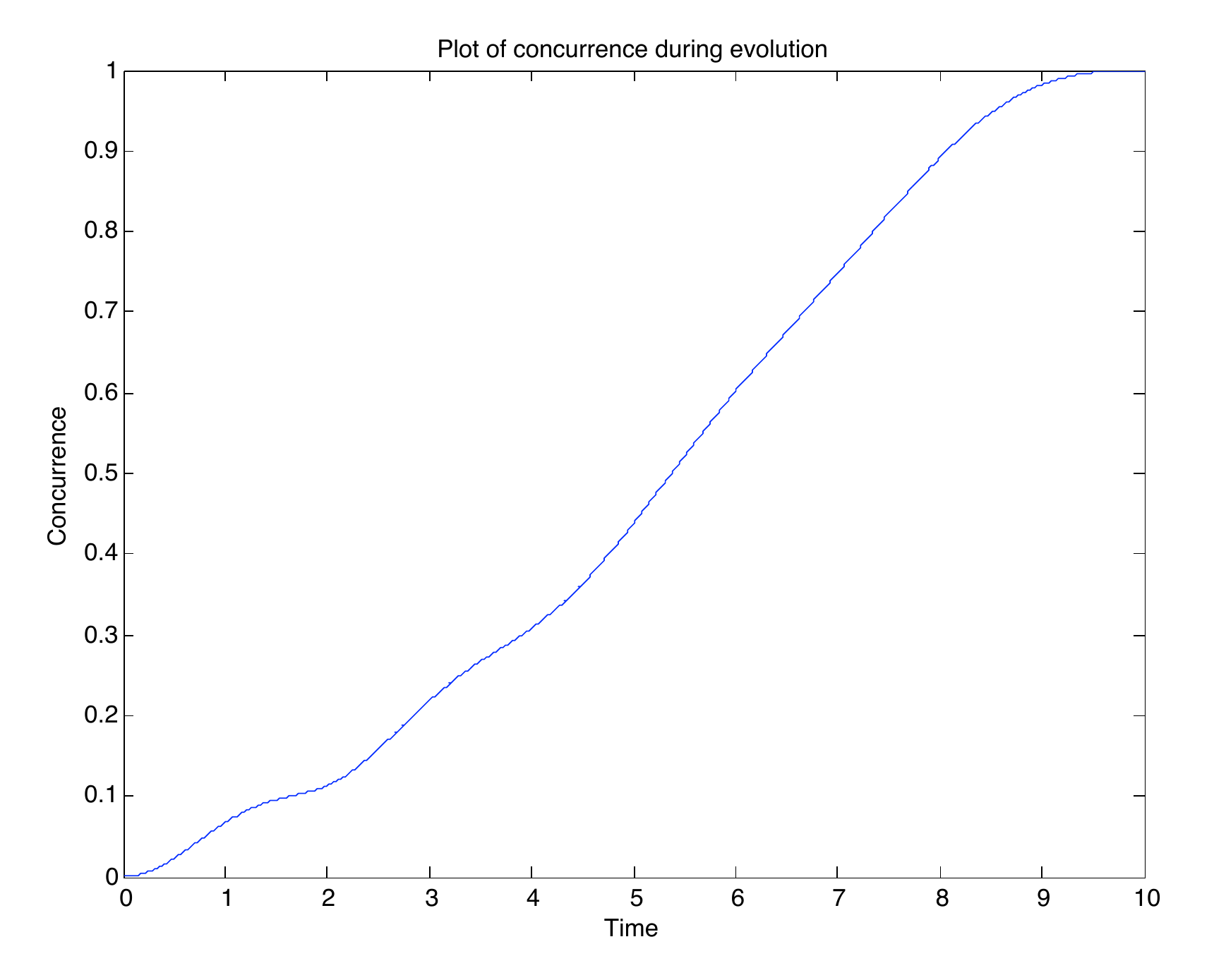} }\\
\subfloat[$h_i$=1]{
\label{fig:2QconcJ-1H1}
\includegraphics[width=0.8\textwidth]{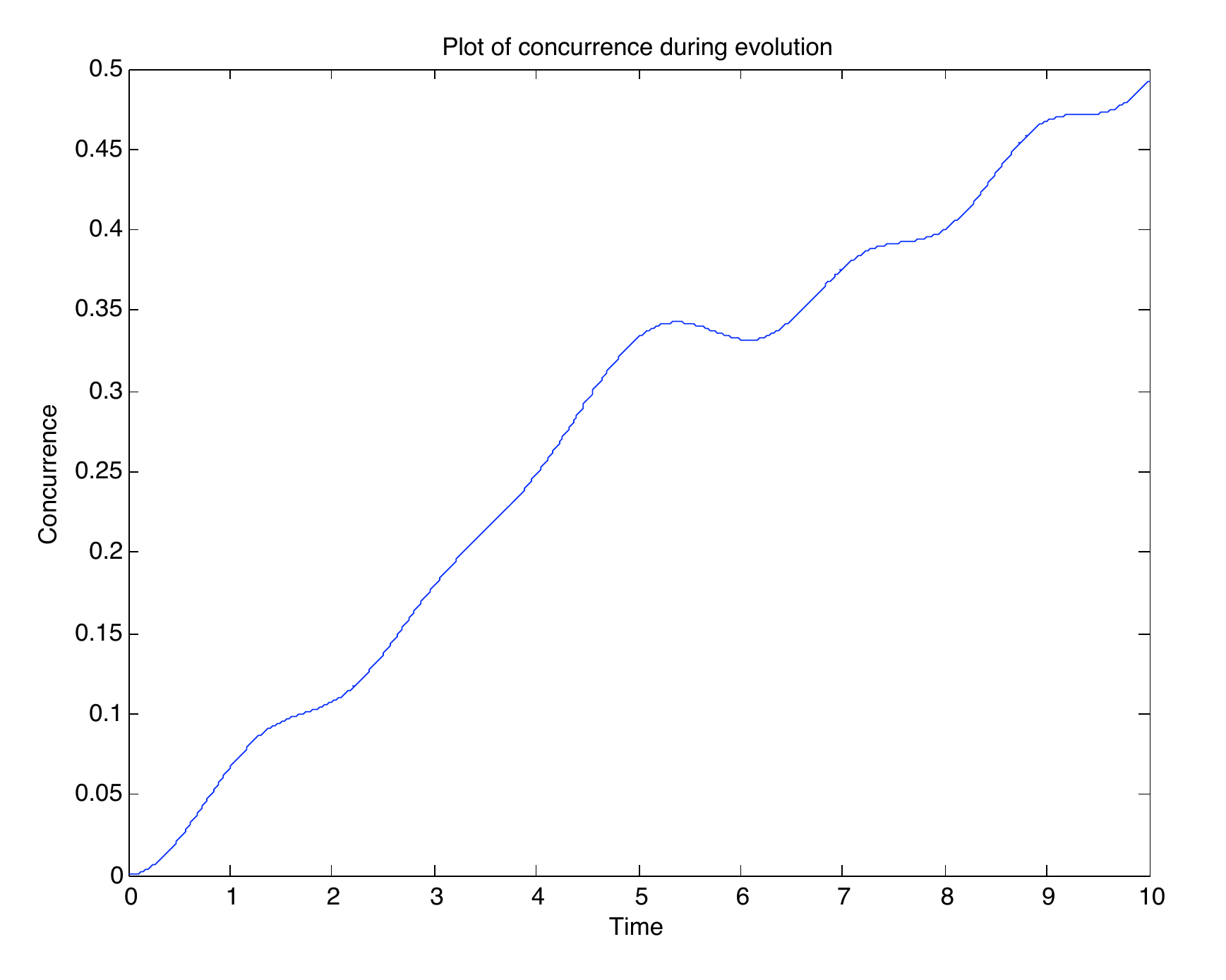} }
\textbf{\caption{Plots of Concurrence for Fixed Value of J=-1 and Varying Local Hamiltonian ($h_i$)}}
\end{figure}

\begin{figure}
\ContinuedFloat 
\centering 
\subfloat[$h_i$=0]{
\label{fig:2QconcJ0H0} 
\includegraphics[width=0.8\textwidth]{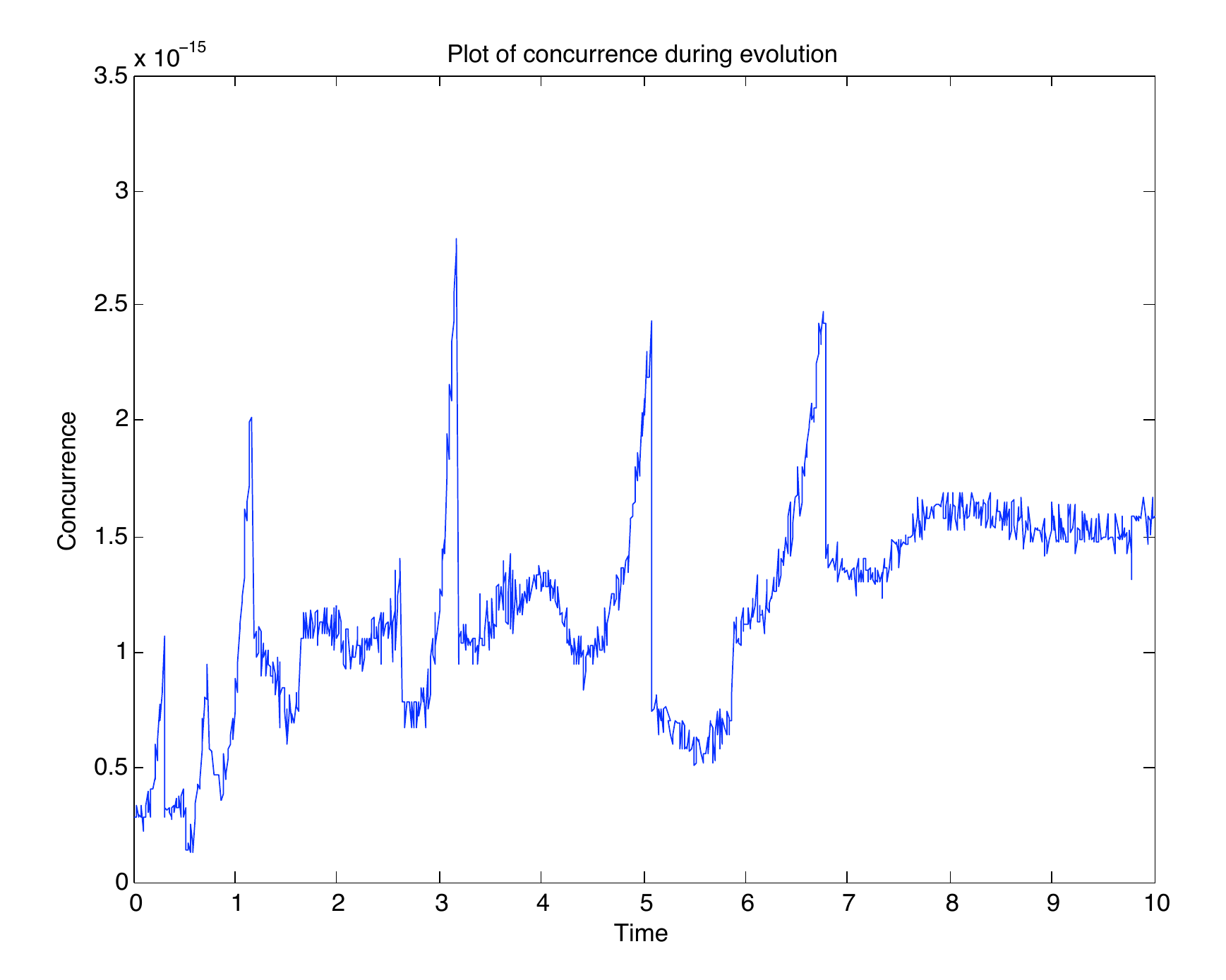} }\\
\subfloat[$h_i$=1]{
\label{fig:2QconcJ0H1}
\includegraphics[width=0.8\textwidth]{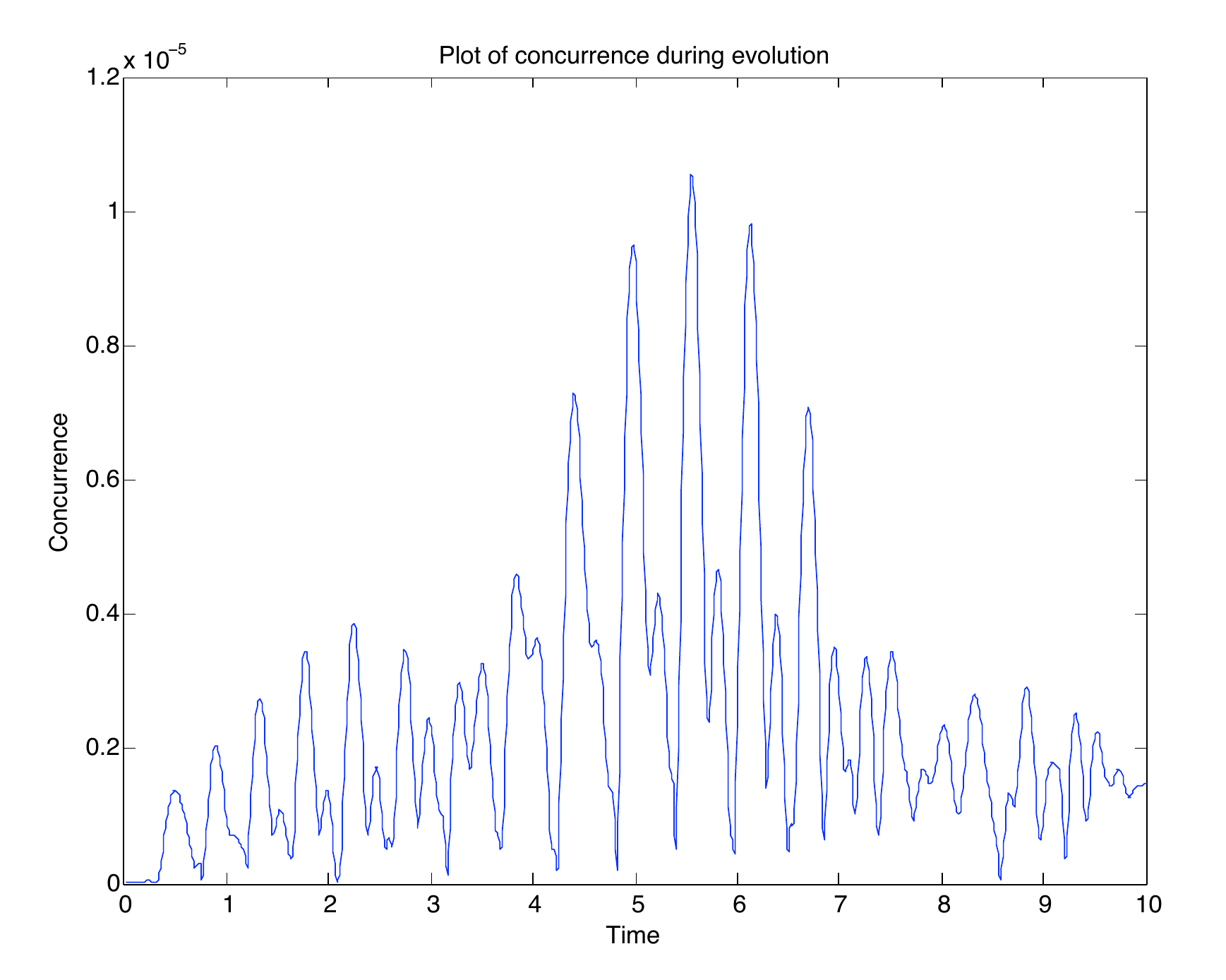} }
\textbf{\caption{Plots of Concurrence for Fixed Value of J=0 and Varying Local Hamiltonian ($h_i$)}}
\end{figure}

\begin{figure}
\ContinuedFloat
\centering 
\subfloat[$h_i$=0]{
\label{fig:2QconcJ1H0} 
\includegraphics[width=0.8\textwidth]{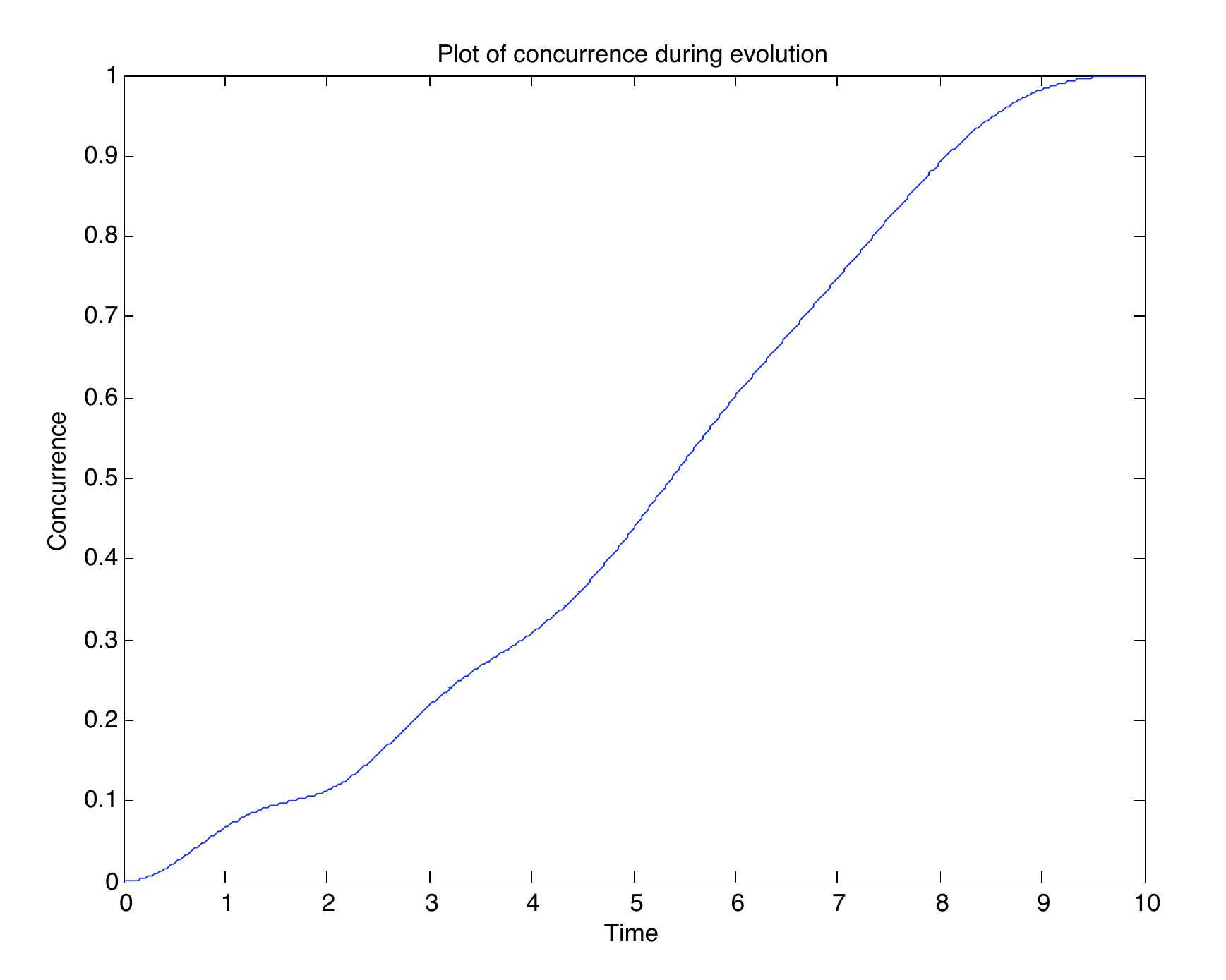} }\\
\subfloat[$h_i$=1]{
\label{fig:2QconcJ1H1}
\includegraphics[width=0.8\textwidth]{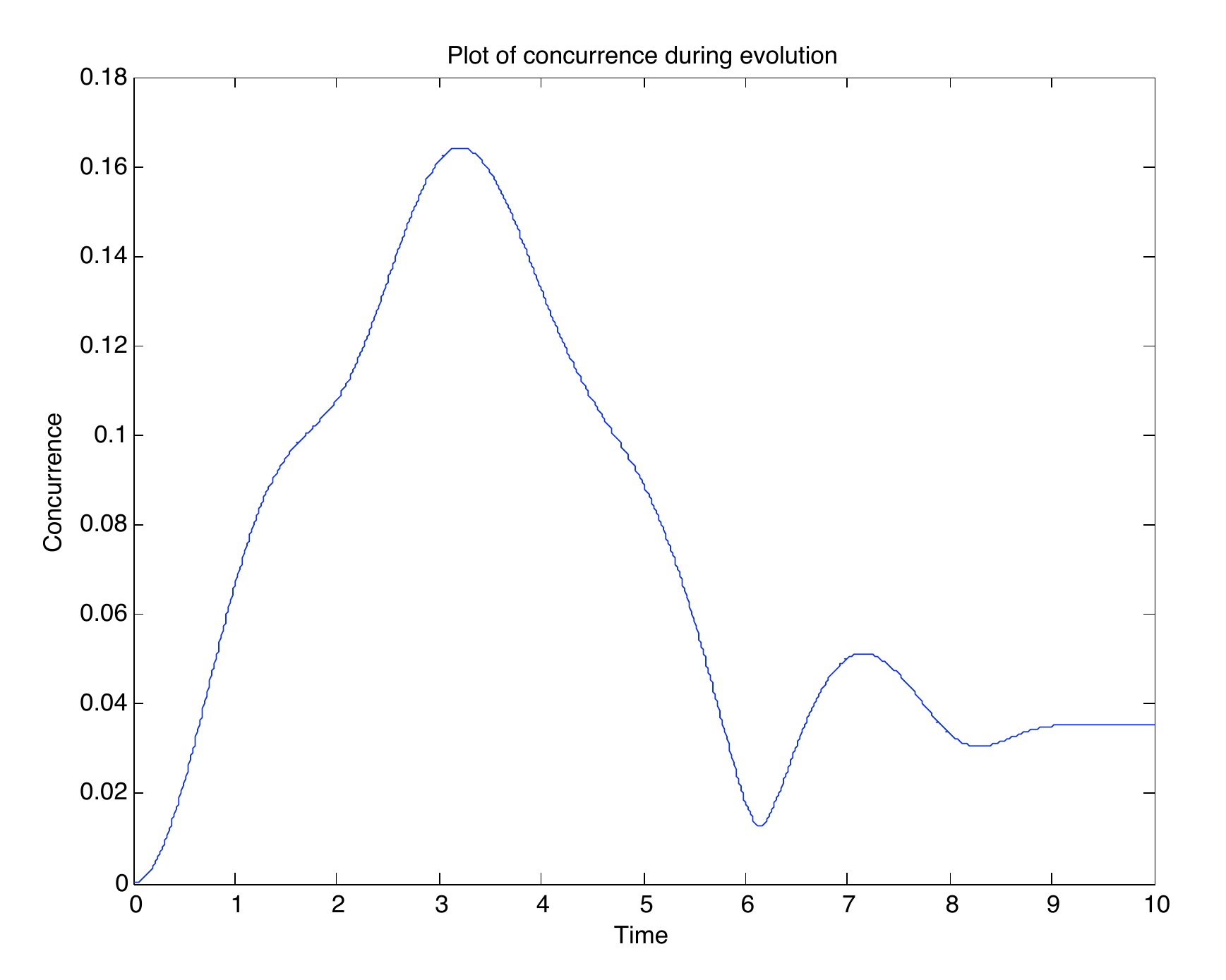} }
\textbf{\caption{Plots of Concurrence for Fixed Value of J=1 and Varying Local Hamiltonian ($h_i$)}}
\end{figure}

The concurrence is a measure of entanglement between the qubits, it ranges between 0 (not entangled) and 1 (fully entangled). The concurrence is calculated over time from the state vector of the system, if the state vector has the following coefficients:

\begin{equation}
\vert\Psi\rangle = c_{\vert 00\rangle} \vert 00\rangle+c_{\vert 01\rangle} \vert 01\rangle+c_{\vert 10\rangle} \vert 10\rangle+c_{\vert 11\rangle} \vert 11\rangle \label{equ:CState}
\end{equation}

The entanglement is calculated in the following way:

\begin{equation}
C= 2(\vert c_{\vert 00\rangle}\vert\cdot\vert c_{\vert 11\rangle}\vert - \vert c_{\vert 01\rangle}\vert\cdot\vert c_{\vert 10\rangle}\vert)
\label{equ:Conc}
\end{equation}

The plots of concurrence during the evolutions can be found in Figures \ref{fig:2QconcJ-1H0} to \ref{fig:2QconcJ1H1}. When there is no interaction between the qubits (J=0), it is clear that they are completely isolated and work independently. The plots of concurrence for this situation in Figures \ref{fig:2QconcJ0H0} and \ref{fig:2QconcJ0H1} are simply due to the truncation errors in the program, with the magnitudes of concurrence that are present, they can be regarded as not entangled at all.

Studying Figures \ref{fig:2QconcJ-1H0} and \ref{fig:2QconcJ-1H1} where the interaction between the qubits means that it is more favourable for the qubits to be anti-aligned indicated that when no local Hamiltonian is applied the qubits only interact with each other. This is indicated in Figure \ref{fig:2QconcJ-1H0} as the concurrence reaches almost 1, which is the point of maximal entanglement. When a local Hamiltonian is applied, it is equally favourable for the qubits to be anti-aligned as well as in state $\vert 11\rangle$ where both are aligned in the field direction. The concurrence reaches a value of 0.5 which can be said to be only half entangled. There is good reasoning for this, it is thought that if two qubits are entangled that by measuring one, information is gained of the state of the other. In the case where the qubits are maximally entangled (no local Hamiltonians), by measuring the first qubit it can be determined that the second will be of opposite parity. In the case of half entanglement, after measuring the first qubit, if the readout is 0, then the second qubit due to entanglement will be anti-aligned. But if the readout is 1, than it cannot be determined if the second qubit is in state 0 or 1. In basic terms this means that only half the measurements on qubit one will reveal the state of qubit two, therefore half-entangled.

In the scenario where it is more favourable for the qubits to be aligned due to their Ising interaction (J=1), without a local Hamiltonian the concurrence reaches a value of 1. Whereby if the first qubit is measured, the second qubit will almost always be in the same state, this is observable in Figure \ref{fig:2QconcJ1H0}. More complications arise when the local Hamiltonian is applied to both qubits, the entanglement appears to rise at the same rate as before but suddenly drops off a third of the way through the simulation and ends with a very low concurrence of about 0.04. It looks as if the application of the local Hamiltonians breaks the entanglement of the qubits. From Table \ref{table:2q} it is clear that the most probable state is $\vert 11\rangle$ although this appears to be fully determined by the local Hamiltonians and not the interaction between qubits. Therefore, by measuring the first qubit as 1, the second qubit will most probably also be one, but this will mainly be due to the applied bias and not the interaction between the two qubits.

\subsection{3 Qubit Simulation}
The three qubit simulation is the lowest possible system where frustration occurs. This is one of the many outcomes of the Ising spin model that can be applied to Spin Glasses (highly frustrated materials). If atoms in a lattice are arranged to be highly connected with nearest neighbours in some triangular form, it is easy to find the ground state of the system if the interactions between the atoms are ferromagnetic, as it is energetically favourable for all the atoms to have their magnetic moments aligned in the same direction. This ease is not reflected in the antiferromagnetic state.

Taking a classical three particle example, where each particle is always connected to the other two, in a triangular arrangement: If anti-alignment is favourable, both other particles should have opposite spin to the first, although as they are also connected directly, do they accept being in the same alignment or does one of them flip? If so which one, and once flipped do the other two experience the same issue? The easy answer is that the system is frustrated, and that any arrangement of the three particles where one is the opposite parity to the other two can be found at any one time. The process of flipping orientation is completely random and the system will never be satisfied with the arrangement, so the probability of finding any one of these arrangements should all be the same. 

Replacing these particles with qubits, the probability of measuring each of the states will be identical to the classical picture. Although when using qubits, due to their quantum nature, before measurement they will be in a superposition of all ground states and measurement will cause their wave-function to collapse into one of the classical states. This is clear from Table \ref{table:3q} as all states have an equal probability of being measured bar the two states where all qubits have identical orientations.

\begin{table}
\centering
\begin{tabular}{c||c|c||c|c}
{} & \multicolumn{2}{c||}{$h_i$=0, $J_{ij}$=-1} & \multicolumn{2}{c}{$h_i$=1, $J_{ij}$=-1}\\
{} & E & P  & E & P \\
\hline
$\vert 000\rangle$ & 3 & 0.000082 & 6 & 0.000002\\
$\vert 001\rangle$ & -1 & 0.166504 & 0 & 0.000291\\
$\vert 010\rangle$ & -1 & 0.166504 & 0 & 0.000291\\
$\vert 011\rangle$ & -1 & 0.166504 & -2 & 0.330406\\
$\vert 100\rangle$ & -1 & 0.166504 & 0 & 0.000291\\
$\vert 101\rangle$ & -1 & 0.166504 & -2 & 0.330406\\
$\vert 110\rangle$ & -1 & 0.166504 & -2 & 0.330406\\
$\vert 111\rangle$ & 3 & 0.000082 & 0 & 0.007053\\
\end{tabular}
\textbf{\caption{Table of Obtained Final Eigenvalues  and Probabilities for All Three Qubit States with Antiferromagnetic Interactions ($\Omega$=0.1)\label{table:3q}}}
\end{table}

\begin{figure} 
\centering 
\subfloat[Eigenvalues]{
\label{fig:3QeigJ-1H0} 
\includegraphics[width=0.8\textwidth]{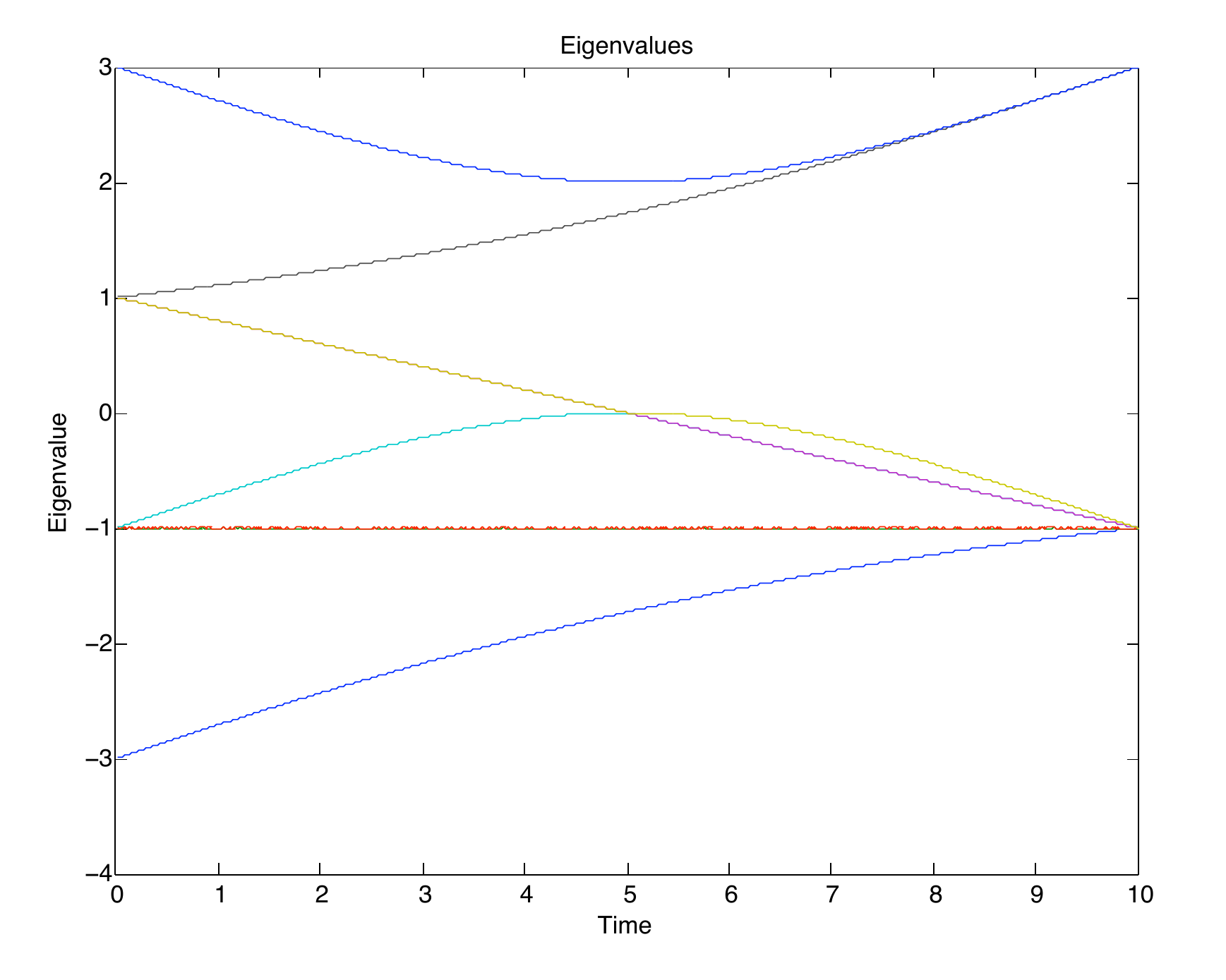} }\\
\subfloat[Probabilities]{
\label{fig:3QprobJ-1H0}
\includegraphics[width=0.8\textwidth]{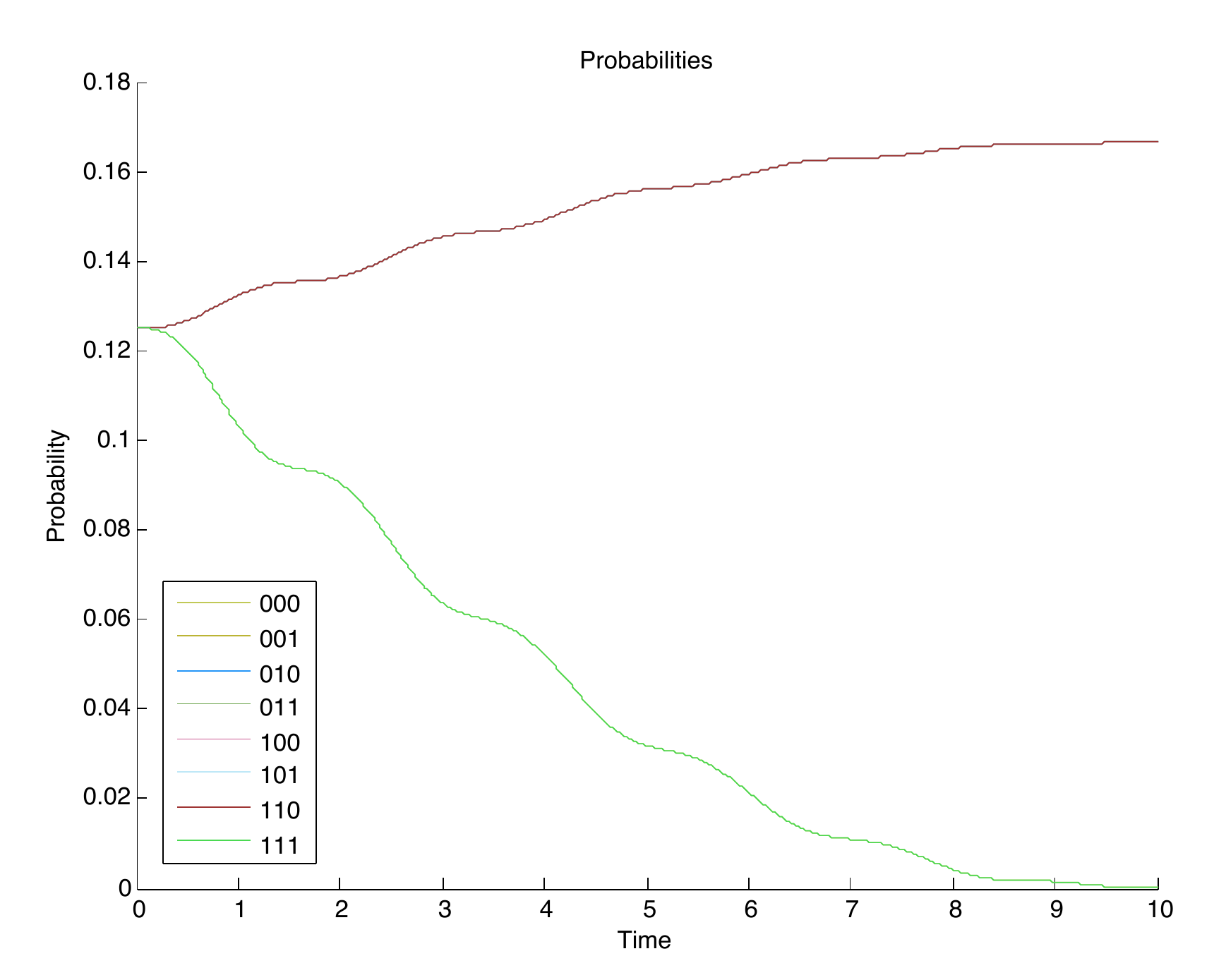} }
\textbf{\caption{Plots of Probabilities and Eigenvalues as Functions of Time for a Three Qubit System with Antiferromagnetic Interactions ($\Omega$=0.1, $J_{ij}$=-1, $h_i$=0)}}
\end{figure}

\begin{figure}
\centering 
\subfloat[Eigenvalues]{
\label{fig:3QeigJ-1H1} 
\includegraphics[width=0.8\textwidth]{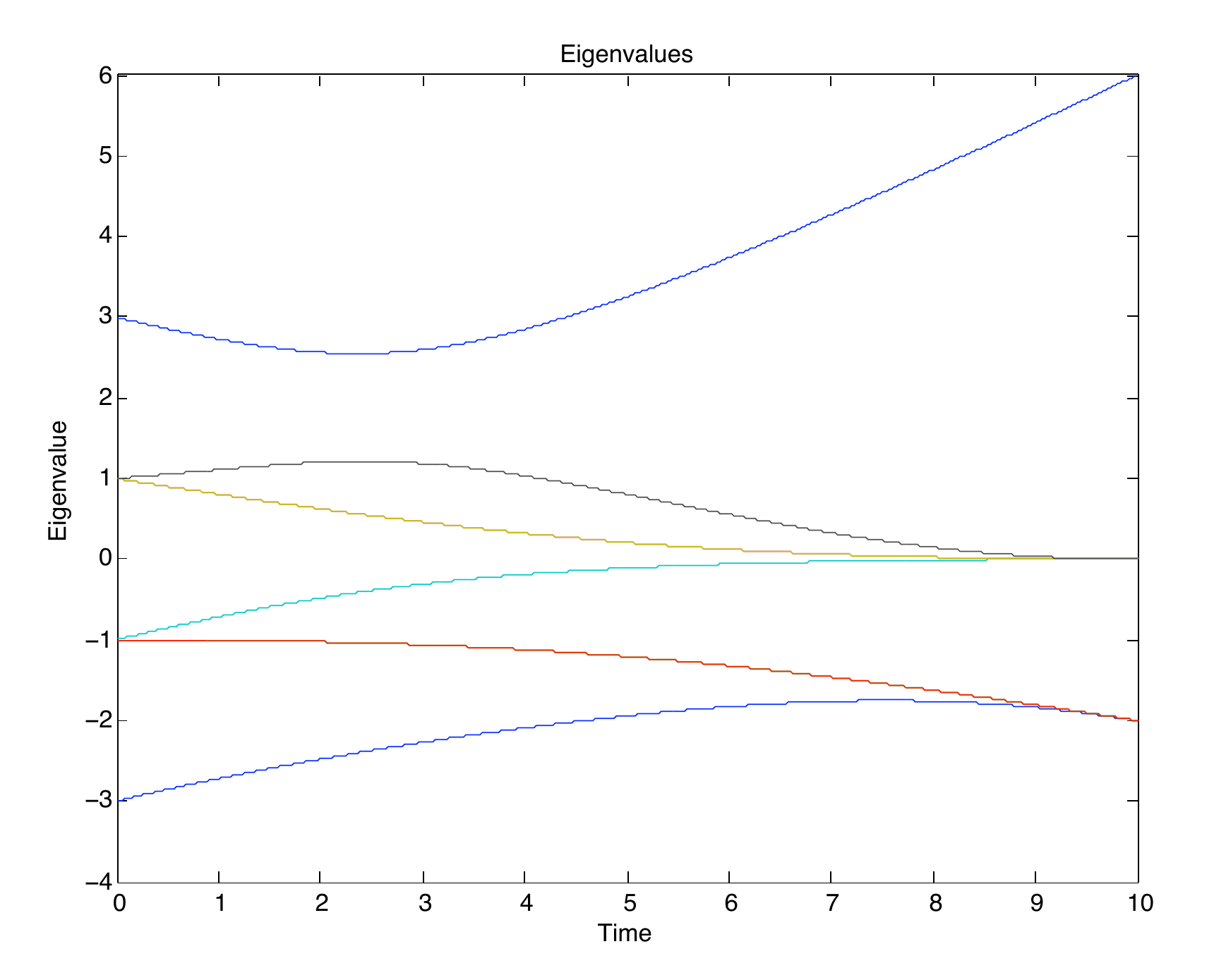} }\\
\subfloat[Probabilities]{
\label{fig:3QprobJ-1H1}
\includegraphics[width=0.8\textwidth]{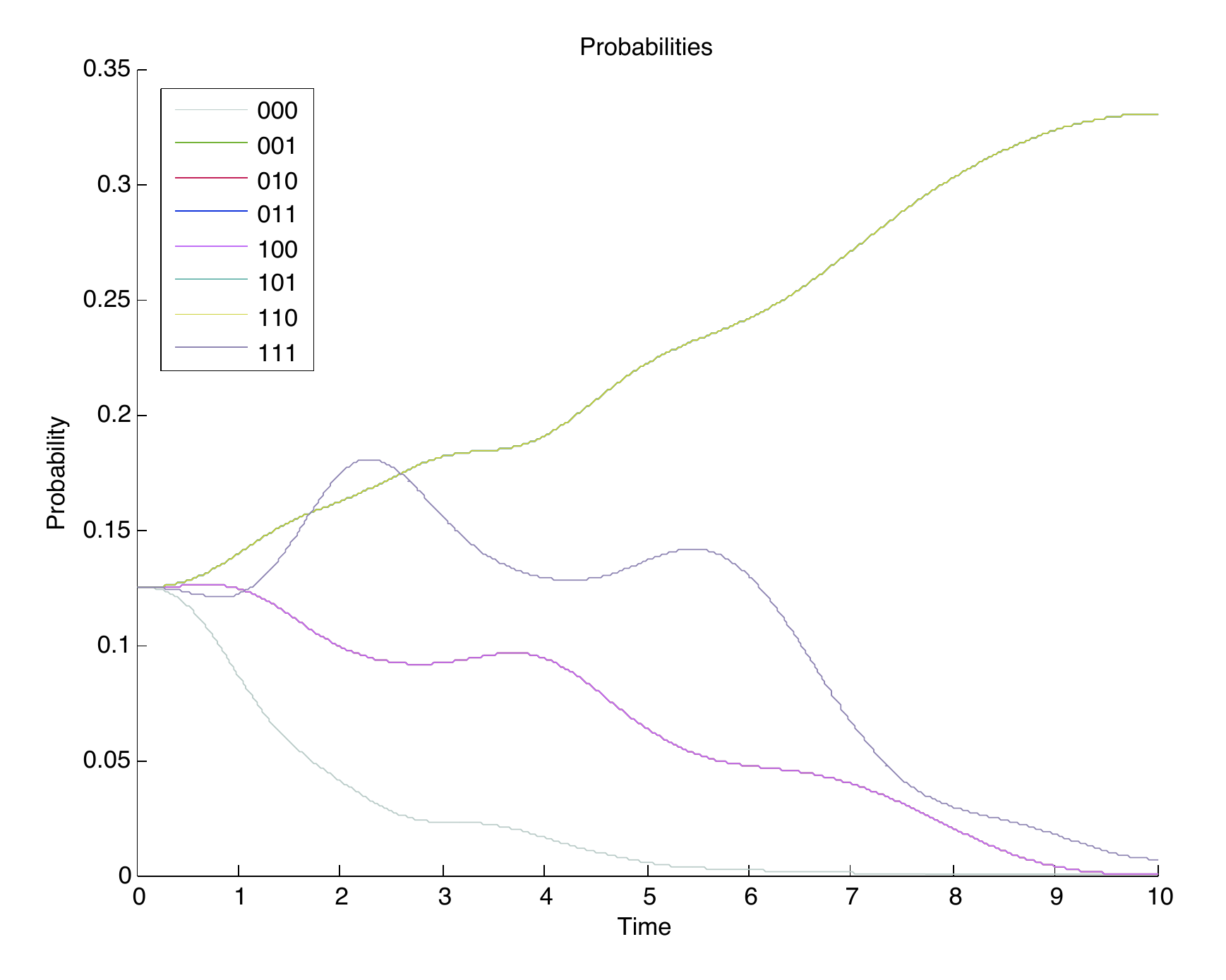} }
\textbf{\caption{Plots of Probabilities and Eigenvalues as Functions of Time for a Three Qubit System with Antiferromagnetic Interactions. ($\Omega$=0.1, $J_{ij}$=-1, $h_i$=1)}}
\end{figure}

After applying the local Hamiltonian to all the qubits, three states become most favourable ($\vert 011\rangle, \vert 101\rangle$ and $\vert 110\rangle$). All of them include two 1s and one 0 and can be regarded as identical due to symmetry. If the triangle of qubits were to be rotated you would not be able to distinguish one state from another. The reason for these states having the lowest energies is due to the fact that having two 1s lowers the energy of the system as two of the qubits are aligned with the local Hamiltonian direction which is energetically more favourable. Even though the third qubit is anti-aligned to the local Hamiltonian and raises the energy of the system, by being of this orientation more of the Ising interactions are satisfied, thus lowering the energy overall. The eigenvalues and probabilities of states listed in Table \ref{table:3q} can be found from the plots in Figures \ref{fig:3QeigJ-1H0} to \ref{fig:3QprobJ-1H1}.

\subsection{4 Qubit Simulation} \label{subsect:4Q}
\begin{figure} 
\centering 
\includegraphics{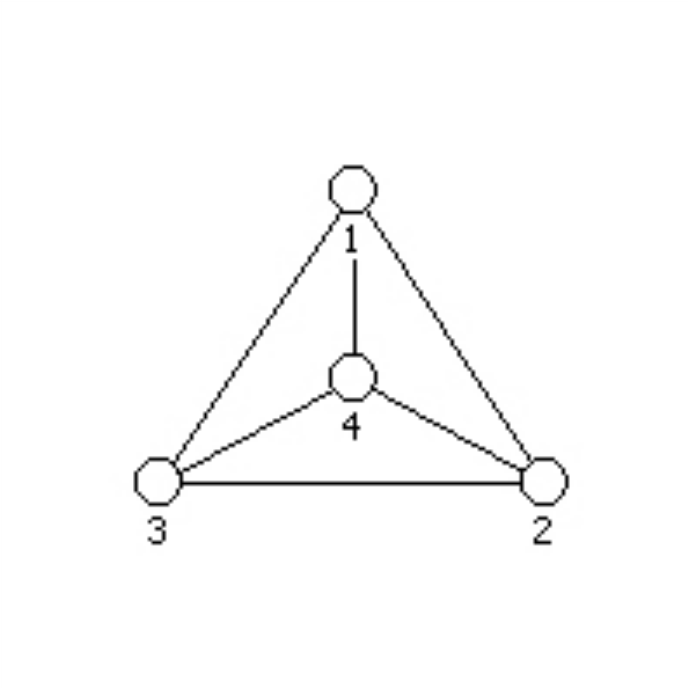}
\textbf{\caption{Fully Connected Cubic Planar 4 Node Graph G=(V,E), where Vertices V Represent Qubits and Edges E Ising Interactions\label{fig:4Q}}}
\end{figure}

\begin{figure} 
\centering 
\includegraphics{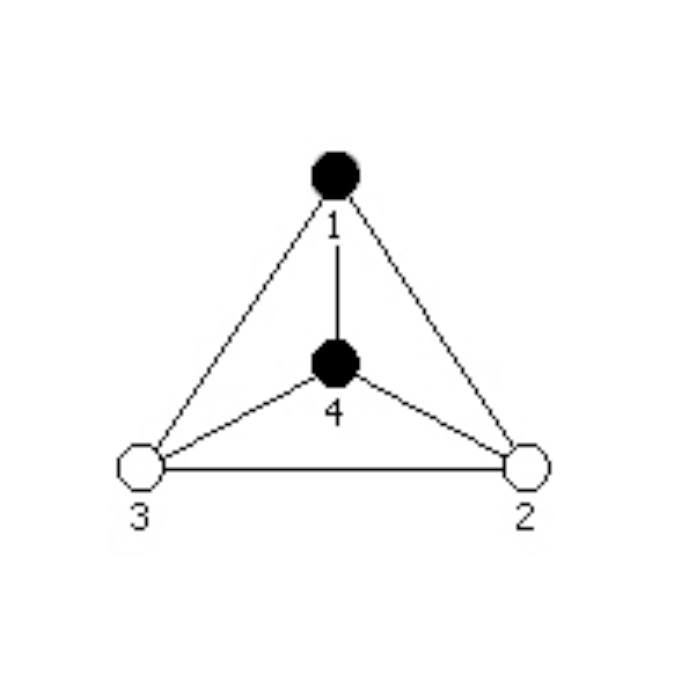}
\textbf{\caption{An Example of a Ground State in a 4 Qubit System that Does Not Represent a Max Independent Set\label{fig:4Q_exp} }}
\end{figure}

The 4 qubit simulation is the lowest possible sized system for the application of a cubic planar graph. All qubits are linked to one another as shown in Figure \ref{fig:4Q}. Meaning that this system is also the lowest sized system for the application of the specifically manufactured MIS problem. This is an interesting system to initially test as the ground state of this system without the application of an external Hamiltonian may in fact contain coupled pairs of qubits with the same parity that are opposite to other pairs (Figure \ref{fig:4Q_exp}). This in turn reveals no information about the MIS as no set of qubits of the same parity are completely isolated from one another deeming them non-independent. For this reason, the local Hamiltonian must be introduced on all qubits, to act as some form of penalty function, so that only qubits of one orientation can be aligned with their nearest neighbours.


\begin{table}
\centering
\begin{tabular}{c||c|c||c|c||c|c}
{$h_i$=0} & \multicolumn{2}{c||}{$\Omega$=1} & \multicolumn{2}{c||}{$\Omega$=0.1} & \multicolumn{2}{c}{$\Omega$=0.01}\\
{$J_{ij}$=-1} & E & P  & E & P  & E & P \\
\hline
$\vert 0000\rangle$ & 6 & 0.006784 & 6 & 0.000001 & 6 & 0.000000\\
$\vert 0001\rangle$ & 0 & 0.046917 & 0 & 0.000977 & 0 & 0.000007\\
$\vert 0010\rangle$ & 0 & 0.046917 & 0 & 0.000977 & 0 & 0.000007\\
$\vert 0011\rangle$ & -2 & 0.101849 & -2 & 0.165172 & -2 & 0.164195\\
$\vert 0100\rangle$ & 0 & 0.046917 & 0 & 0.000977 & 0 & 0.000007\\
$\vert 0101\rangle$ & -2 & 0.101849 & -2 & 0.165172 & -2 & 0.164195\\
$\vert 0110\rangle$ & -2 & 0.101849 & -2 & 0.165172 & -2 & 0.164195\\
$\vert 0111\rangle$ & 0 & 0.046917 & 0 & 0.000977 & 0 & 0.000007\\
$\vert 1000\rangle$ & 0 & 0.046917 & 0 & 0.000977 & 0 & 0.000007\\
$\vert 1001\rangle$ & -2 & 0.101849 & -2 & 0.165172 & -2 & 0.164195\\
$\vert 1010\rangle$ & -2 & 0.101849 & -2 & 0.165172 & -2 & 0.164195\\
$\vert 1011\rangle$ & 0 & 0.046917 & 0 & 0.000977 & 0 & 0.000007\\
$\vert 1100\rangle$ & -2 & 0.101849 & -2 & 0.165172 & -2 & 0.164195\\
$\vert 1101\rangle$ & 0 & 0.046917 & 0 & 0.000977 & 0 & 0.000007\\
$\vert 1110\rangle$ & 0 & 0.046917 & 0 & 0.000977 & 0 & 0.000007\\
$\vert 1111\rangle$ & 6 & 0.006784 & 6 & 0.000001 & 6 & 0.000000\\
\end{tabular}
\textbf{\caption{Table of Obtained Final Eigenvalues and Probabilities for All 4 Qubit States in the Absence of an External Hamiltonian\label{table:4q-noH}}}
\end{table}

Table \ref{table:4q-noH} contains final probabilities of all admissible states in the absence of an external Hamiltonian acting on all qubits. The ground states of the system with no external Hamiltonian contains two 1s and two 0s arranged in six possible combinations. These are clearly not independent sets as neither of the two `sets' are independent. It is simple to deduce that the independent set for this 4 qubit system only contains one qubit that is anti-aligned to the other three, where the qubit that is anti-aligned is of opposite parity to the applied Hamiltonians direction to minimise the energy of the system. After introducing the external Hamiltonian (local Hamiltonian acting equally on all qubits) the results obtained are available in Table \ref{table:4q}.

\begin{table}
\centering
\begin{tabular}{c||c|c||c|c||c|c}
{$h_i$=1} & \multicolumn{2}{c||}{$\Omega$=1} & \multicolumn{2}{c||}{$\Omega$=0.1} & \multicolumn{2}{c}{$\Omega$=0.01}\\
{$J_{ij}$=-1} & E & P  & E & P  & E & P \\
\hline
$\vert 0000\rangle$ & 10 & 0.007518 & 10 & 0.000000 & 10 & 0.000000\\
$\vert 0001\rangle$ & 2 & 0.025720 & 2 & 0.000025 & 2 & 0.000000\\
$\vert 0010\rangle$ & 2 & 0.025720 & 2 & 0.000025 & 2 & 0.000000\\
$\vert 0011\rangle$ & -2 & 0.092610 & -2 & 0.080024 & -2 & 0.081754\\
$\vert 0100\rangle$ & 2 & 0.025720 & 2 & 0.000025 & 2 & 0.000000\\
$\vert 0101\rangle$ & -2 & 0.092610 & -2 & 0.080024 & -2 & 0.081754\\
$\vert 0110\rangle$ & -2 & 0.092610 & -2 & 0.080024 & -2 & 0.081754\\
$\vert 0111\rangle$ & -2 & 0.025720 & -2 & 0.129636 & -2 & 0.124055\\
$\vert 1000\rangle$ & 2 & 0.025720 & 2 & 0.000025 & 2 & 0.000000\\
$\vert 1001\rangle$ & -2 & 0.092610 & -2 & 0.080024 & -2 & 0.081754\\
$\vert 1010\rangle$ & -2 & 0.092610 & -2 & 0.080024 & -2 & 0.081754\\
$\vert 1011\rangle$ & -2 & 0.025720 & -2 & 0.129636 & -2 & 0.124055\\
$\vert 1100\rangle$ & -2 & 0.092610 & -2 & 0.080024 & -2 & 0.081754\\
$\vert 1101\rangle$ & -2 & 0.025720 & -2 & 0.129636 & -2 & 0.124055\\
$\vert 1110\rangle$ & -2 & 0.025720 & -2 & 0.129636 & -2 & 0.124055\\
$\vert 1111\rangle$ & 2 & 0.012711 & 2 & 0.000119 & 2 & 0.000000\\
\end{tabular}
\textbf{\caption{Table of Obtained Final Eigenvalues and Probabilities for All 4 Qubit States in the Presence of an External Hamiltonian\label{table:4q}}}
\end{table}

As mentioned above, the independent set for this system has a total cardinality of 1. Meaning that the independent set comprises only of a single qubit. It is known that the qubit will be of opposite parity to all other qubits as well as the local Hamiltonian that it is subjected to, indicating states that represent independent sets comprise of three 1s and one 0. Where the qubit of state 0, represents the MIS. From the collected data in Table \ref{table:4q} it is clear that the states that represent the correct independent sets in fact have the highest final probabilities although share their eigenvalue with the states that have two qubits of each orientation (Figure \ref{fig:4QeigsJ-1H1}). This means that four correct states and six incorrect states have the identical ground state energy. Using the results where $\Omega$=0.1, the probability of obtaining a correct answer is just above 0.5. To increase this probability, the parameters within the simulation must be adjusted. Measures must be taken to reduce the energy of the correct states and increase the energy of the incorrect states. Table \ref{table:Eigs} has a list of equations that calculate the energy eigenvalues of each type of state.

\begin{table}
\centering
\begin{tabular}{c|c}
State & Eigenvalue\\
\hline\hline
$\vert 0000\rangle$ (x1) & 4h - 6J \\
$\vert 0001\rangle$ (x4) & 2h \\
\hline
$\vert 0011\rangle$ (x6) & -2J\\
$\vert 0111\rangle$ (x4) & -2h\\
$\vert 1111\rangle$ (x1) & -4h - 6J\\
\end{tabular}
\textbf{\caption{Table of Equations for the Calculation of Eigenvalues of Various States (J=-1)\label{table:Eigs}}}
\end{table}
 
\begin{figure} 
\centering 
\includegraphics[width=\textwidth]{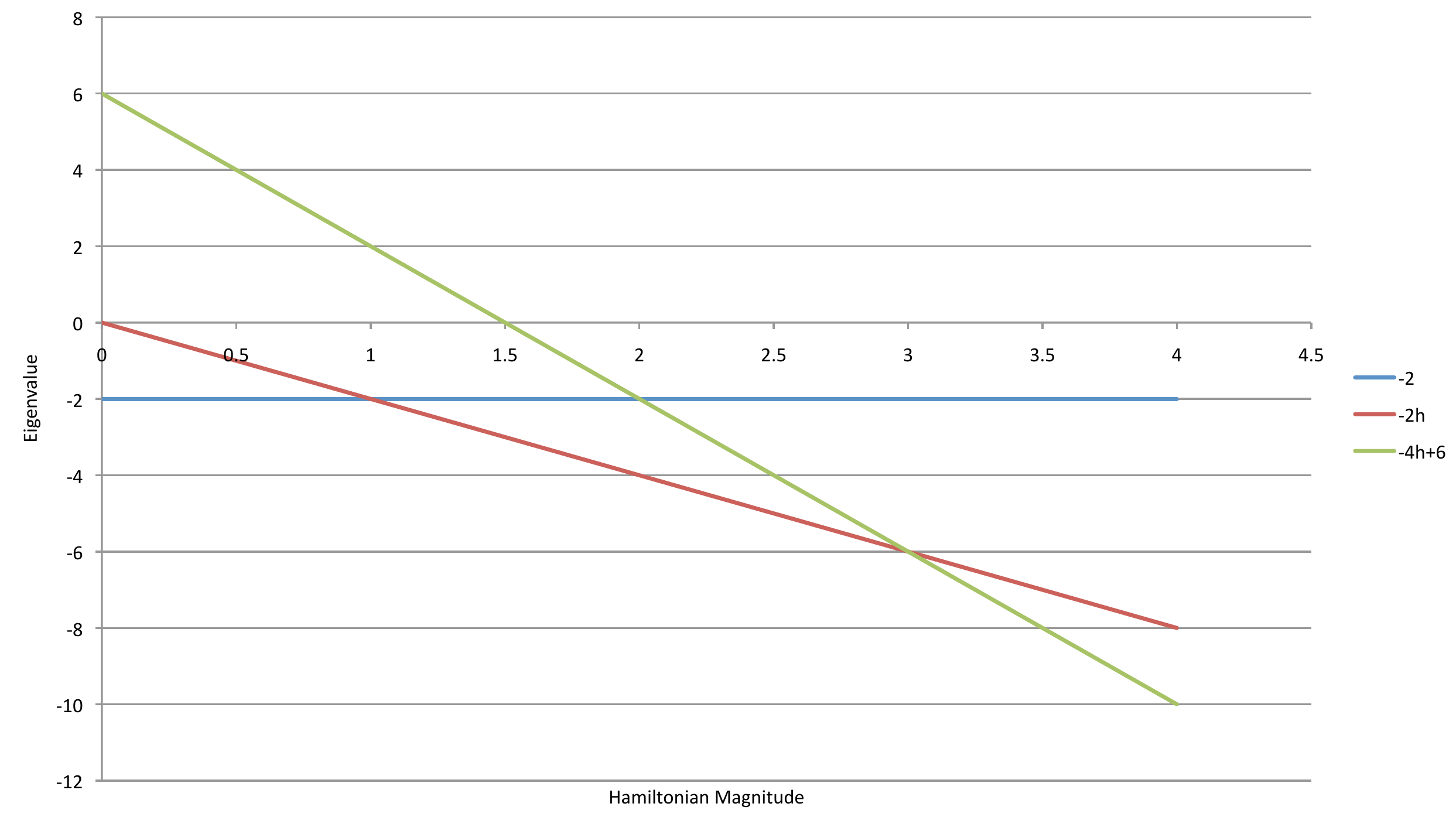}
\textbf{\caption{Eigenvalues as Functions of the Magnitude of the Applied Local Hamiltonians (J=1)\label{fig:eigsVh}}}
\end{figure}

In Table \ref{table:Eigs} the top two combinations of states have been separated with a horizontal line as no matter how much the magnitudes of the variables $h_i$ and $J_{ij}$ are increased, their total energy will only increase. Leaving only three other combinations of states. To lower the eigenvalue of the needed combination (three 1s and one 0) to be less than the conflicting combination (two 1s and two 0s) it is clear that the local Hamiltonian acting on the qubits must be increased. Although increasing it too much will cause the state $\vert 1111\rangle$ to become more energetically favourable as the Hamiltonian strength will overwhelmingly supercede the interaction strength and cause all qubits to align in the Hamiltonians orientation. Traditionally the magnitude of $J$ is equal to 1 (and negative for antiferrromagnetic interactions), continuing this trend only the parameter of $h$ must be adjusted. Plotting the eigenvalues that may be obtained against values of $h$ gives Figure \ref{fig:eigsVh}. Clearly the correct states are more energetically favourable when: $1<h_i<3$, as they have the lowest eigenvalue between this range.

\begin{figure} 
\centering 
\includegraphics[width=\textwidth]{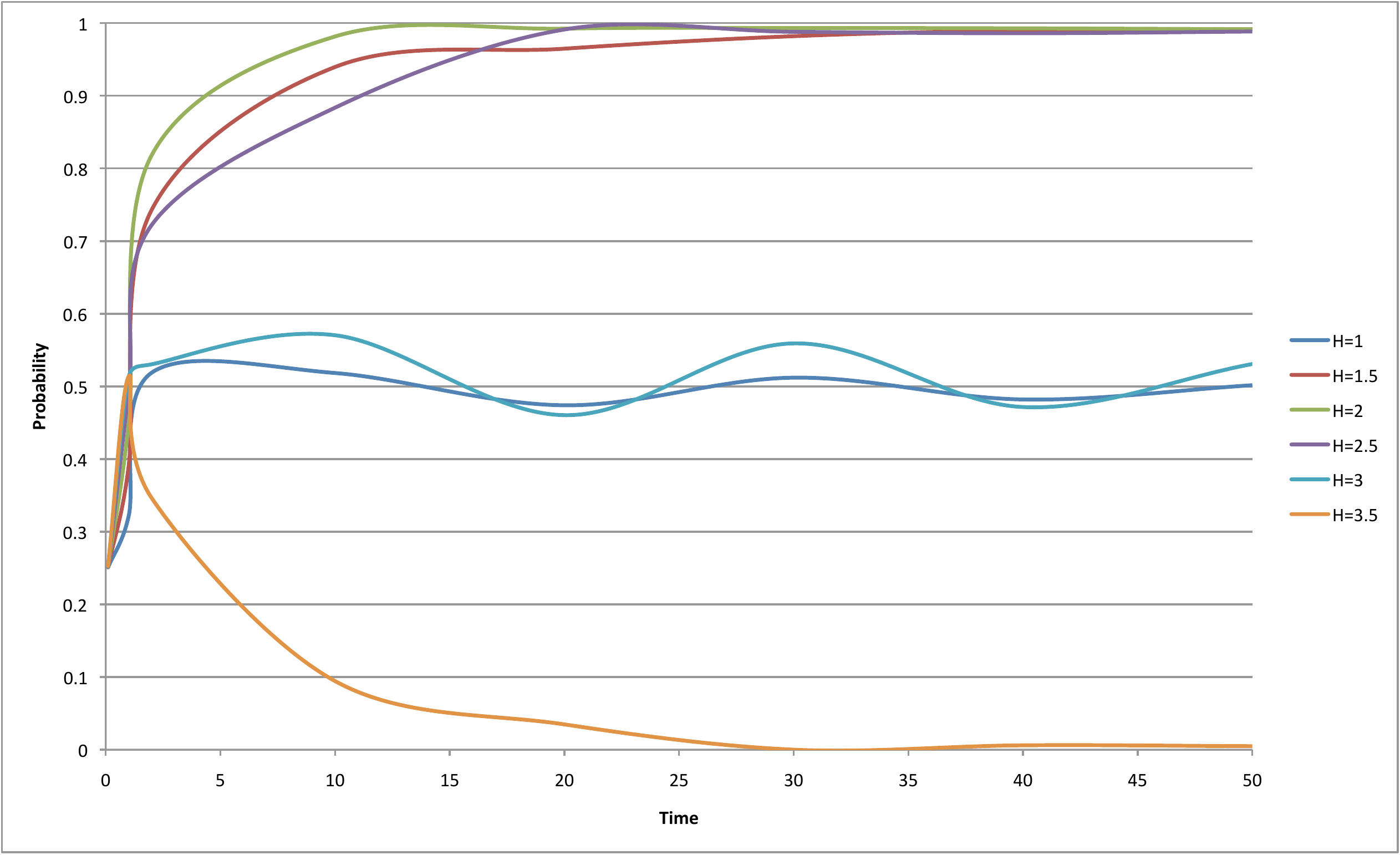}
\textbf{\caption{Combined Probabilities of Independent Sets as Functions of Time for Varying Magnitudes of Local Hamiltonians (J=-1)\label{fig:probsVt} }}
\end{figure}

\begin{figure} 
\centering 
\includegraphics[width=0.8\textwidth]{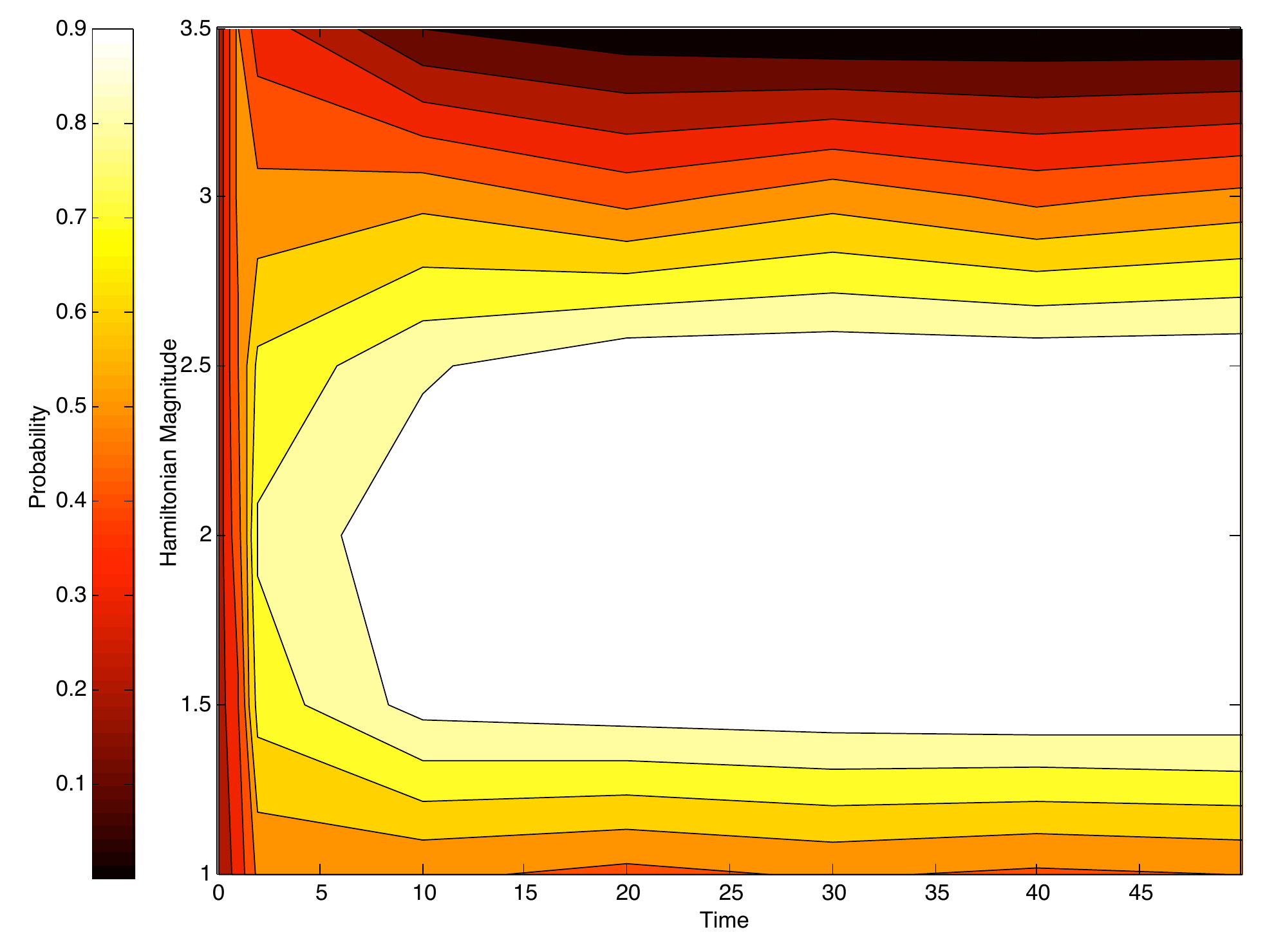} 
\textbf{\caption{Probabilities for Obtaining an Independent Set for Varying Times and Hamiltonians in Contour Form (J=-1)\label{fig:probsVh} } }
\end{figure}

Knowing how the system will react to Hamiltonians of various magnitude, it is clear that further simulations are needed to investigate the probability of obtaining a state that represents a MIS for varying strengths of local Hamiltonian and for a range of simulated times. The obtained data is plotted in Figures \ref{fig:probsVt} and \ref{fig:probsVh}.

\begin{table}
\centering
\begin{tabular}{c||c|c|c|c|c|c}
{} & \multicolumn{6}{c}{Hamiltonian Magnitude}\\
{Time} & 1 & 1.5 & 2 & 2.5 & 3 & 3.5\\
\hline
0.1 & 0.2508 & 0.2512 & 0.2516 & 0.252 & 0.2524 & 0.2528\\
1 & 0.3216 & 0.3908 & 0.45 & 0.4916 & 0.5128 & 0.5136\\
2 & 0.5188 & 0.7428 & 0.818 & 0.722 & 0.5304 & 0.3465\\
10 & 0.5184 & 0.9396 & 0.9816 & 0.8836 & 0.5704 & 0.0944\\
20 & 0.474 & 0.9648 & 0.9924 & 0.9912 & 0.4604 & 0.0348\\
30 & 0.512 & 0.982 & 0.9932 & 0.988 & 0.5592 & 0\\
40 & 0.482 & 0.99 & 0.9928 & 0.986 & 0.4716 & 0.006\\
50 & 0.5016 & 0.9916 & 0.9916 & 0.9884 & 0.5308 & 0.0048\\
\end{tabular}
\textbf{\caption{Combined Final Probability of Correct Four States for Varying Time and Hamiltonian Magnitude\label{table:4Qprobs}}}
\end{table}

From Figures \ref{fig:probsVt} and \ref{fig:probsVh} it can be deduced that the adiabatic theorem is still a major factor in determining the state that the system evolves to. In short time periods the probability of finding a correct state is almost constant across all Hamiltonian magnitudes, with almost no dependance on the magnitude of the external Hamiltonian. This is also deducible from Table \ref{table:4Qprobs}, as for a time of 0.1, the probability varies very little from 0.25 for obtaining a correct final state under all tested Hamiltonian magnitudes. Although for this system out of sixteen possible states there are four that can be regarded as correct, exactly a quarter of the total number. Meaning that within this time period, all states right and wrong have almost identical final probabilities.

As expected for Hamiltonian magnitudes between one and three the probabilities of measuring a state that represents a MIS becomes almost one for large simulation times (from 20 units of time and greater). With these magnitudes the eigenvalues of the correct states split from the others with a reasonable gap. The states that start with the four lowest eigenvalues all converge to a single final eigenvalue meaning that the final state of the system is a linear superposition of these four states. The changes in eigenvalue progression are very interesting for the various Hamiltonian magnitudes, these are viewable in Figures \ref{fig:4QeigsJ-1H1} to \ref{fig:4QeigsJ-1H3-5}.

\begin{figure} 
\centering 
\subfloat[$h_i$=1]{
\label{fig:4QeigsJ-1H1} 
\includegraphics[width=0.8\textwidth]{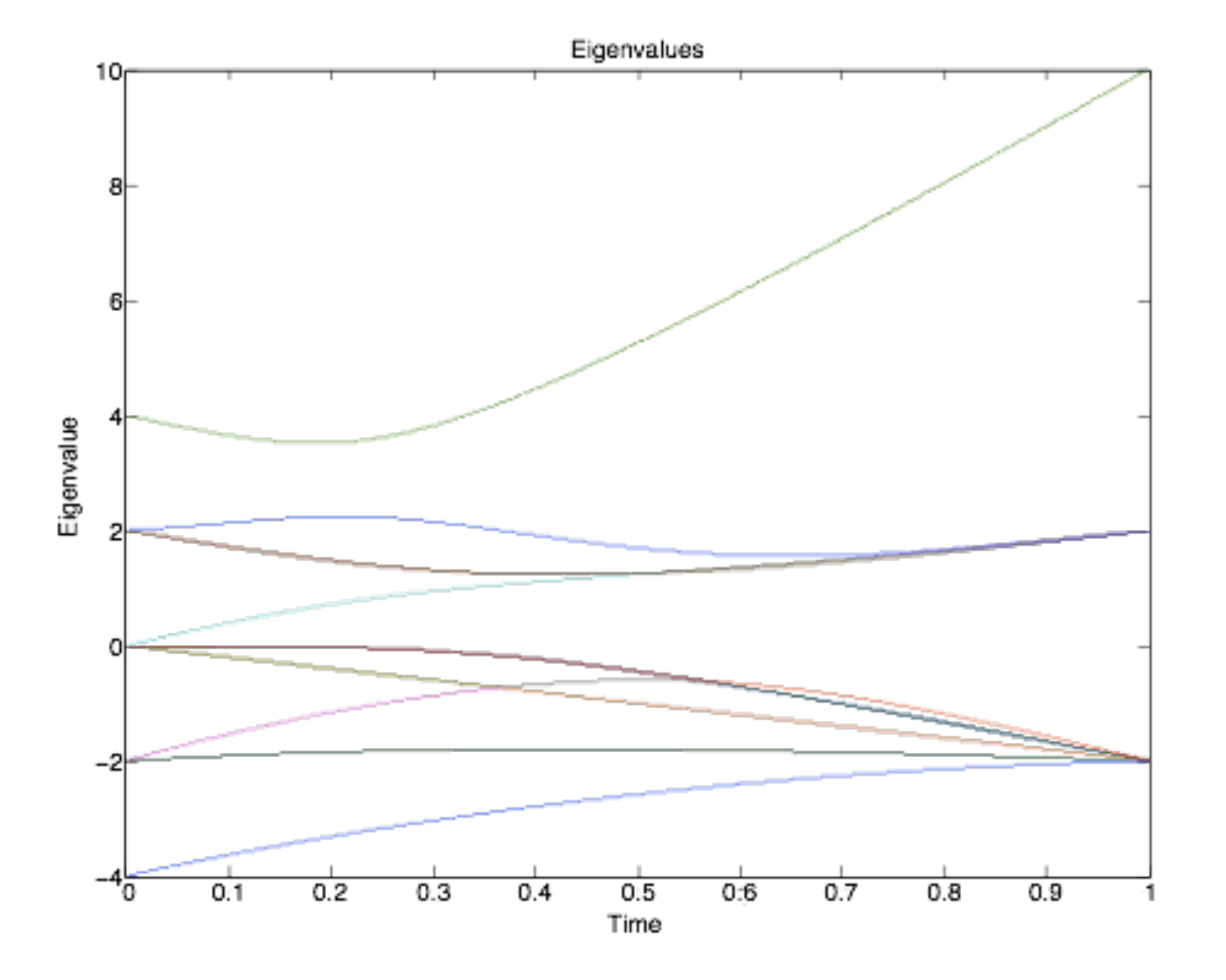} }\\
\subfloat[$h_i$=1.5]{
\label{fig:4QeigsJ-1H1-5}
\includegraphics[width=0.8\textwidth]{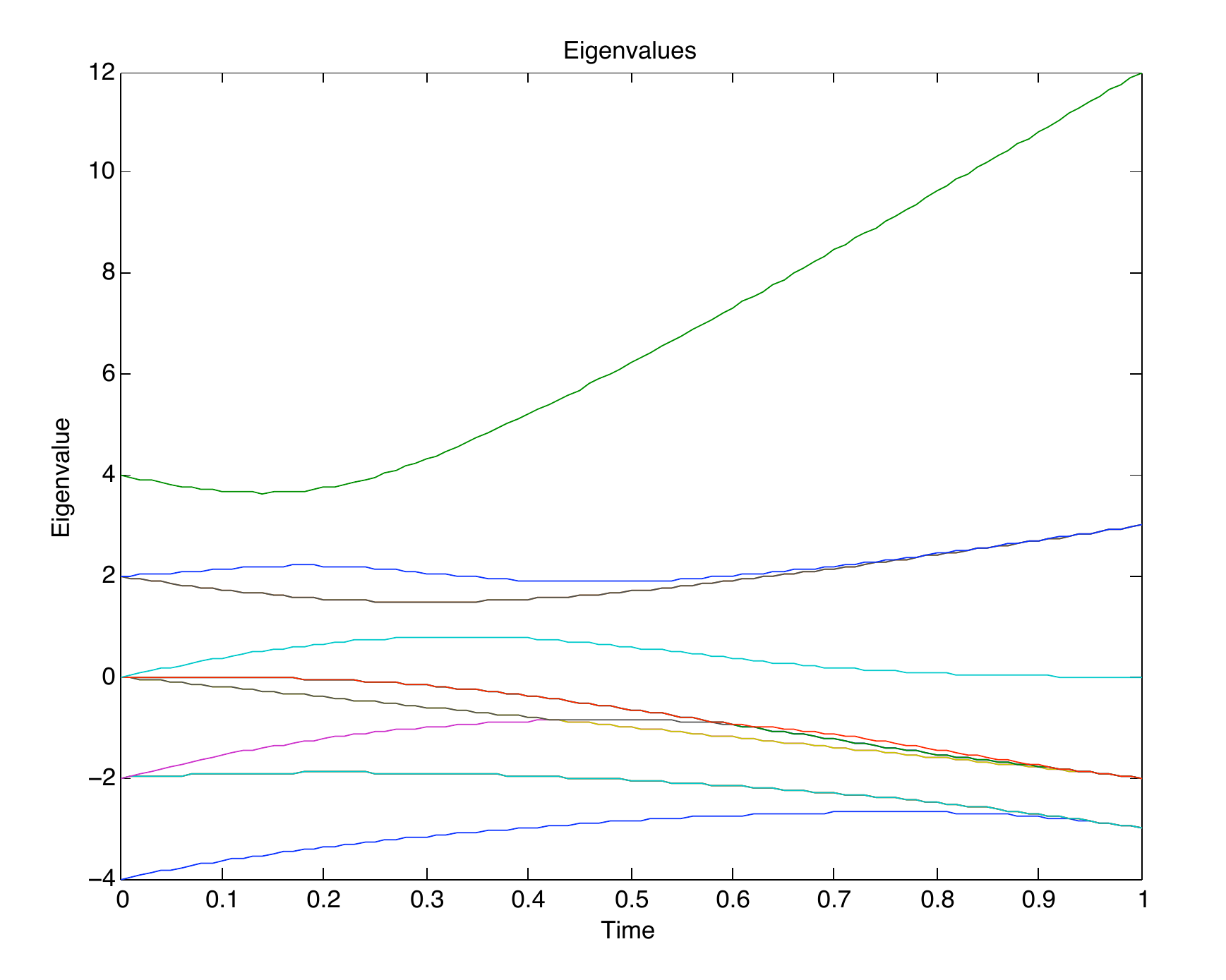} }
\textbf{\caption{Plots of Eigenvalues for Fixed Value of J=-1 and Varying local Hamiltonian ($h_i$)}}
\end{figure}

\begin{figure}
\ContinuedFloat 
\centering 
\subfloat[$h_i$=2]{
\label{fig:4QeigsJ-1H2} 
\includegraphics[width=0.8\textwidth]{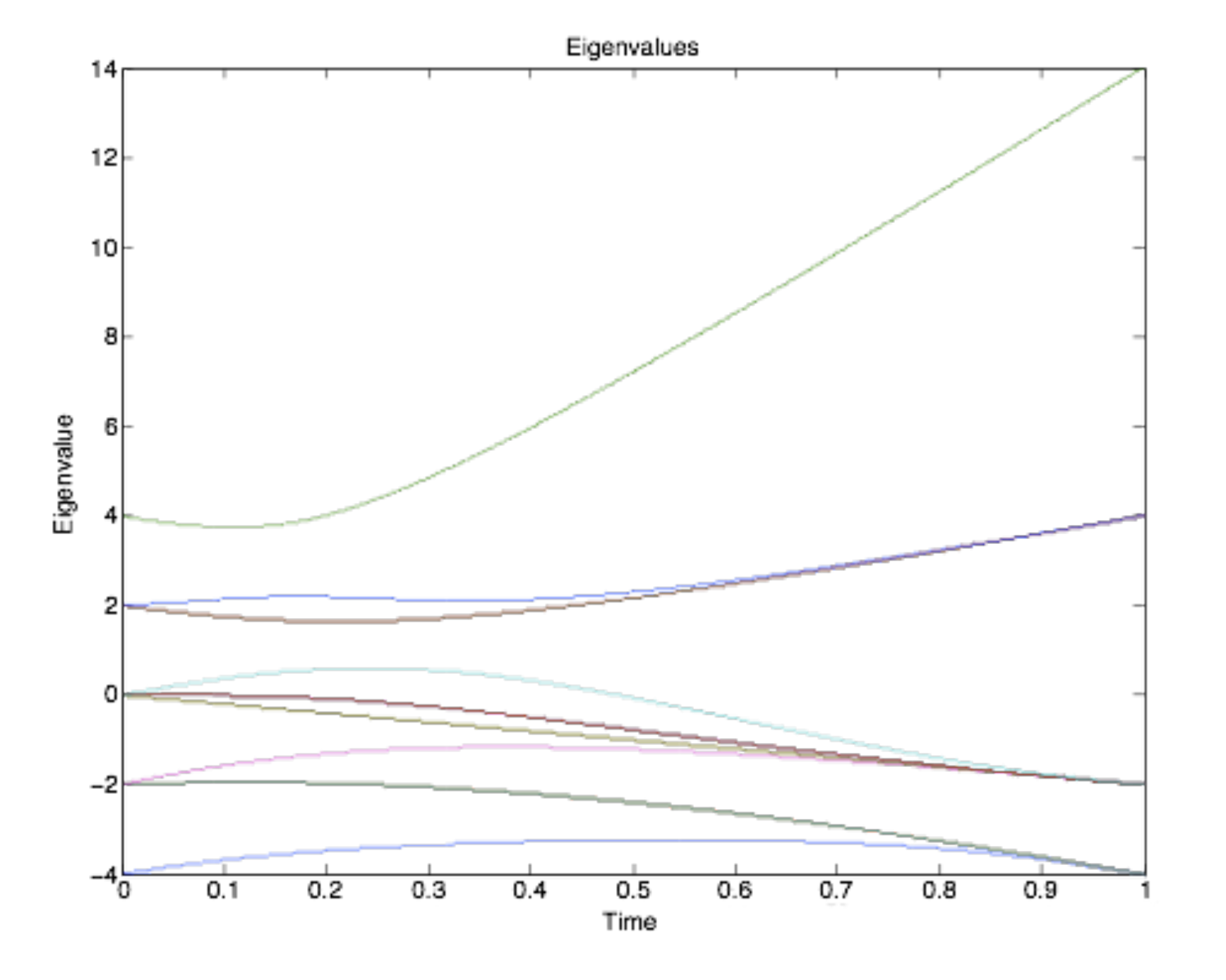} }\\
\subfloat[$h_i$=2.5]{
\label{fig:4QeigsJ-1H2-5}
\includegraphics[width=0.8\textwidth]{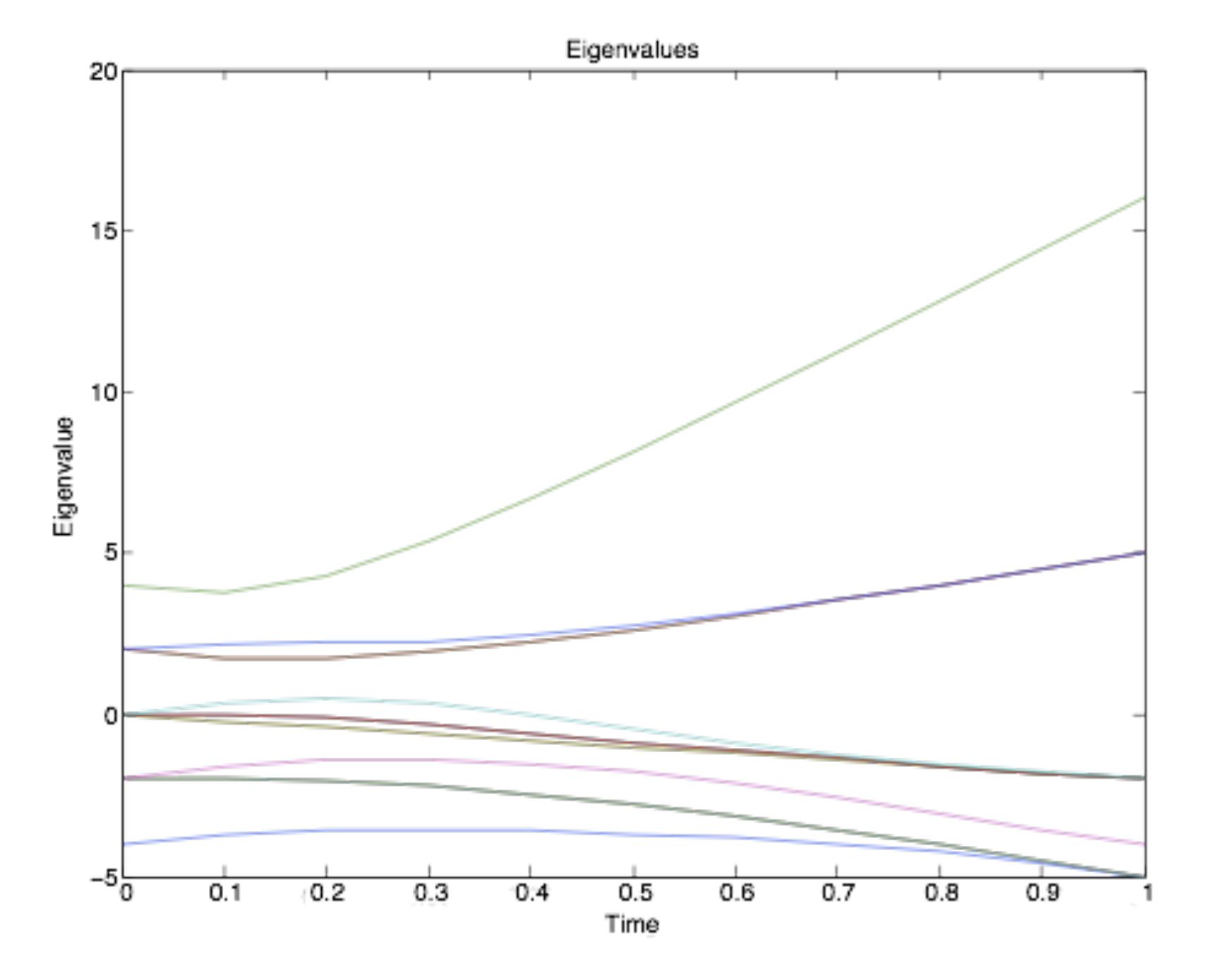} }
\textbf{\caption{Plots of Eigenvalues for Fixed Value of J=-1 and Varying Local Hamiltonian ($h_i$)}}
\end{figure}

\begin{figure}
\ContinuedFloat
\centering 
\subfloat[$h_i$=3]{
\label{fig:4QeigsJ-1H3} 
\includegraphics[width=0.8\textwidth]{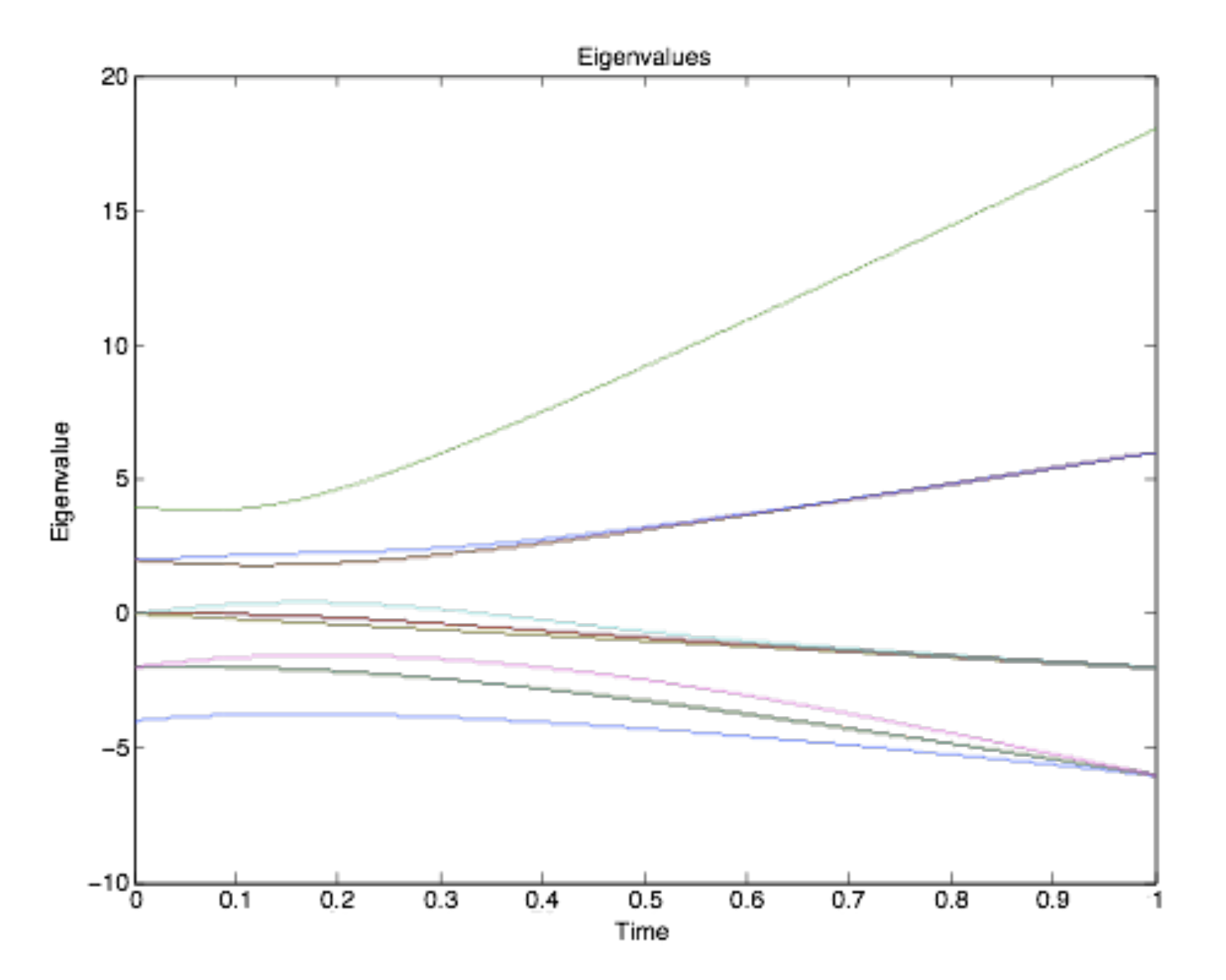} }\\
\subfloat[$h_i$=3.5]{
\label{fig:4QeigsJ-1H3-5}
\includegraphics[width=0.8\textwidth]{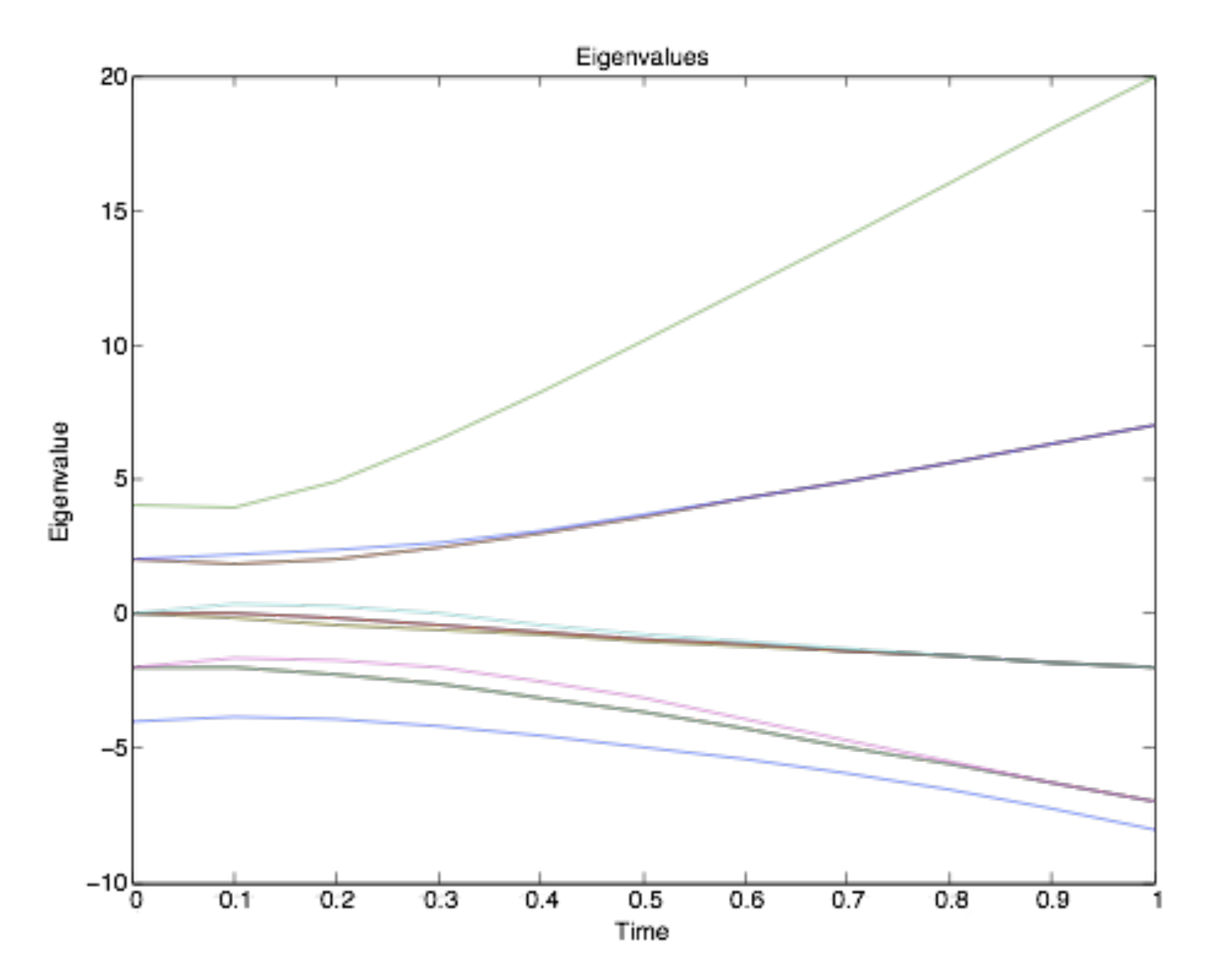} }
\textbf{\caption{Plots of Eigenvalues for Fixed Value of J=-1 and Varying Local Hamiltonian ($h_i$)}}
\end{figure}

Figure \ref{fig:4QeigsJ-1H1} is the situation with $h_i=1$ for all $i$. It is clear that a large set of states both correct and incorrect all evolve to become energetically equal. At the end of the simulation where all states converge, the probability of entering any of these states should be almost equal, although due to symmetry, the incorrect states have less probability individually. This is because the total probability of the incorrect independent sets (ground states) is equal to the total probability of the correct independent sets for sufficiently long simulation times (see Table \ref{table:4q}). 

Continuing to Figure \ref{fig:4QeigsJ-1H1-5} where $h_i=1.5$ for all $i$ the eigenvalues for the states that do not represent independent sets have split from those that do by only one energy unit. Meaning that the probability of tunnelling into more excited states will be far smaller for sufficiently slow simulations. Similarly in Figure \ref{fig:4QeigsJ-1H2}, where $h_i=2$ for all $i$ the eigenvalues are still separated but with a larger gap, yet increasing the local Hamiltonian further to 2.5 (Figure \ref{fig:4QeigsJ-1H2-5}) the gap reduces again to only one unit of energy. Meaning there is symmetry in the eigenvalue gaps in this system due to the magnitude of the Hamiltonians about a centre which is when $h_i=2$ for all $i$. Yet even with a smaller gap, it seems that there is very little difference between probabilities obtained using $h_i$ = 1.5, 2 and 2.5 (see Figure \ref{fig:probsVt}). Finally, Figure \ref{fig:4QeigsJ-1H3-5} is the plot of eigenvalues when $h_i=3.5$ for all $i$, in this situation there is only a single ground state, with a well-defined gap between the ground state and all other states throughout the evolution. This ground state as expected is the state where all qubits are aligned simultaneously with the Hamiltonian direction, as the lowering in energy due to the Hamiltonian magnitude is far greater in this situation than that of the Ising interactions. Therefore the probability of measuring a state that represents a MIS at the end of the simulation is dramatically reduced, as this relies on the system randomly tunnelling into a correct state.

\subsection{8 Qubit Simulation}
\subsubsection{8 Qubit Cube} \label{subsubsect:8Qc}

\begin{figure}
\centering 
\includegraphics{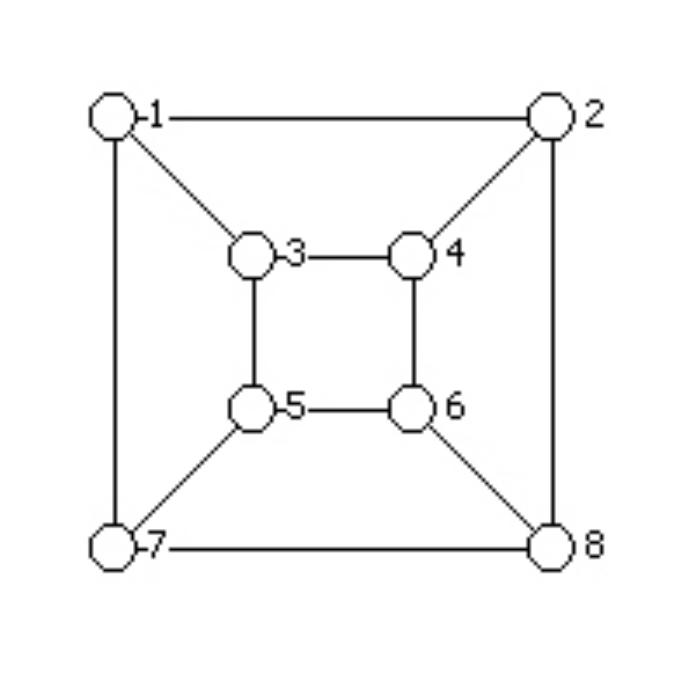}
\textbf{\caption{8 Qubit Cube in Cubic Planar 8 Node Graph G=(V,E) Form, where Vertices V Represent Qubits and Edges E Ising Interactions\label{fig:8Q} }}
\end{figure}

\begin{figure} 
\centering 
\includegraphics{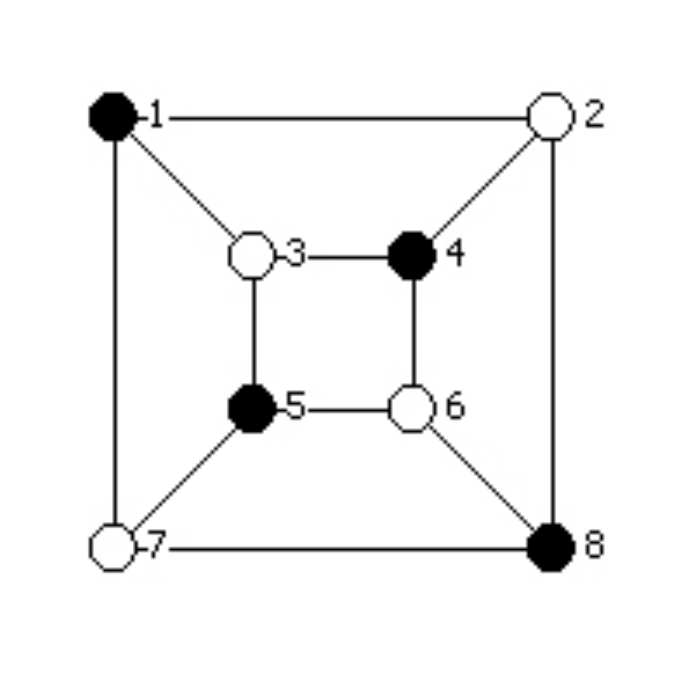}
\textbf{\caption{Example of the Max Independent Set of a Cube\label{fig:8Qset} }}
\end{figure}

The 8 qubit cube is studied as it is the simplest 8 qubit system due to various symmetries. The cube can be drawn in cubic planar form by simply taking a cube, which by default each qubit is linked to three others, and squashing it into a 2 dimensional plane. The qubits are situated at the corners of the cube as shown in Figure \ref{fig:8Q}. 

Every link in the system can be satisfied with antiferromagnetic interactions. Every qubit can be of opposite orientation to all of its nearest neighbours. Meaning that exactly half of the qubit will be of each orientation. An example of the exact MIS is viewable in Figure \ref{fig:8Qset} where filled qubits are of one orientation and the unfilled are the other. It is not important which ones are which due to the symmetries, by just rotating the system about its centre by ninety degrees the filled and unfilled orientations swap.

As all of the ferromagnetic Ising interactions in this system can be satisfied simultaneously, no local Hamiltonian needs to be applied to the system in order for the ground state of this system to be representative of a MIS. Although this is a special case and not representative of all cubic planar systems, therefore this system is investigated to find a threshold Hamiltonian under which the system's ground state no longer represents a solution to the MIS problem.

As mentioned before, due to symmetries there are only two degenerate ground states for this system, without an application of a local Hamiltonian on each qubit. These are $\vert 01100110\rangle$ and the opposite parity $\vert 10011001\rangle$. To reach these, the system would be prepared in a superposition of all admissible states and the evolution would involve slowly increasing the Ising interactions whilst decreasing the magnitude of the initial Hamiltonian. The system by the end of the evolution would be a superposition of the two ground states (mentioned above) and according to the Copenhagen interpretation of quantum mechanics, measurement would collapse the system's wavefunction into one of these states. As the two degenerate ground states are of completely opposite parity, it would appear that the system is in a fully entangled state, so by measuring only one of the qubits the state of all others could be determined (if the ground state is known prior to evolution). Even though this entanglement is obvious from the two admissible ground states, the concurrence of a system greater than 2 qubits cannot practically be determined\cite{Wootters:1998}. Obtaining one of these states would obviously depend on the sufficiently slow adiabatic evolution and the robustness of the system to thermal excitation into a more energetic state.

\begin{table}
\centering
\begin{tabular}{c||c|c|c|c}
{} & \multicolumn{4}{c}{Hamiltonian Magnitude}\\
{Time} & 0 & 1 & 2 & 3\\
\hline
0.1 & 0.0082 & 0.0082 & 0.0082 & 0.0082\\
1 & 0.0608 & 0.0522 & 0.0342 & 0.0182\\
2 & 0.233 & 0.2024 & 0.0958 & 0.0144\\
10 & 0.8374 & 0.8496 & 0.8704 & 0.0186\\
20 & 0.9678 & 0.9698 & 0.9656 & 0.0104\\
33.33 & 0.9822 & 0.9822 & 0.9822 & 0.0322\\
\end{tabular}

\textbf{\caption{Combined Final Probability of Correct Two States for Varying Time and Hamiltonian Magnitude for a Cube of Qubits\label{table:8Qprobs}}}
\end{table}

\begin{figure} 
\centering 
\includegraphics[width=\textwidth]{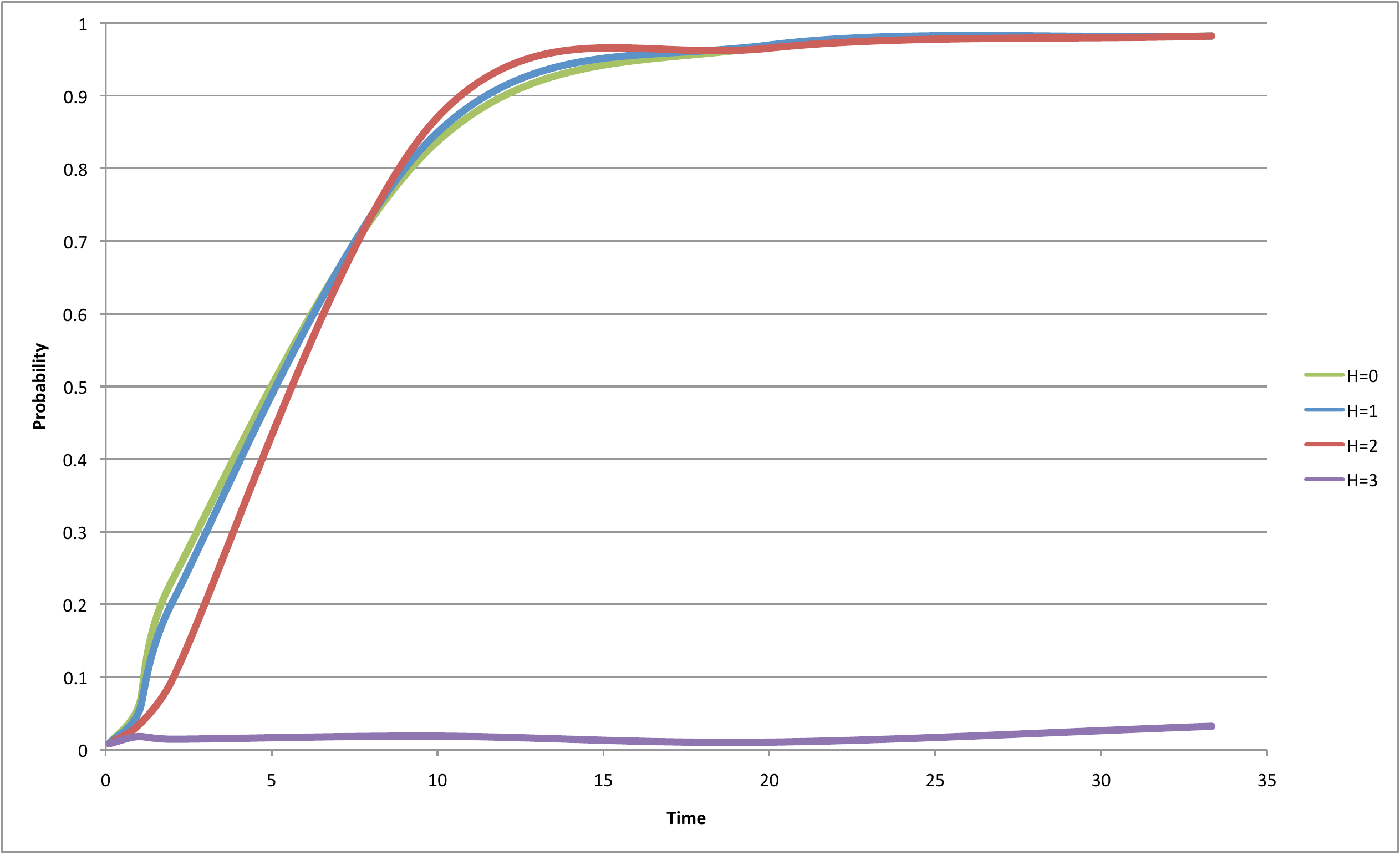}
\textbf{\caption{Total Probability of Obtaining an State Representing the Max Independent Set After Measurement (J=-1)\label{fig:8Qprobs} }}
\end{figure}

The obtained probabilities for various Hamiltonian magnitudes and simulation times are available in Table \ref{table:8Qprobs}. By raising the Hamiltonian magnitude to three, the most probable state changes to the state where all qubits are aligned with their local Hamiltonian direction, which is the same for all qubits. Yet there is very little to distinguish between the probabilities obtained for identical simulation times for Hamiltonian magnitudes from zero to three, as seen in Figure \ref{fig:8Qprobs}. 

Plots of the eigenvalues and probabilities of states for all of the four situations with varying local Hamiltonian magnitudes can be found in Figures \ref{fig:8QsEigsH0} to \ref{fig:8QsProbsH3}. It is obvious that the minimum gap between the two lowest eigenvalues will always be zero due to there being a pair of degenerate ground states, deeming Equation \ref{equ:Min} for the relationship between minimum gap and evolution time of even less use. The equation suggests that the run time for a system with degenerate ground states should be infinite. There appears to be no criteria on using this equation when the degenerate ground states individually represent correct answers to the encoded problem. If the gap between the ground state eigenvalue and first incorrect state eigenvalue is considered, there is more sense to the equation. Exciting the system to the first incorrect state will induce errors in the final measured value. The argument against such consideration is that one of the final degenerate ground states starts the evolution with its energy equal to a selection of incorrect states, where there is no energy gap. Although this would not affect the situation as at the start of the evolution the system is initialised to a non-degenerate ground state that is not energetically equivalent to an incorrect state. During the evolution, if the system were to be excited into the state that eventually evolves to become a degenerate ground state, this would increase the probability of measuring a state that represents a correct answer. To enter an incorrect state, the system would have to be raised from the initialised ground state to the first excited state and again to the second excited state, which is energetically equivalent to the difference between the initialised ground state and the second excited state. By considering Equation \ref{equ:Min} with the gap between initialised ground state and lowest incorrect state, it is clear why the probabilities for identical simulation times but varying local Hamiltonians are similar in Table \ref{table:8Qprobs}. By observing the eigenvalue evolutions, the new suggested minimum gap is 2 (and not 0) for all local Hamiltonian magnitudes and this is not due to the magnitude of the local Hamiltonian applied to all qubits, but a consequence of the initial Hamiltonian. Meaning that the minimum eigenvalue gap across all simulations in Table \ref{table:8Qprobs} is constant, giving good justification for the almost identical probabilities obtained across various Hamiltonian magnitudes.

\begin{figure} 
\centering 
\subfloat[$h_i$=0]{
\label{fig:8QsEigsH0} 
\includegraphics[width=0.8\textwidth]{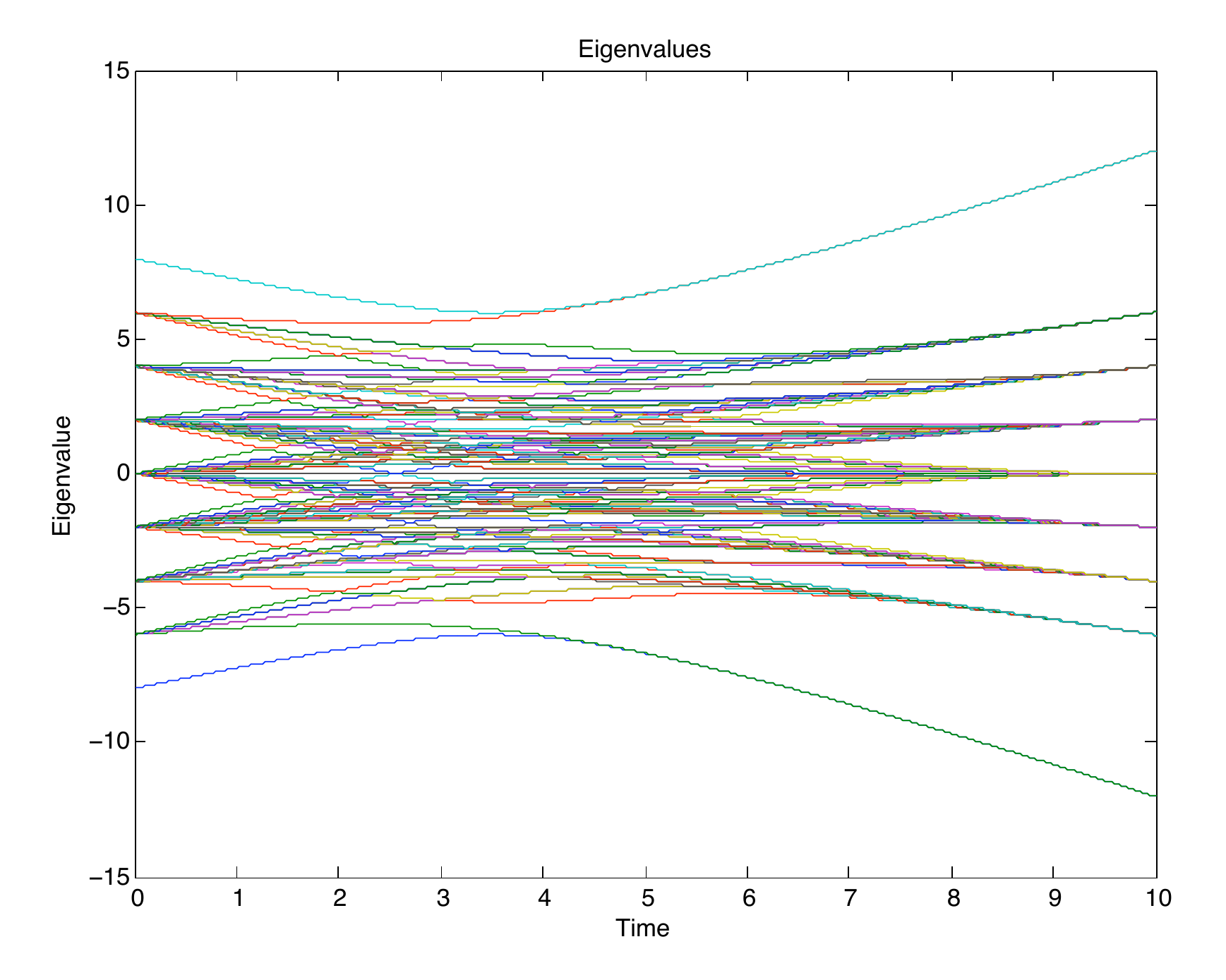} }\\
\subfloat[$h_i$=0]{
\label{fig:8QsProbsH0}
\includegraphics[width=0.8\textwidth]{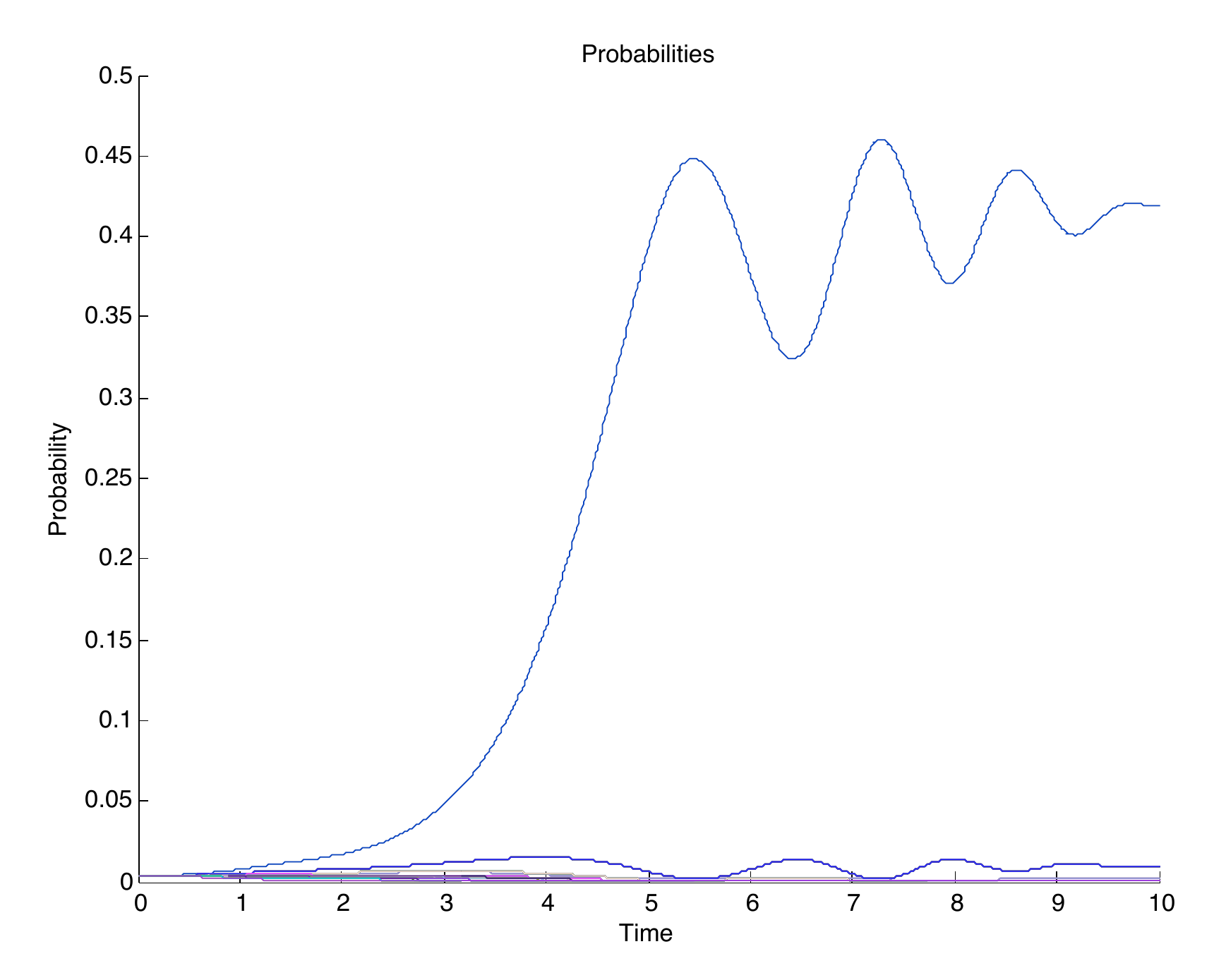} }
\textbf{\caption{Plots of Eigenvalues and Probabilities Whilst J=-1, $h_i$=0 and $\Omega=0.1$ (Legends Have Been Omitted as they Include 256 Entries)}}
\end{figure}

\begin{figure}
\ContinuedFloat 
\centering 
\subfloat[$h_i$=1]{
\label{fig:8QsEigsH1} 
\includegraphics[width=0.8\textwidth]{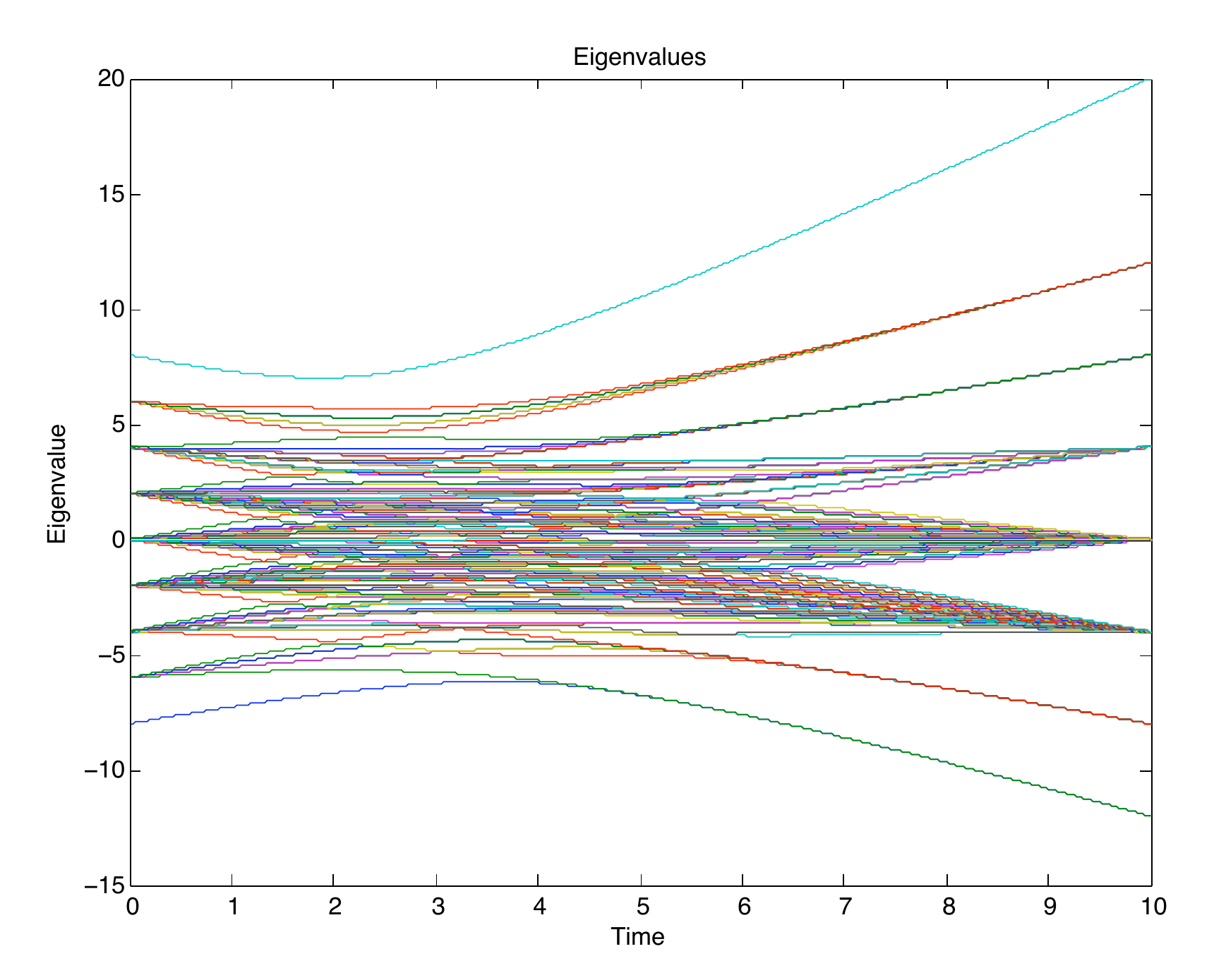} }\\
\subfloat[$h_i$=1]{
\label{fig:8QsProbsH1}
\includegraphics[width=0.8\textwidth]{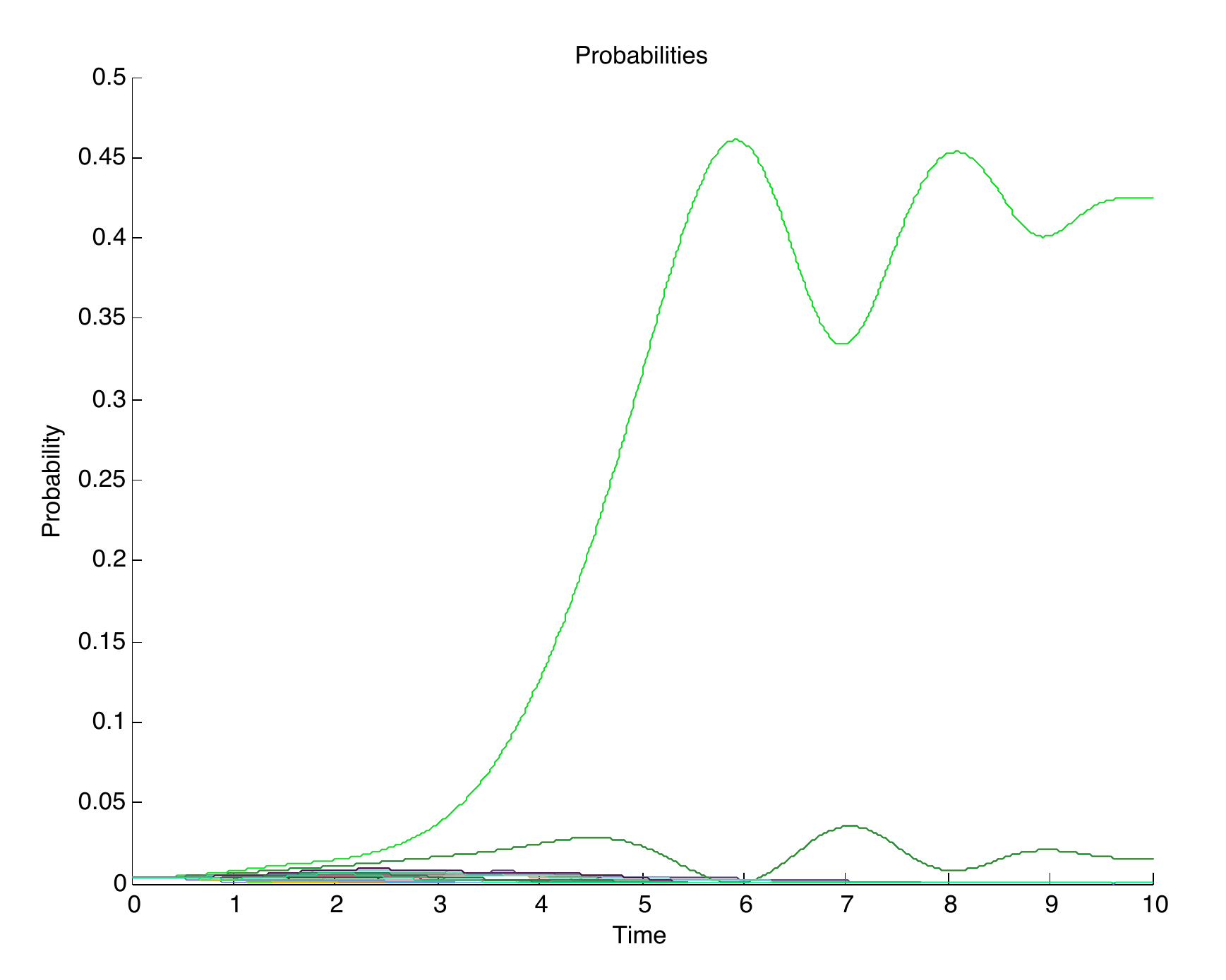} }
\textbf{\caption{Plots of Eigenvalues and Probabilities Whilst J=-1, $h_i$=1 and $\Omega=0.1$ (Legends Have Been Omitted as they Include 256 Entries)}}
\end{figure}

\begin{figure}
\ContinuedFloat
\centering 
\subfloat[$h_i$=2]{
\label{fig:8QsEigsH2} 
\includegraphics[width=0.8\textwidth]{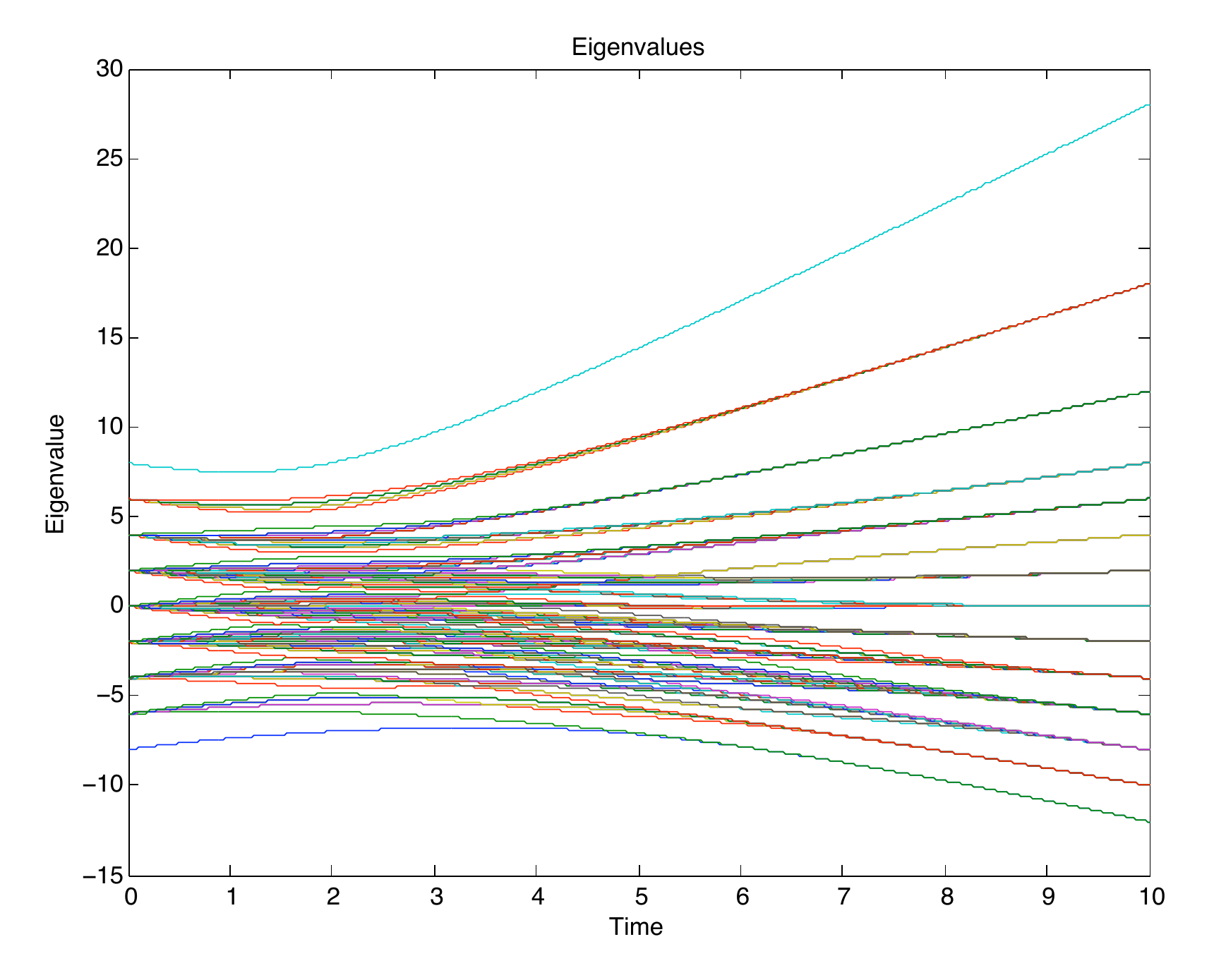} }\\
\subfloat[$h_i$=2]{
\label{fig:8QsProbsH2}
\includegraphics[width=0.8\textwidth]{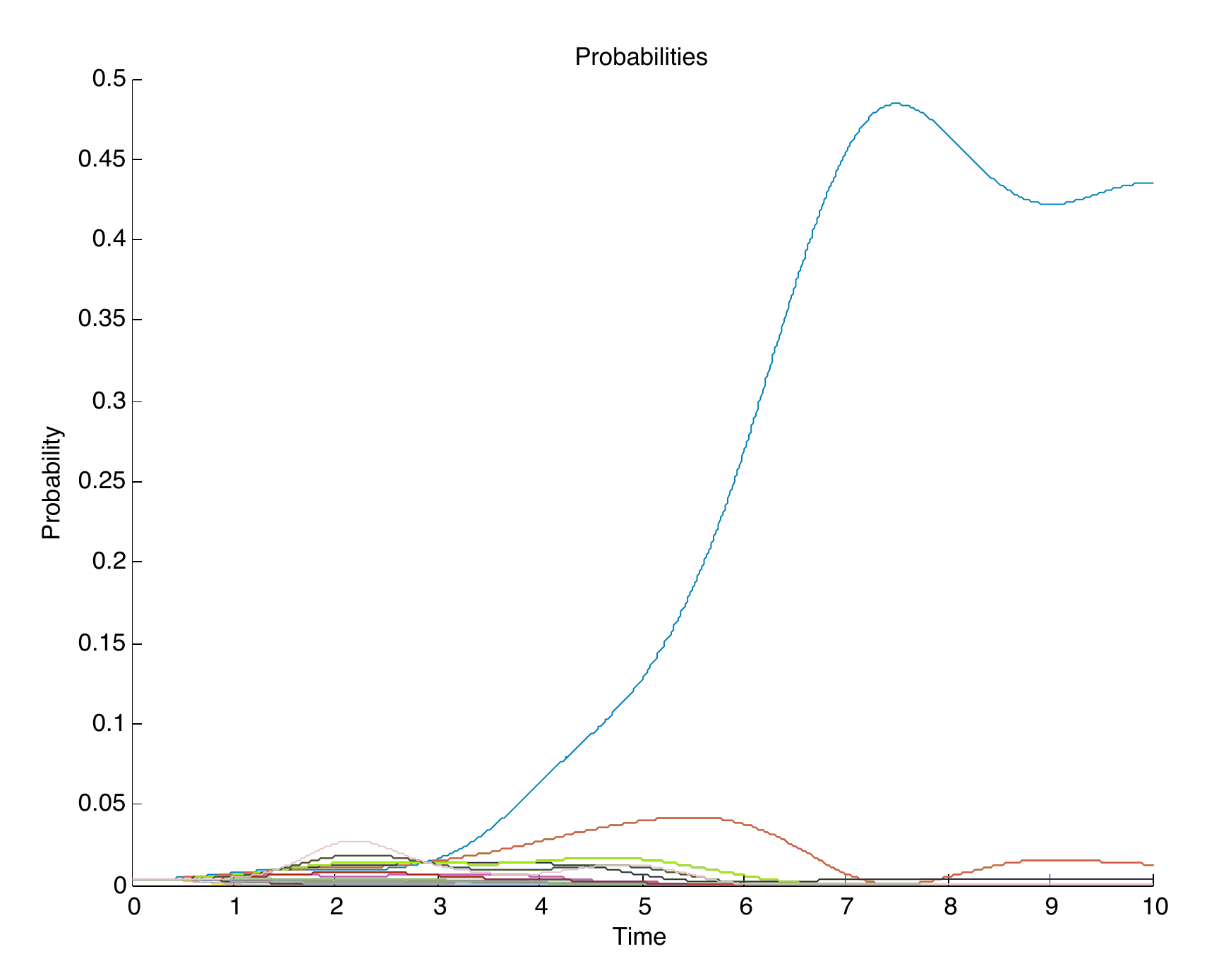} }
\textbf{\caption{Plots of Eigenvalues and Probabilities Whilst J=-1, $h_i$=2 and $\Omega=0.1$ (Legends Have Been Omitted as they Include 256 Entries)}}
\end{figure}
 
 \begin{figure}
\ContinuedFloat
\centering 
\subfloat[$h_i$=3]{
\label{fig:8QsEigsH3} 
\includegraphics[width=0.8\textwidth]{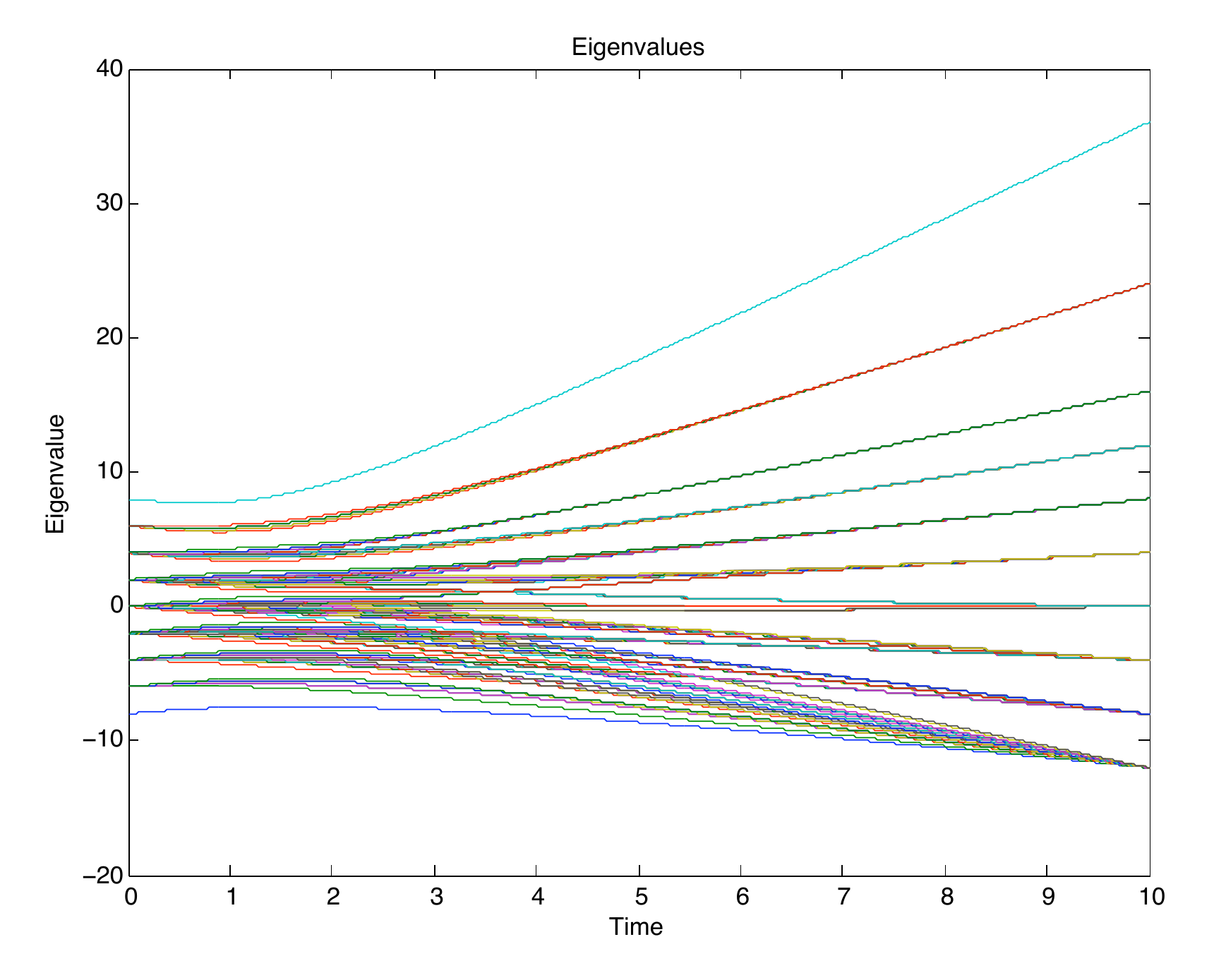} }\\
\subfloat[$h_i$=3]{
\label{fig:8QsProbsH3}
\includegraphics[width=0.8\textwidth]{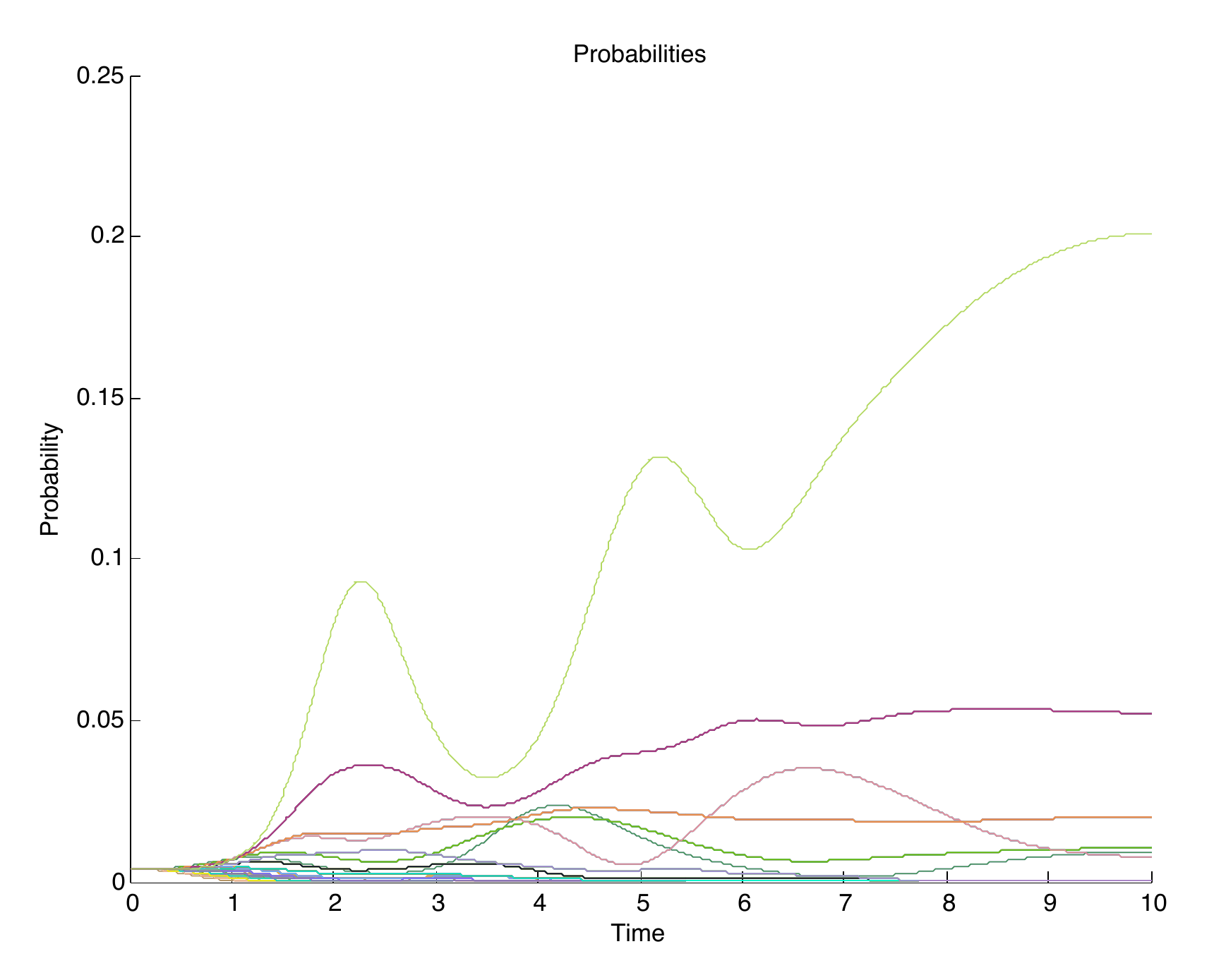} }
\textbf{\caption{Plots of Eigenvalues and Probabilities Whilst J=-1, $h_i$=3 and $\Omega=0.1$ (Legends Have Been Omitted as they Include 256 Entries)}}
\end{figure}
 
\subsubsection{Random 8 Qubit} \label{subsubsect:8QRand}

\begin{figure} 
\centering 
\includegraphics{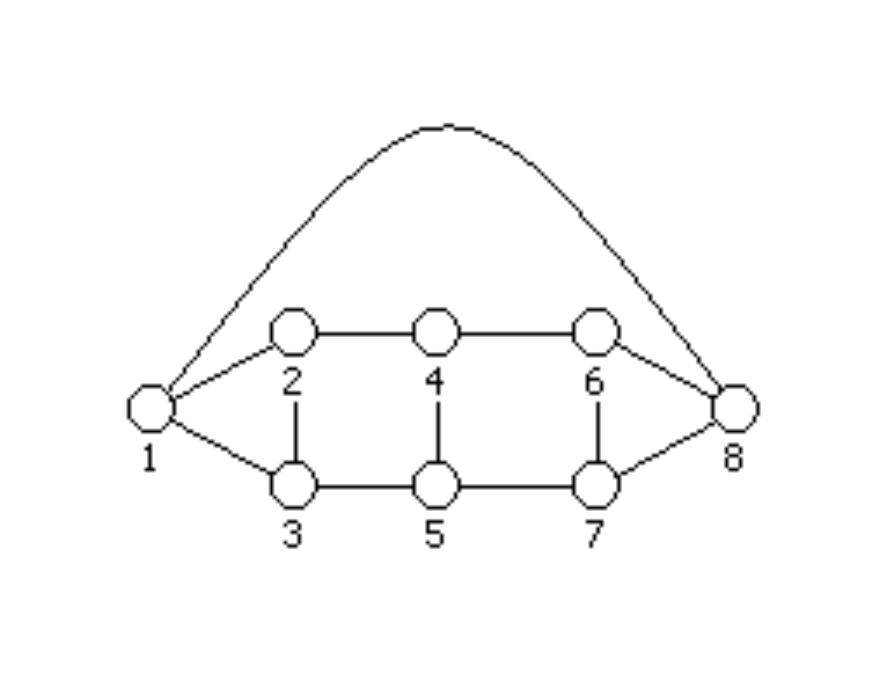}
\textbf{\caption{Random 8 Qubit Cubic Planar 8 Node Graph G=(V,E) Form, Where Vertices V Represent Qubits and Edges E Ising Interactions\label{fig:8Qa} }}
\end{figure}

\begin{figure} 
\centering 
\subfloat[Set 1]{
\label{fig:8QrSet1} 
\includegraphics[width=\textwidth]{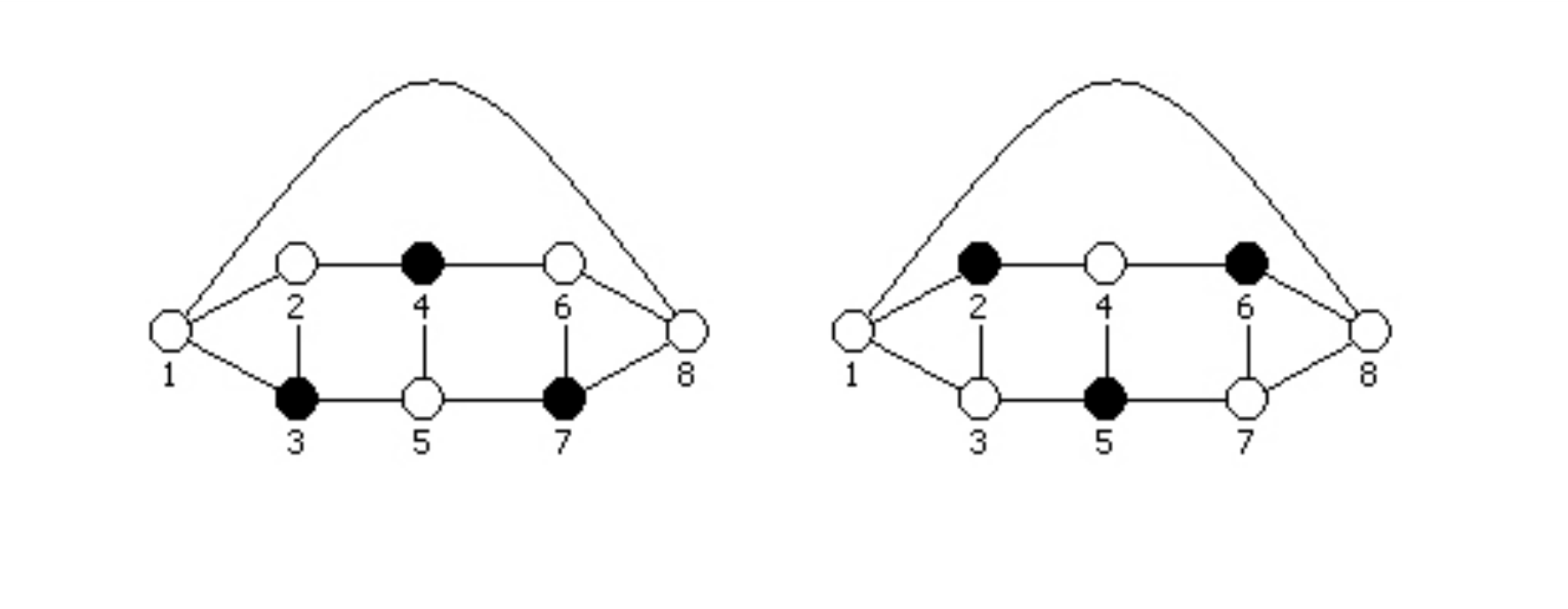} }\\
\subfloat[Set 2]{
\label{fig:8QrSet2}
\includegraphics[width=\textwidth]{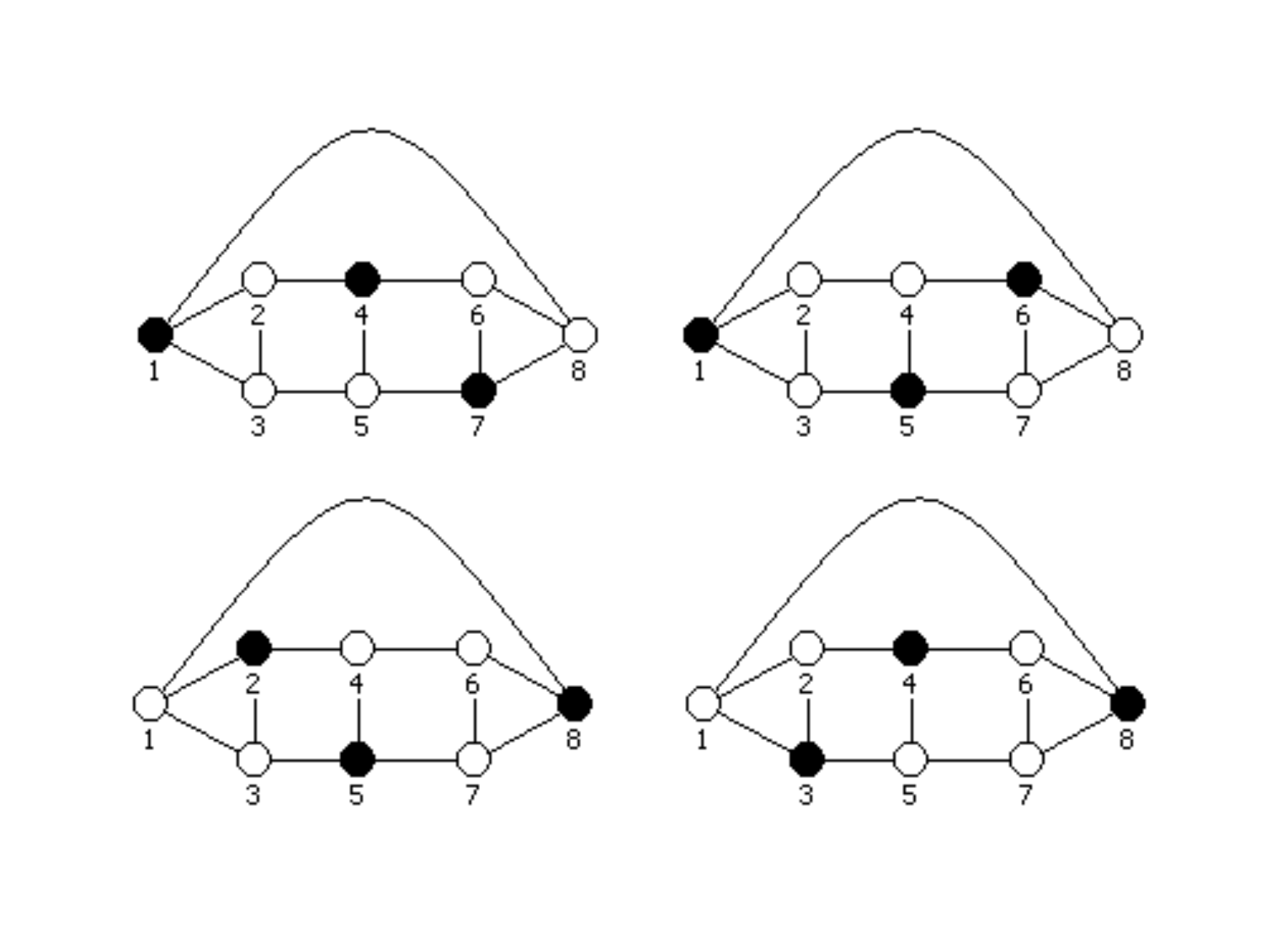} }
\textbf{\caption{Graphs of All Possible Ground States Separated Into Two Sets By Symmetry}}
\end{figure}

The random 8 qubit simulation in Figure \ref{fig:8Qa} has been specifically devised and investigated to confirm previously established relationships in Sections \ref{subsect:4Q} and \ref{subsubsect:8Qc}. By using the brute force method to find all possible MISs for this system, six different possibilities were found. Many related by symmetry as shown in Figures \ref{fig:8QrSet1} and \ref{fig:8QrSet2}.  

As there are a possible 256 ($2^n$) states that this system can occupy, so only the probabilities of the six MISs have been recorded in Table \ref{table:8Qr}, where $P_1$ is the individual final probability of the two states in set 1 and $P_2$ is the individual final probabilities of the four states in set 2. $P_T$ is the total final probability of all 6 states that represent MISs.

\begin{table}
\centering
\begin{tabular}{c||c|c|c|c|c|c|c}
\multicolumn{8}{l}{T=1}\\
H & 0 & 0.25 & 0.5 & 0.75 & 1 & 1.25 & 1.5\\
\hline
$P_1$ & 0.0098 & 0.0112 & 0.0126 & 0.0139 & 0.0149 & 0.0157 & 0.0162\\
$P_2$ & 0.0114 & 0.0131 & 0.0146 & 0.0160 & 0.0171 & 0.0178 & 0.0182\\
\hline
$P_T$ & 0.0652 & 0.0748 & 0.0836 & 0.0918 & 0.0982 & 0.1026 & 0.1052\\
\multicolumn{8}{l}{}\\
H & 1.75 & 2 & 2.25 & 2.5 & 2.75 & 3 & 3.25\\
\hline
$P_1$ & 0.0163 & 0.0161 & 0.0155 & 0.0148 & 0.0138 & 0.0127 & 0.0116\\
$P_2$ & 0.0181 & 0.0177 & 0.0169 & 0.0159 & 0.0147 & 0.0134 & 0.0121\\
\hline
$P_T$ & 0.1050 & 0.1030 & 0.0986 & 0.0932 & 0.0864 & 0.0790 & 0.0716\\
\multicolumn{8}{l}{}\\
\multicolumn{8}{l}{T=10}\\
H & 0 & 0.25 & 0.5 & 0.75 & 1 & 1.25 & 1.5\\
\hline
$P_1$ & 0.0006 & 0.0062 & 0.0291 & 0.0818 & 0.1535 & 0.2093 & 0.2269\\
$P_2$ & 0.0068 & 0.0063 & 0.0011 & 0.0049 & 0.0318 & 0.0729 & 0.1072\\
\hline
$P_T$ & 0.0285 & 0.0376 & 0.0626 & 0.1832 & 0.4342 & 0.7102 & 0.8826\\
\multicolumn{8}{l}{}\\
H & 1.75 & 2 & 2.25 & 2.5 & 2.75 & 3 & 3.25\\
\hline
$P_1$ & 0.2132 & 0.1885 & 0.1612 & 0.1149 & 0.0575 & 0.0238 & 0.0046\\
$P_2$ & 0.1254 & 0.1356 & 0.1425 & 0.1253 & 0.0710 & 0.0203 & 0.0025\\
\hline
$P_T$ & 0.9280 & 0.9194 & 0.8924 & 0.7310 & 0.3990 & 0.1288 & 0.0192 \\
\multicolumn{8}{l}{}\\
\multicolumn{8}{l}{T=25}\\
H & 0 & 0.25 & 0.5 & 0.75 & 1 & 1.25 & 1.5\\
\hline
$P_1$ & 0.0000 & 0.0004 & 0.0022 & 0.0310 & 0.1585 & 0.2892 & 0.3051\\
$P_2$ & 0.0002 & 0.0000 & 0.0005 & 0.0080 & 0.0453 & 0.0787 & 0.0931\\
\hline
$P_T$ & 0.0008 & 0.0009 & 0.0065 & 0.0940 & 0.4982 & 0.8932 & 0.9826\\
\multicolumn{8}{l}{}\\
H & 1.75 & 2 & 2.25 & 2.5 & 2.75 & 3 & 3.25\\
\hline
$P_1$ & 0.2849 & 0.2631 & 0.2403 & 0.2057 & 0.1224 & 0.0109 & 0.0000\\
$P_2$ & 0.1037 & 0.1147 & 0.1263 & 0.1399 & 0.1164 & 0.0102 & 0.0000\\
\hline
$P_T$ & 0.9846 & 0.9850 & 0.9858 & 0.9710 & 0.7104 & 0.0626 & 0.0000\\
\end{tabular}
\textbf{\caption{Table of Obtained Final Probabilities for Independent Set States in Specified Circumstances\label{table:8Qr}}}
\end{table}

From Table \ref{table:8Qr} it is clear that the simulation has found the correct MISs, in some cases with a total probability of measurement of 0.9848 for 6 states out of a possible 256. This is good reason to believe that the simulation has worked, considering that if all states in this system were equally probable, the total probability of 6 states would amount to a value of 0.0234.

\begin{figure} 
\centering 
\subfloat[Probabilities as Functions of Time]{
\label{fig:8QrProbsT} 
\includegraphics[width=\textwidth]{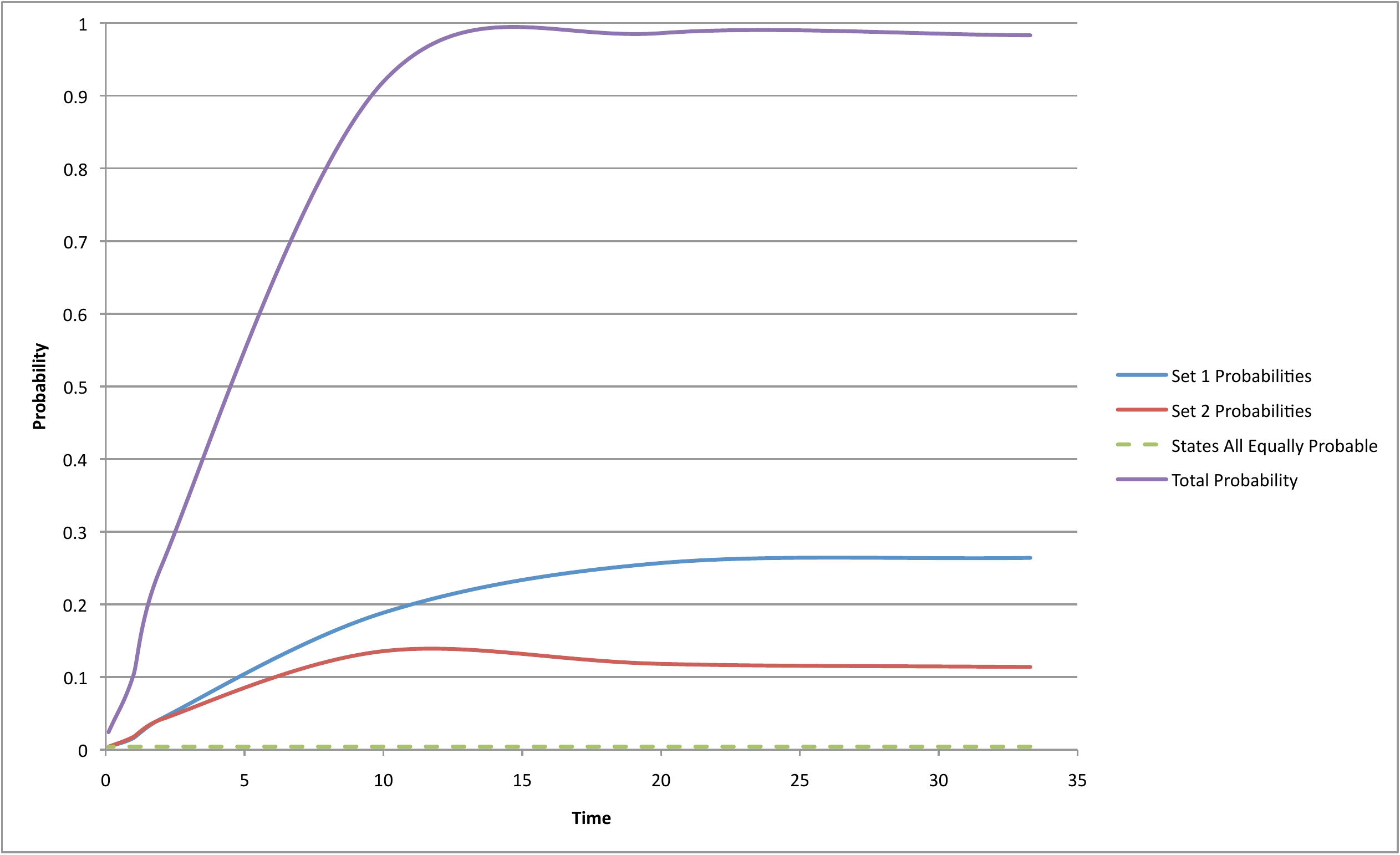} }\\
\subfloat[Probabilities as Functions of Final Hamiltonian Magnitude]{
\label{fig:8QrProbsH}
\includegraphics[width=\textwidth]{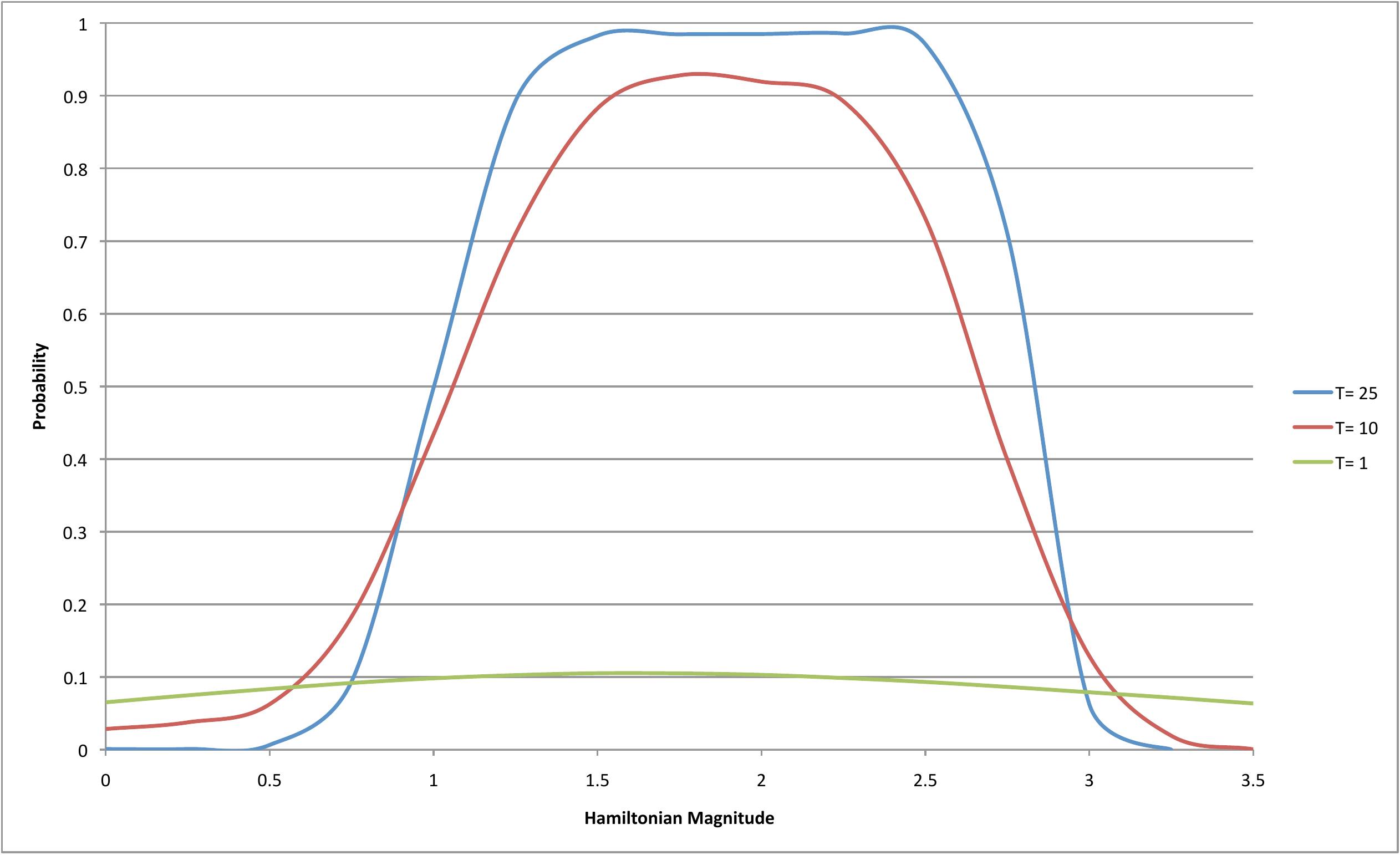} }
\textbf{\caption{Plots of the Probabilities Obtained for the Random 8 Qubit Graph}}
\end{figure}

Plots of the individual set probabilities and total probabilities have been compiled as functions of time and local Hamiltonian magnitudes. These are viewable as Figures \ref{fig:8QrProbsT} and \ref{fig:8QrProbsH}. The plots of probability as functions of total time of the evolution in Figure \ref{fig:8QrProbsT} reveal that the adiabatic theorem still holds strong. Similarly to simulation in previous systems, the probability saturates to a value very close to 1 at large units of time, around 10 to 15 units. Interestingly the individual sets appear to have more symmetry than initially expected. The two MISs in set 1 converge to have a probability of about 0.25 each at saturation and ideally in total this would amount to exactly half of the available probability of states. The four MISs in set 2 appear to converge to a value of 0.125 each and their cumulative total would also amount to a value of 0.5 in an ideal system. By the end of the evolution all of the states that represent MISs have an identical eigenvalue of -10 and the states would be in some superposition with entanglement throughout the system. This entanglement is the most likely the cause for the interesting symmetry between states, although as previously explained, no method yet exists to calculate the entanglement between more than two qubits, and leaves this phenomena as a matter for further investigation.

Figure \ref{fig:8QrProbsH} shows the symmetry across various magnitudes of local Hamiltonian. There are two effects occurring within this plot, firstly non-MISs becoming more favourable and secondly the adiabatic theorem. When all lines converge either side of the centre near Hamiltonian values 0.75 and 3, these are the points at which other states (non-MISs) become more favourable. Interestingly at this point, the values for probability at T=1 are the highest, this is induced due to the adiabatic theorem. With these local Hamiltonians ($0.75>H>3$), other states become more energetically favourable, so evolving the system slowly the correct MISs measurement becomes less probable, although evolving the system too quickly will cause almost all states to have an almost equal probability of measurement. Towards the centre of the plot, there is clearly some symmetry across the values of local Hamiltonian, although by simply evolving the system even slower will cause almost all probabilities to become close to 1, for a large range of Hamiltonian values, at a minimum between 1.5 and 2.

\section{Conclusion} \label{sect:Con}

This project has exploited the isomorphic mapping of the ground states of multi-qubit systems (with anti-ferromagnetic coupling) to the well known MIS problem. This permitted an investigation into the magnitude of the required local Hamiltonian for the development of a general algorithm. It is clear that some energy penalty must be introduced via the local Hamiltonian to all qubits, so states that involve qubit pairs of the same orientation cannot appear in both parities. As in these cases neither parity can represent an independent set.

Significant progress has been made on understanding the basic nature of the MIS problem. By observing the eigenvalue evolutions and the probabilities of measuring the ground state of the system under varying magnitudes of local Hamiltonian, a better assumption of the Hamiltonian that must be applied to yield a ground state that represents a MIS on all possible new systems can be made. For the 4 qubit system in Section \ref{subsect:4Q}, it was found that a value for the local Hamiltonian between 1 and 3 was appropriate and values less than 1 or greater than 3 cause the ground state of the system to be a state other than one that represents a MIS. Furthermore, it was found that at Hamiltonian magnitudes of exactly 1 and 3, the various possible states that represent MISs shared their eigenvalue with states that did not, therefore values between 1.5 and 2.5 are considered to be more acceptable. The simplest 8 qubit example that is referred to as a cube in Section \ref{subsubsect:8Qc} was a special case, where without an acting Hamiltonian the system has a perfect ground state where all Ising interactions can be satisfied. The use of this system, introduced the possibility of applying a Hamiltonian where it is not necessary for the ground state to represent a MIS, this is useful to find a maximum threshold Hamiltonian. Essentially the point at which the energy penalty introduced by the local Hamiltonian (acting on all qubits) causes a ground state that does not represent a MIS to become most probable at measurement after the evolution. The threshold was found to be when the Hamiltonian magnitude was 3. This is good evidence that applying a Hamiltonian to find a MIS can work with any system even if not entirely necessary, allowing a general algorithm to be established for this problem. The random 8 qubit system in Section \ref{subsubsect:8QRand} simply confirmed previous results, although due to the sharpening of the peak by the adiabatic theorem in Figure \ref{fig:8QrProbsH}, it appears that the symmetry in local Hamiltonian is not exactly centred about the value of 2. A basic Figure that can be read from the plot is about 1.75. A solid relationship has yet to be established for the ratio of coupling strength to local Hamiltonian magnitude, although if the system is evolved sufficiently slowly a range of values can be used, which in turn can yield some very interesting results (see Section \ref{subsect:RandFinal}).

The scalability of this problem to larger systems is expected to work from results obtained in Section \ref{subsect:MISP}, although due to computational restrictions within the student version of Matlab no system over 8 qubits can be simulated with the general program that has been written for this project. Implementation of the general algorithm on larger systems should not cause any issues, this is mainly due to the restrictions in place on the type of problem that can be simulated. For instance, the fact that the systems have been restricted to be cubic. Meaning that every qubit can only be linked to three others, this in turn allows for the scalability of the algorithm to larger systems, as only the three nearest neighbour interactions of every qubit need to be satisfied.

Very little of this project has been spent on the time necessary to evolve a system adiabatically. The theorem has been shown to hold, as in all cases the results became increasingly accurate with larger evolution time. There are various reasons for not spending much time on this subject, mainly the fact that the simulation remains to be calibrated, meaning that if any analysis were to be made, there could be no comparison made to any physically realisable system of qubits. Secondly, the main research topic of this paper was to investigate the implementation of an NP-Hard problem to AQC with the hope of developing a general algorithm for a specific problem.

Finally, the MIS problem is the minimisation version of the NP-Complete problem known as the Independent Set problem. Although, this pair of counterpart problems is very unique for a single reason; solving the NP-Hard problem and finding the cardinality of the MIS for any system will also solve all instances of the NP-Complete counterpart. By simply determining the max $k$ value of a graph G=(V,E), determining if $k$ is greater of less than some nominal value is trivial. Hence solving all instances of the NP-Complete counterpart.

In conclusion, to obtain a visual representation and the cardinality of a MIS of any given cubic planar system, one must apply a final Hamiltonian in an adiabatic evolution that contains only interactions that favour anti-alignement ($J_{ij}= -1$) with an applied Hamiltonian of magnitude 2 ($h_i=2$) for all qubits. This is in agreement with the mathematical analysis in Section \ref{subsect:MISP}.

\section{Recommendations for Further Research} \label{sect:Further}

There are many areas of AQC that remain to be explored. Below is an extensive list of various effects that could in theory either quicken or improve the results of AQC.

\subsection{Quantum Annealing} \label{subsect:Anneal}

With reference to the 8 qubit cube simulation in Section \ref{subsubsect:8Qc}, the probabilities gained by the standard method are expected to increase by doubling the magnitude of the initial Hamiltonian, in effect introducing some sort of quantum annealing. The relationship in Equation \ref{equ:Min} that gives a sufficient evolution time so that the system is evolved adiabatically shows that the time is dependent on the minimum gap between the two lowest eigenvalues. By doubling the magnitude of the initial Hamiltonian, the gap at the start of the simulation is also doubled, and in many situations the initial gap is in fact the minimum gap over the whole simulation. With this increase of initial eigenvalue gap, the simulations in Table \ref{table:8Qprobs} have been re-run to yield the new results in Table \ref{table:8Qprobsnew}.

\begin{table}
\centering
\begin{tabular}{c||c|c|c}
{} & \multicolumn{3}{c}{Hamiltonian Magnitude}\\
{Time} & 0 & 1 & 2\\
\hline
0.1 & 0.0080 (0.0082) & 0.0084 (0.0082) & 0.0084 (0.0082)\\
1 & 0.1337 (0.0608) & 0.1063 (0.0522) & 0.0543 (0.0342)\\
2 & 0.4245 (0.233) & 0.3228 (0.2024) & 0.1002 (0.0958)\\
10 & 0.9458 (0.8374) & 0.9334 (0.8496) & 0.9455 (0.8704)\\
20 & 0.9848 (0.9678) & 0.9822 (0.9698) & 0.9842 (0.9656)\\
\end{tabular}
\textbf{\caption{Combined Final Probability of Correct Two States for Varying Time and Hamiltonian Magnitude for a Cube of Qubits with the Initial Hamiltonian Doubled in Magnitude where Brackets Indicate Previous Values\label{table:8Qprobsnew}}}
\end{table}

In almost all cases the new probabilities exceed those of previous simulations. Indicating that greater gaps in eigenvalues do indeed reduce the possibility of Landau-Zener tunnelling to excited states. The system with no applied local Hamiltonian on the qubits has a minimum gap of 4 due to the initial Hamiltonian but increases towards the end of the simulation, so the probability of measuring a correct state does in fact increase dramatically. The system where the local Hamiltonian has a magnitude of 1 on all qubits, the gap is 4 at both start and end of the simulation, as the gap does not increase more towards the end of the evolution, the probability of measuring a correct final state does increase but not as much as in the scenario of no local Hamiltonian. The simulation where the local Hamiltonian applied to all qubits has a magnitude of 2, still has a minimum gap of 2 due to the final Hamiltonian, although as the simulation started with a larger gap, the probability of measuring a correct final state still increases. This small exercise has shown that the eigenvalue gap between correct and incorrect states does affect the system undergoing adiabatic evolution, although it does seem that the dependence of total run time is not solely on the minimum gap, but the evolution of the gap.

Therefore further research needs to be conducted on the possibilities of increasing Hamiltonian Magnitudes to yield greater eigenvalue gaps.

\subsection{Variable Rate Evolution}

Similarly to Section \ref{subsect:Anneal}, due to probabilities increasing (although minimum gap being static in various scenarios) an investigation into the use of variable rate evolution may yield greater speed-up of computation, yet the question of at which point the speed-up should be applied (and for how long) during the evolution remains to be answered. 

\subsection{Random Local Hamiltonian Magnitudes} \label{subsect:RandFinal}

For the random 8 qubit simulation, there is a clear range of local Hamiltonians that give extremely high probabilities for the correct states (viewable in Figure \ref{fig:8QrProbsH}). To induce eigenvalue gaps between correct states, using a randomly distributed set of local Hamiltonian within some finite range of magnitude on every qubit should prove effective.

\begin{table}
\centering
\begin{tabular}{c||c|c}
State & T= 10 & T=25 \\
\hline
\hline
$\vert 11001101\rangle$ & 0.1281 (0.2132) & 0.1077 (0.2403) \\
$\vert 10110011\rangle$ & 0.1701 (0.2132) & 0.2034 (0.2403) \\
\hline
$\vert 01101101\rangle$ & 0.1301 (0.1254) & 0.0997 (0.1263) \\
$\vert 01110011\rangle$ & 0.1375 (0.1254) & 0.1176 (0.1263) \\
$\vert 10110110\rangle$ & 0.1837 (0.1254) & 0.2302 (0.1263) \\
$\vert 11001110\rangle$ & 0.1881 (0.1254) & 0.2212 (0.1263) \\
\hline
\hline
Total & 0.9376 (0.9280) & 0.9798 (09858)
\end{tabular}
\textbf{\caption{Table of Probabilities Obtained for All States Representing Max Independent Sets After Application of Randomly Distributed Local Hamiltonian Magnitudes\label{table:8Qrandomlocal}}}
\end{table}

The random magnitudes for the various qubits were:
\begin{equation*}
\small{h_1= 1.78, h_2= 1.8, h_3= 1.85, h_4= 1.73, h_5= 1.77, h_6= 1.82, h_7= 1.9, h_8= 1.69}
\end{equation*}

These were completely randomly selected, although care was taken to make sure the values did not differ much from the centre of Hamiltonian magnitude symmetry found in Section \ref{subsubsect:8QRand}. The values in brackets in Table \ref{table:8Qrandomlocal} are those of previous bests taken from Table \ref{table:8Qr}, therefore for T= 10, the values in brackets are previously from the local Hamiltonian magnitude of 1.75 and for T= 25 the previous magnitude being 2.25. It is clear that for T= 10, the probability of measuring a correct final state has in fact increased to a Figure better than that of the previous maximum, although for T= 25, the value has slightly decreased, even though it is itself an impeccably high probability.

This study could be taken further to variations of the coupling strengths between qubits. Furthermore, this effect could be used to model errors in control parameters, as with all devices there is a finite internal error.

\subsection{More Highly Connected Graphs}

Looking at systems that are more highly connected than the standard cubic type is an option. This could lead to the development of a general algorithm across all types of graph for the MIS problem. By simply repeating the analysis in Section \ref{subsect:MISP} for highly connected graphs, a simple mathematical proof may be deduced. This has been performed for all combinations in a system where all qubits have four interactions with nearest neighbours. The inequality developed to find the maximum Hamiltonian magnitude is: $-3H+4J < -5H-4J$. This gives an upper bound of $H<4$. Study of the previous upper bound inequality for the cubic system along with this inequality gives a general upper bound inequality for all systems as: $-(n-1)H+nJ < -(n+1)H-nJ$, where $n$ is the number of nearest neighbour interactions on each qubit. Therefore a general upper bound is given by $H<n$. Although a lower bound inequality must be developed and simulations of highly connected systems are appropriate.

This research may also extend to the implementation of highly connected graphs onto a two dimensional planar chip. Optimisation algorithms do exist for the drawing of cubic planar graphs on regular lattices of qubits (see reference \cite{Saidur:1999}), although very little research has been conducted on problems where qubits are required to have a greater number of couplings. With chips that have been manufactured to date within the industry, the maximum number of couplings achieved is four, with edge and corner qubits having less. If this trend is set to continue, research into the some form of algorithm to implement problems on currently manufactured chips where qubits require more than four couplings is necessary. This will inevitably involve the use of strong ferromagnetic Ising interactions between neighbouring qubits that will effectively cause two qubits to act as one, with their combined remaining couplings accessible for needed problem interactions.

\subsection{Other NP-Hard Problems}

\begin{figure}
\centering 
\includegraphics{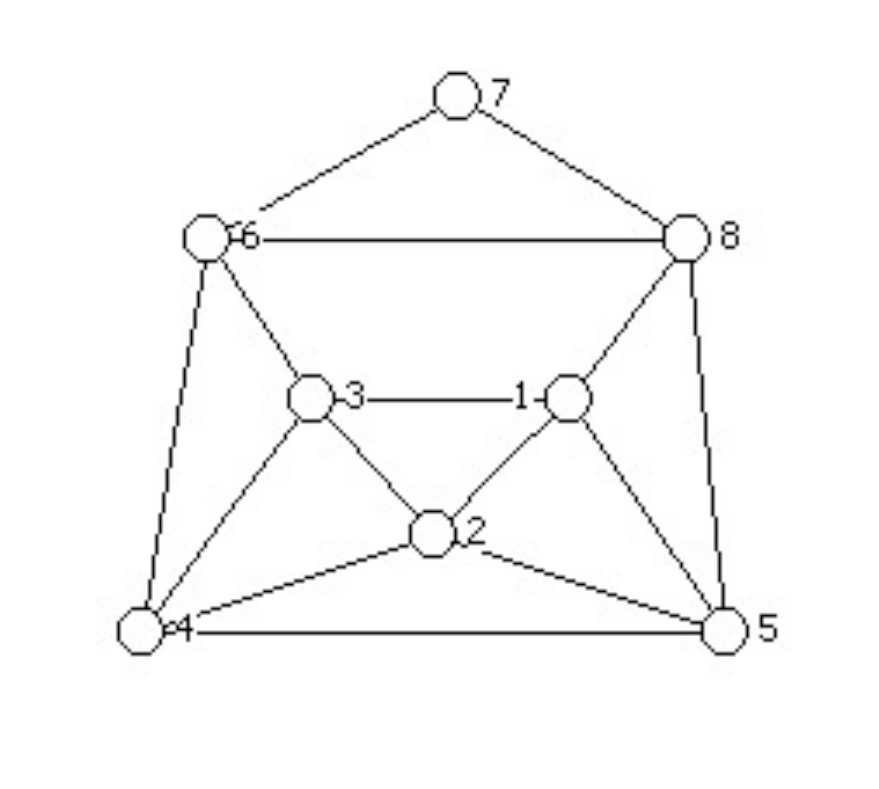}
\textbf{\caption{8 Qubits in a Randomly Selected Formation to Demonstrate an Adaptation of the Exact Cover Problem\label{fig:exact}} }
\end{figure}

There are many variations of NP-Hard problems, and new problems can always be devised (whether there is any use for these newly devised problems is another matter). However extensive lists are easily accessible\footnote{For example: http://www.csc.kth.se/\~{}viggo/wwwcompendium/node276.html} and algorithms for these other problems may be devised.

A good example problem is an adaptation of the Exact Cover problem. Given a number of qubits, each of which can take a value 0 or 1, and a set of clauses that each involves three qubits, whereby one qubit in every clause is of opposite orientation to the other two. Satisfying every clause will be an example of the adapted Exact Cover. Without researching the problem further and by carefully selecting the clauses so that no two clauses overlap by more than one qubit. The problem appears to be analogous to finding the ground state of a frustrated system, by connecting sets of three qubit in each clause into triangles and applying antiferromagnetic Ising interactions on every coupling, finding the ground state will give an adapted Exact Cover.

As 8 qubits is the maximum that can be simulated, the following clauses of three qubits were randomly selected: 123, 346, 245, 158 and 678. These clauses allow for the system to be drawn in planar form, although it is important to note that this system is not cubic, as seen in Figure \ref{fig:exact}.

After coding the system and running the simulation with an $\Omega$ of 0.1, there were many states that ended the simulation with probabilities close to that of the maximum. After checking a selection of the highest probability state it was clear that they all satisfied all encoded clauses. A selection of those are as follows: $\vert 11010110\rangle , \vert 11010101\rangle , \vert 11010100\rangle , \vert 11010011\rangle , \vert 11010011\rangle$ , etc. By including further clauses less states will end the simulation with high probability.

This is quite a crude example, and only shown to work on a very specific system. By researching this further on a variety of systems, using a variety of clauses, only then can a general algorithm for this problem be fully developed.

\newpage
\bibliographystyle{unsrt}
\bibliography{References}

\section{Appendix} \label{sect:App}
\pagenumbering{roman}
\setcounter{page}{1}

\subsection{Programming Code}
\subsubsection{Instructions to User}

This code requires a Matlab installation of version 7.4 or greater. The file `SystemSolver.m' should be saved to a desired directory and this directory must be selected as the `current directory' in the Matlab window.

Initially the user must input data into the M-file depending on the type of system. The areas for this are designated in the code with examples of typical inputs. Running the M-file can be performed in one of two ways, if the M-file is open, the `run' button can be pressed in this window. Otherwise typing 'run SystemSolver' in the main Matlab window will perform the same operation.

After running the the user will be posed a series of questions. Firstly, a value of $\Omega$ must be input by giving a typical number and pressing the `enter' key. If no number is given, the code will use a default value of 0.1. Secondly the number of qubits in the system is input, this is obviously also dependent on the user and input in the same fashion. Finally, the simulation will ask the user to indicate the magnitude of the local Hamiltonian acting on the qubits, in most cases this will be the same for all qubits and can be input in this manner. Otherwise the user can indicate the individual local Hamiltonians before running the simulation in the M-file itself, the area for this is also clearly labelled with examples.

Once theses inputs have been successfully completed the solver will continue to solve the system and output three to four plots and all final probabilities with corresponding states in the main Matlab window. 

There are two known bugs that may randomly crop up. Firstly, the program writes the system of ODEs to a separate M-File (named Hamiltonian\_ODE.m) in the current directory. If this file fails to be completed before the program goes on the solve the system of ODEs stored in that file, the simulation will be interrupted. Running the simulation a second time will bypass this issue. Secondly, to plot all of the probabilities as a function of time, the plotting function is contained in a loop. The colour of each state's probability is determined by three random number generators giving individual values for Red, Green and Blue. An error may occasionally crop up at this point if two or more lines have identical colours, although this rests solely on all three random number generator generating identical numbers for all three inputs, all of which have three decimal places. The likelihood of this occurring is minimal although in some instances it could happen.

\newpage
\landscape
\begin{multicols}{2}
\subsubsection{The Code}
\singlespacing
{\tiny
clear all\newline
global Om\newline
Om = input('Please state a value of Omega:');       \% Input\newline
if isempty(Om)\newline
    Om = 0.1;\newline
end\newline
if Om==0;\newline
    tspan=[0:0.01:10];\newline
else\newline
    tspan=[0:0.01:1/Om];\newline
end\newline
Nq= input('Please enter the number of Qubits: ');    \% Input\newline
totalh=input('Indicate local bias value: ');\newline
Jij= zeros(Nq);\newline
Hi= zeros(Nq,1)+totalh; \%can change if local biases occur\newline
\newline
\%List all local Hamiltonians below e.g. Hi(1)=2 if they must differ from\newline
\%one another.\newline
\%Hi(1)=1.5;\newline
\newline
\%List all interactions below, e.g: Jij(1,3)=7\newline
\% Jij(1,2)=-1; \%4 qubit cubic\newline
\% Jij(1,3)=-1;\newline
\% Jij(1,4)=-1;\newline
\% Jij(2,3)=-1;\newline
\% Jij(2,4)=-1;\newline
\% Jij(3,4)=-1;\newline
\newline
\% Jij(1,2)=-1;  \%cube\newline
\% Jij(1,3)=-1;\newline
\% Jij(1,7)=-1;\newline
\% Jij(2,4)=-1;\newline
\% Jij(2,8)=-1;\newline
\% Jij(3,4)=-1;\newline
\% Jij(3,5)=-1;\newline
\% Jij(4,6)=-1;\newline
\% Jij(5,6)=-1;\newline
\% Jij(5,7)=-1;\newline
\% Jij(6,8)=-1;\newline
\% Jij(7,8)=-1;\newline
\newline
\% Jij(1,2)=-1; \%random graph\newline
\% Jij(1,3)=-1;\newline
\% Jij(1,4)=-1;\newline
\% Jij(2,5)=-1;\newline
\% Jij(2,6)=-1;\newline
\% Jij(3,4)=-1;\newline
\% Jij(3,8)=-1;\newline
\% Jij(4,5)=-1;\newline
\% Jij(5,7)=-1;\newline
\% Jij(6,7)=-1;\newline
\% Jij(6,8)=-1;\newline
\% Jij(7,8)=-1;\newline
\newline
\%Finding all the SIGMAz's and SIGMAx's for the system\newline
I=[1 0; 0 1];\newline
single\_sigmaz=[1 0; 0 -1];\newline
single\_sigmax=[0 1; 1 0];\newline
sum\_sigmax=0;\newline
\newline
for s=1:Nq\newline   
    for q=Nq:-1:1    \newline
        if q==s\newline
            a= single\_sigmaz;\newline
            y=single\_sigmax; \newline
        else\newline
            a=I;\newline
            y=I;\newline
        end\newline
        \newline
        if q==Nq\newline          
            if q==s\newline
                b= single\_sigmaz;\newline
                z=single\_sigmax;\newline
            else\newline
                b=I;\newline
                z=I;\newline
            end\newline
        else\newline
            z=kron(y,z);\newline
            b=kron(a,b);\newline
        end\newline
    end\newline
    sum\_sigmax=sum\_sigmax+z;\newline
    sigmax(:,:,s)=z;\newline
    sigmaz(:,:,s)=b;\newline
end\newline
\newline
\%Finding the ground state vector of the initial Hamiltonian\newline
\%sum\_sigmax can print to check\newline
[V,D] = eig(sum\_sigmax);\newline
Eig\_Vector=eig(D);\newline
MinEig=min(Eig\_Vector);\newline
size\_EV=size(Eig\_Vector);\newline
\newline
for w=1:size\_EV(1)\newline
    if Eig\_Vector(w)==MinEig\newline
        z=w;\newline
    end\newline
end\newline
\newline
x0= V(:,z);\newline
\newline
\%Needs to find an easy form of writing the Hamiltonains\newline
\%\%\%\%\%\%\%\%\%\%\%\%\%\%\%\%\%\%\%\%\%\%\%\%\%\%\%\%\%\newline\newline
Hb= sum\_sigmax;\newline
\newline
Vector=0;\newline
for aa=1:Nq\newline
    Vector=Vector+(Hi(aa)*sigmaz(:,:,aa));\newline
end\newline
\newline
Matrix=0;\newline
for ab=1:Nq\newline
    for ac=1:Nq\newline
        Matrix=Matrix+(Jij(ab,ac)*(sigmaz(:,:,ab)*sigmaz(:,:,ac)));\newline
    end\newline
end\newline
\newline
Hp= Vector-Matrix;\newline
\newline
\%\%\%\%\%\%\%\%\%\%\%\%\%\%\%\%\%\%\%\%\%\%\%\%\%\%\%\%\%\newline\newline
sizeHb=size(Hb);\newline
sizeHp=size(Hp);\newline
\newline
\%error checks on sizes of the Hamiltonians\newline
if sizeHb~=sizeHp\newline
    error ('Matrix dimensions do not agree');\newline
end\newline
\newline
if sizeHb(1)~=sizeHb(2)\newline
    error ('Matrix must be square');\newline
elseif sizeHp(1)~=sizeHp(2)\newline
    error ('Matrix must be square');\newline
end\newline
\newline
p=sizeHb(1);\newline
q=log(p)/log(2);\newline
if mod(q,1)~=0 \%0r ceil(q) ~= floor(q)\newline
    error('not an admissable number of ODE''s')\newline
elseif q~=Nq\newline
    error('Number of qubits doesn''t match')\newline
end\newline
\newline
\%Open file within which we'll write ODEs, always Hamiltonian\_ODE\newline
\%Writing general ODEs in correct Matlab form\newline
\newline
fid = fopen('Hamiltonian\_ODE.m', 'w+');\newline
\newline
fprintf(fid, 'function ode\_system = Hamiltonian\_ODE(t,x) $\backslash$ n$\backslash$ n\newline
global Om $\backslash$ n$\backslash$ n');\newline
for i=1:1:p;\newline
    fprintf(fid, 'ode\%d = -i*(',i);\newline
    for j=1:1:p;\newline
        fprintf(fid, '((\%d - Om*t*\%d + Om*t*\%d)*x(\%d))', Hb(i,j), Hb(i,j), Hp(i,j),j);\newline
        if j~=p\newline
            fprintf(fid,'+');\newline
        end\newline
    end\newline
    fprintf(fid, '); $\backslash$ n');\newline
end\newline
\newline
fprintf(fid, ' $\backslash$ node\_system=[');\newline
for k=1:1:p;\newline
    fprintf(fid,'ode\%d',k);\newline
    if k~=p\newline
        fprintf(fid,';');\newline
    end\newline
end\newline
fprintf(fid, '];$\backslash$ n');\newline
fclose(fid);\newline
\newline
[t,x]=ode45(@Hamiltonian\_ODE,tspan,x0);\newline
    \newline
x\_abs=abs(x);\newline
Sizex=size(x\_abs);\newline
\newline
\%error checking below\newline
\%vectors are same length\newline
for e=1:1:p-1\newline
    if length(x(e))~=length(x(e+1));\newline
        error ('absolute vectors are not the same length');\newline
    end\newline
end\newline
\newline
\% show that sum(coefficients\^2)=1 for averages\newline
    \newline
check=0;\newline
for f=1:1:p\newline
    temp\_check= x\_abs(:,f).\^2;\newline
    check=check+temp\_check;\newline
end\newline
\newline
if max(check) > 1.01\newline
    error ('trig rules broken');\newline
elseif min(check) < 0.98\newline
    error ('trig rules broken');\newline
end\newline
\newline
aq=x\_abs(Sizex(1),:).\^2;\newline
ax=max(aq);\newline
\newline
hold on \%Plot probabilities\newline
    figure(1)\newline
    for c=1:1:p\newline
        h(c)= plot(tspan, (x\_abs(:,c).\^2),'color',[rand rand rand]);\newline
        Bins = dec2bin(c-1,q);\newline
        Binaries(c)= cellstr(Bins);\newline
        \newline
        \% Can introduce for larger systems only printing states with\newline
        \% probabilities over a certain threshold. Uncomment 3 lines\newline
        \% below and comment out 4 lines beneath that. The threshold\newline
        \% is set to 0.1 below but can be changed.\newline
        \newline
\%         if aq(c)>(ax-0.1)    \%possible outcomes below a certain p\newline
\%         sprintf('State: \%s, has final probability: \%f', Bins, aq(c))\newline
\%         end\newline
\newline
        \%Or simply use the following:\newline
        \newline
        sprintf('State: \%s, has final probability: \%f', Bins, aq(c))\newline
        if aq(c)==ax\newline
            sprintf('State: \%s, has the maximum final probability of: \%f', Bins, aq(c))\newline
        end\newline
        \newline
    end\newline
    \newline
   legend(h, Binaries)\newline
    title('Probabilities')\newline
    xlabel('Time')\newline
    ylabel('Probability')\newline
\newline
hold off\newline
\newline
    r=1;\newline
    for t=tspan\newline
        d(r,:)= eig((1-Om*t)*Hb+Om*t*Hp);\newline
        r=r+1;\newline
    end\newline
    \newline
   figure(2) \%Plot Eigenvalues\newline
   plot(tspan, d);\newline
   title('Eigenvalues')\newline
   xlabel('Time')\newline
   ylabel('Eigenvalue')\newline
   \newline
   \%Find minimum gap between two lowest Eigenvalues\newline
   \newline
   gap= d(:,2)-d(:,1);\newline
   min\_gap= min(gap);\newline
   min\_gap1= num2str(min\_gap);\newline
   \newline
  figure(3) \%Plot minimum gap\newline
  plot(tspan, gap)\newline
  xlabel('Time')\newline
  ylabel('Lowest two eigenvalue gap')\newline
  text(0.2,0.2, ['Minimum gap= ' min\_gap1])\newline
  \newline
  \%\%\%\%\%\%\% For Individual Systems \%\%\%\%\%\%\%\%\%\%\%\%\newline\newline
  if p==2 \%Plot trajectory through Bloch Sphere\newline
    \newline
    \%finding values of theta \& phi\newline
\newline
    theta=2*acos(x\_abs(:,1));\newline
    c= x(:,2)./x(:,1);\newline
    phi= imag(log(c));\newline
\newline
    [X,Y,Z]=sph2cart(phi,-theta+(pi/2),1);\newline
\newline
    \%plotting a sphere to use as a basis:\newline
    i=0;\newline
    for time=[0:0.01:1]\newline
        i=i+1;    \newline
        rtheta(i)=2*pi*time;\newline
    end\newline
\newline
    L=length(rtheta);\newline
    ftheta1=ones(L,1).*(pi/6);\newline
    ftheta2=ones(L,1).*(2*pi/6);\newline
    ftheta3=ones(L,1).*(pi/2);\newline
    ftheta4=ones(L,1).*(4*pi/6);\newline
    ftheta5=ones(L,1).*(5*pi/6);\newline
    rtheta1=rtheta';\newline
\newline
    figure(4)\newline
    hold on\newline
    plot3([cos(ftheta1).*sin(rtheta1)], [sin(ftheta1).*sin(rtheta1)], [cos(rtheta1)],'b--')\newline
    plot3([cos(ftheta2).*sin(rtheta1)], [sin(ftheta2).*sin(rtheta1)], [cos(rtheta1)],'b--')\newline
    plot3([cos(ftheta3).*sin(rtheta1)], [sin(ftheta3).*sin(rtheta1)], [cos(rtheta1)],'b--')\newline
    plot3([cos(ftheta4).*sin(rtheta1)], [sin(ftheta4).*sin(rtheta1)], [cos(rtheta1)],'b--')\newline
    plot3([cos(ftheta5).*sin(rtheta1)], [sin(ftheta5).*sin(rtheta1)], [cos(rtheta1)],'b--')\newline
    plot3([cos(rtheta1).*sin(ftheta1)], [sin(rtheta1).*sin(ftheta1)], [cos(ftheta1)],'b--')\newline
    plot3([cos(rtheta1).*sin(ftheta2)], [sin(rtheta1).*sin(ftheta2)], [cos(ftheta2)],'b--')\newline
    plot3([cos(rtheta1).*sin(ftheta3)], [sin(rtheta1).*sin(ftheta3)], [cos(ftheta3)],'b--')\newline
    plot3([cos(rtheta1).*sin(ftheta4)], [sin(rtheta1).*sin(ftheta4)], [cos(ftheta4)],'b--')\newline
    plot3([cos(rtheta1).*sin(ftheta5)], [sin(rtheta1).*sin(ftheta5)], [cos(ftheta5)],'b--')\newline
    \%include a plot of the actual results:\newline
    plot3(X, Y, Z,'r','LineWidth',3)\newline
 \newline
    xlabel('X')\newline
    ylabel('Y')\newline
    zlabel('Z')\newline
    title('Trajectory through Bloch Sphere')\newline
    hold off\newline
    \newline
  elseif p==4 \%Plot the concurrence for 2 qubit system\newline
      \newline
      ent= 2.*((x(:,1).*x(:,4))-(x(:,2).*x(:,3)));\newline
      \newline
      figure(4)\newline
      plot(tspan, abs(ent))\newline
      xlabel('Time')\newline
      ylabel('Concurrence')\newline
      title('Plot of concurrence during evolution')\newline     
      \newline
  end\newline}
  \end{multicols}

\end{document}